\def\versionno{ fjfrs2 (=fact-alg) - version 10.0 -- by jf -- 26.08.08 }

\newcommand{\void}[1]{}

\newcommand\eqpic[4]{\begin{eqnarray}
                   \begin{picture}(#2,#3){}\end{picture}\nonumber\\
                   \raisebox{-#3pt}{ \begin{picture}(#2,#3) #4 \end{picture} }
                   \label{#1} \\~\nonumber \end{eqnarray} }
\newcommand\eqpicc[4]{\begin{eqnarray}
                   \begin{picture}(#2,#3){}\end{picture}\nonumber\\
                   \raisebox{-#3pt}{ \begin{picture}(#2,#3) #4 \end{picture} }
                   \nonumber \\~\label{#1} \end{eqnarray} }

\newcommand\eqpiczz[5]{\begin{eqnarray}
                   { \begin{picture}(#2,#3) #4 \end{picture} }\nonumber\\
                   \raisebox{-#3pt}{ \begin{picture}(#2,#3) #5 \end{picture} }
                   \nonumber \\~\label{#1} \end{eqnarray} }

\void{
\catcode`\@=11
\newif\if@fewtab\@fewtabtrue
{\count255=\time\divide\count255 by 60
\xdef\hourmin{\number\count255}
\multiply\count255 by-60\advance\count255 by\time
\xdef\hourmin{\hourmin:\ifnum\count255<10 0\fi\the\count255}}
\def\ps@draft{\let\@mkboth\@gobbletwo
    \def\@oddfoot{\hbox to 7 cm{\tiny \versionno
       \hfil}\hskip -7cm\hfil\rm\thepage \hfil {\tiny\draftdate}}
    \def\@oddhead{}
    \def\@evenhead{}\let\@evenfoot\@oddfoot}
\def\draftdate{\number\month/\number\day/\number\year\ \ \ \hourmin }

\def\citen#1{\if@filesw \immediate\write \@auxout {\string\citation{#1}}\fi%
\@tempcntb\m@ne \let\@h@ld\relax \def\@citea{}%
\@for \@citeb:=#1\do {\@ifundefined {b@\@citeb}%
    {\@h@ld\@citea\@tempcntb\m@ne{\bf ?}%
    \@warning {Citation `\@citeb ' on page \thepage \space undefined}}%
    {\@tempcnta\@tempcntb \advance\@tempcnta\@ne
    \setbox\z@\hbox\bgroup\ifcat0\csname b@\@citeb \endcsname \relax
    \egroup \@tempcntb\number\csname b@\@citeb \endcsname \relax
    \else \egroup \@tempcntb\m@ne \fi \ifnum\@tempcnta=\@tempcntb
    \ifx\@h@ld\relax \edef \@h@ld{\@citea\csname b@\@citeb\endcsname}%
    \else \edef\@h@ld{\hbox{--}\penalty\@highpenalty
    \csname b@\@citeb\endcsname}\fi
    \else \@h@ld\@citea\csname b@\@citeb \endcsname \let\@h@ld\relax \fi}%
\def\@citea{,\penalty\@highpenalty\hskip.13em plus.13em minus.13em}}\@h@ld}
\def\@citex[#1]#2{\@cite{\citen{#2}}{#1}}%
\def\@cite#1#2{\leavevmode\unskip\ifnum\lastpenalty=\z@\penalty\@highpenalty\fi%
  \ [{\multiply\@highpenalty 3 #1%
  \if@tempswa,\penalty\@highpenalty\ #2\fi}]}   %
\makeatother 
\catcode`\@=12
}

\newcounter{defthm}

\def\Alpha         {\aleph}
\def\be            {\begin{equation}}
\def\bea           {\begin{equation}\begin{array}l\displaystyle}
\def\beaa          {\begin{equation}\begin{array}{ll}\displaystyle}
\def\ee            {\end{equation}}

\def\enL           {\\[.5em]\displaystyle}
\newcommand\erf[1] {(\ref{#1})}
\newcommand\labl[1]{\label{#1}\ee}
\def\nxt           {\raisebox{.08em}{\rule{.44em}{.44em}}\hspace{.4em}}
\newcommand\sect[1]{\section{#1}\setcounter{equation}0\setcounter{defthm}0}
\def\sse           {\scriptsize}

\newcommand\epicture[2] {\end{picture}\\{}\\[#1.#2em]\end{array}}

\newcommand\Includeourbeautifulpicture[3] {{\begin{picture}(0,0)(0,0)
            \scalebox{.38}{\includegraphics{pic_fjfrs2_#1#2#3.eps}} \end{picture}}}
\newcommand\Includeournicelargepicture[3] {{\begin{picture}(0,0)(0,0)
            \scalebox{.28}{\includegraphics{pic_fjfrs2_#1#2#3.eps}} \end{picture}}}
\newcommand\IncludeourniceMediumpicture[3] {{\begin{picture}(0,0)(0,0)
            \scalebox{.48}{\includegraphics{pic_fjfrs2_#1#2#3.eps}} \end{picture}}}
\newcommand\Includeournicemediumpicture[3] {{\begin{picture}(0,0)(0,0)
            \scalebox{.55}{\includegraphics{pic_fjfrs2_#1#2#3.eps}} \end{picture}}}

\newcommand\Includeournicesmallpicture[3] {{\begin{picture}(0,0)(0,0)
            \scalebox{.65}{\includegraphics{pic_fjfrs2_#1#2#3.eps}} \end{picture}}}

\def\bl             {B\hspace*{-.5pt}\ell}
\def\Bl             {{B\hspace*{-.5pt}\ell\hspace*{.5pt}}^{\mathrm c}_{\phantom|}}
\def\BlxBr          {B_l{\times}\overline{B_r}}
\def\C              {\ensuremath{\Cc}}
\def\CAA            {\ensuremath{\Cc_{\!A|A}}}
\def\Cdot           {\,{\cdot}\,}
\def\cft            {conformal field theory}
\def\Circ           {\,{\circ}\,}
\def\Cong           {\,{\cong}\,}
\def\cor            {C\hspace*{-.9pt}o\hspace*{-.6pt}r}
\def\Cor            {C\hspace*{-.9pt}o\hspace*{-.6pt}r^{\mathrm c}}
\def\corA           {C\hspace*{-.9pt}o\hspace*{-.6pt}r^{A}} 
\def\corP           {C\hspace*{-.9pt}o\hspace*{-.6pt}r'}    

\def\CxCb           {{\Cc\hspace*{+.05em}{\boxtimes}\hspace*{+.1em}\overline\Cc}}
\def\dimc           {\dim_\Cb}
\def\disc           {disk}
\def\dsty           {\displaystyle }
\def\eps            {\varepsilon}
\def\eq             {\,{=}\,}
\def\fill           {\text{fl}}
\def\hb             {T}  
\def\Hcl            {H_\mathrm{cl}}
\def\hM             {h}
\def\Hom            {\mathrm{Hom}}
\def\HomAA          {\mathrm{Hom}_{\!A|A}}
\def\Hop            {H_\mathrm{op}}
\def\Htft           {\tftC}
\def\id             {\mbox{\sl id}}
\def\idsmall        {\mbox{\footnotesize\sl id}}
\def\iN             {\,{\in}\,}

\def\koerper        {\ensuremath{\Bbbk}}
\def\Mapsto         {\,{\mapsto}\,}
\def\mcc            {\multicolumn2{|c|}}
\def\mcl            {\multicolumn2{|l|}}
\def\Mult           {{\mathcal M}\hspace*{-.9pt}\mbox{\sl lin}}
\def\Obj            {\mathcal{O}bj}
\def\Omeka          {K}  
\def\one            {{\bf1}}
\def\oti            {\,{\otimes}\,}
\def\Oti            {{\otimes}}
\def\otic           {\,{\otimes_\Cb}\,}
\def\ol             {\overline}
\def\qed            {\hfill\checkmark\medskip}

\def\RepVc          {{\mathcal R}\hspace*{-.5pt}\mbox{\sl ep}(\Vc)}
\def\sew            {{\mathscr S}}
\def\sfc            {\mathsf{S}}
\def\tft            {topological field theory}
\def\tftC           {t\hspace*{-1.2pt}f\hspace*{-1.2pt}t_\Cc}
\def\Times          {\,{\times}\,}
\def\To             {\,{\rightarrow}\,}

\def\Tri            {\mathrm{T}}  
\def\triv           {O\hspace*{-.6pt}n\hspace*{-.6pt}e}
\def\Triv           {O\hspace*{-.6pt}n\hspace*{-.6pt}e^{\mathrm c}}

\def\Varphi         {\delta} 
\def\Vect           {{\mathcal V}\hspace*{-.8pt}\mbox{\sl ect}}
\def\wO             {w_\Omeka}
\def\Wor            {{\mathcal{WS}}\hspace*{-.5pt}\mbox{\sl h}}
\def\Worc           {{\mathcal{WS}}\hspace*{-.5pt}\mbox{\sl h}^{\mathrm c}}
\def\Xhat           {\widehat{\Xr}}
\def\xm             {\varpi}  
\def\Xrc            {\Xr^{\mathrm c}_{\phantom;}}
\def\XrC            {\Xr^{\mathrm c}}
\def\Xtil           {\widetilde{\Xr}}
\def\Yrc            {\Yr^{\mathrm c}_{\phantom;}}

\def\Mr             {\mathrm{M}}

\def\Cb             {{\ensuremath{\mathbbm C}}}
\def\Rb             {\mathbb{R}}
\def\Zb             {\mathbb{Z}}

\def\Cc             {\mathcal{C}}

\def\Ic             {\mathcal{I}}
\def\Vc             {\mathcal{V}}

\def\Fr             {\mathrm{F}}
\def\Xr             {\mathrm{X}}
\def\Yr             {\mathrm{Y}}
\def\Zr             {\mathrm{Z}}

\documentclass[12pt]{article}

\usepackage{latexsym,amsmath,amssymb,amsfonts,bbm,xypic}
\usepackage[mathscr]{eucal}
\xyoption{all}
\usepackage{graphicx} \usepackage{rotating}
\usepackage{epstopdf,hyperref} 

\newtheorem{defthm}{\whattheorem}[section]
\newcommand\dt[1]  {\noindent\def\whattheorem{#1}\pagebreak[0]\begin{defthm}{}%
                   \samepage{$\!\!${\rm:}\nopagebreak\\[-1.91em]{}}\end{defthm}}
\newcommand\dtl[2] {\noindent\def\whattheorem{#1}\pagebreak[0]\begin{defthm}{}%
                   \samepage{$\!\!${\rm:}\label{#2}\nopagebreak\\[-1.91em]{}}%
                   \end{defthm}}

\begin{document}

\thispagestyle{empty}


\begin{flushright}  {~} \\[-14mm]
{\sf hep-th/0612306}\\[1mm]{\sf KCL-MTH-06-18}\\[1mm]{\sf ZMP-HH/2006-21}\\[1mm]
{\sf Hamburger$\;$Beitr\"age$\;$zur$\;$Mathematik$\;$Nr.$\;$260}\\[1mm]
{\sf December 2006} \end{flushright}
\begin{center} \vskip 9mm
{\Large\bf UNIQUENESS OF OPEN/CLOSED RATIONAL CFT}\\[3mm]
{\Large\bf WITH GIVEN ALGEBRA OF OPEN STATES}\\[18mm]
{\large
Jens Fjelstad$\;^{0,1}$ \ \ \ J\"urgen Fuchs$\;^1$ \ \ \
Ingo Runkel$\;^2$ \ \ \ Christoph Schweigert$\;^{3}$}
\\[8mm]
 $^0\;$ Cardiff School of Mathematics, Cardiff University\\
       Senghennydd Road, \ GB\,--\,Cardiff\, CF24 4AG\\[3mm]
$^1\;$ Avdelning fysik, \ Karlstads Universitet\\
       Universitetsgatan 5, \ S\,--\,651\,88\, Karlstad\\[3mm]
$^2\;$ Department of Mathematics\\ King's College London, Strand\\
       GB\,--\, London WC2R 2LS\\[3mm]
$^3\;$ Organisationseinheit Mathematik, \ Universit\"at Hamburg\\
       Schwerpunkt Algebra und Zahlentheorie\\
       Bundesstra\ss e 55, \ D\,--\,20\,146\, Hamburg
\end{center}
\vskip 9mm

\begin{quote}{\bf Abstract}\\[1mm]
We study the sewing constraints for rational two-dimensional conformal field 
theory on oriented surfaces with possibly non-empty boundary. The boundary 
condition is taken to be the same on all segments of the boundary.
The following uniqueness result is established: For a solution to the sewing 
constraints with nondegenerate closed state vacuum and nondegenerate two-point 
correlators of boundary fields on the disk and of bulk fields on the sphere, up
to equivalence all correlators are uniquely determined by the one-, two,- and 
three-point correlators on the disk. Thus for any such theory every consistent 
collection of correlators can be obtained by the TFT approach of \cite{tft1,tft5}.
As morphisms of the category of world sheets we include 
not only homeomorphisms, but also sewings; interpreting the correlators
as a natural transformation then encodes covariance both under homeomorphisms
and under sewings of world sheets.

\end{quote} \vfill \newpage


 \tableofcontents \newpage


\sect{Introduction}

To get a good conceptual and computational grasp on two-dimensional \cft\ (CFT) 
has been a challenge for a 
long time. Several rather different aspects need to be comprehended, ranging 
from analytic and algebro-geometric questions to representation theoretic and 
combinatorial issues. Though considerable progress has been made on some of 
these, compare e.g.\ the books \cite{HUan,BAki,FRbe}, it seems fair to say that 
at present the understanding of CFT is still not satisfactory. For example, 
heuristically one expects various models describing physical systems to furnish
`good' CFTs, but a precise mathematical description is often missing.

One early, though not always sufficiently appreciated, insight has been that 
one must distinguish carefully between chiral and full \cft. For instance, in 
chiral CFT the central objects of study are the bundles of conformal blocks
and their sections which are, in general, multivalued, while in full CFT one 
considers correlators which are actual functions of the locations of field 
insertions and of the moduli of the world sheet. It has also been commonly 
taken for granted that at least in the case of {\em rational\/} \cft\ (RCFT), 
every full CFT can be understood with the help of a corresponding chiral CFT, 
e.g.\ any correlation function of the full CFT can be described as a specific 
element in the relevant space of conformal blocks of a chiral CFT. In fact 
(as again realised early \cite{ales}) what is relevant to a full CFT on a world 
sheet $\Xr$ is a chiral CFT on a complex double of $\Xr$, compare e.g.\ 
\cite{bcdcd,bisa2,fuSc6,fffs3}. More recently, it has been established that for
rational CFT this indeed leads to a clean separation of chiral and non-chiral 
aspects and, moreover, that the relation between chiral and full CFT can be 
studied entirely in a model-independent manner when taking the representation 
category \C\ of the chiral symmetry algebra as a starting point. More 
specifically, in a series of papers \cite{tft1,tft4,tft5,ffrs5} it was shown 
how to obtain a consistent set of defining quantities like field contents and 
operator product coefficients, from algebraic structures in the category \C.

At the basis of the results of \cite{tft1,tft4,tft5,ffrs5} lies the idea that 
for constructing a full CFT, in addition to chiral information only a single
further datum is required, namely a simple symmetric special Frobenius algebra\,%
  \footnote{~One must actually distinguish between full CFT on oriented and on
  unoriented (including in particular unorientable) world sheets.
  In the unoriented case the algebra $A$ must in addition come with a reversion
  (a braided analogue of an involution), see \cite{tft2,tft5} for details. In
  the present paper we restrict our attention to the oriented case.} 
$A$ in \C. Given such an algebra $A$, a consistent set of combinatorial data 
determining all correlators, i.e.\ the types of field insertions, boundary 
conditions 
and defect lines, can be expressed in terms of the representation theory of 
$A$ -- boundary conditions are given by $A$-modules and defect lines by\,%
  \footnote{~More generally, for any pair $A$, $B$ of simple symmetric special
  Frobenius algebras the $A$-$B$-bimodules give defect lines separating 
  regions in which the CFT is specified by $A$ and $B$, respectively
  \cite{scfr2,ffrs5}.}
$A$-bimodules, while bulk, boundary and defect fields are particular types of 
(bi)module morphisms. We work in the setting of rational CFT, so that the
category \C\ is a modular tensor category. Exploiting the relationship 
\cite{TUra,KArt} between modular tensor categories and three-dimensional \tft, 
one can then specify each correlator of a full rational CFT, on a world sheet of
arbitrary topology, as an element in the relevant space of conformal blocks, by 
representing it as the invariant of a suitable ribbon graph in a three-manifold.
The correlators obtained this way can be proven \cite{tft5} to satisfy all 
consistency conditions that the correlators of a CFT must obey. Thus, specifying
the algebra $A$ is sufficient to obtain a consistent system of correlators; in 
contrast, in other approaches to CFT only a restricted set of correlators and 
of constraints can be considered, so that only some necessary consistency 
conditions can be checked.  Another feature of our approach is that Morita 
equivalent algebras give equivalent systems of correlators; it can be shown 
that in any modular tensor category there is only a finite number of Morita 
classes of simple symmetric special Frobenius algebras, so that only a 
finite number of distinct full CFTs can share a given chiral RCFT.  

It was also explained in \cite{tft1} how one may extract a simple symmetric 
special Frobenius algebra from a given full \cft\ that is defined on world 
sheets with boundary: it is the algebra of boundary fields for a given
boundary condition (different boundary conditions give rise to Morita 
equivalent algebras). On the other hand, what could not be shown so far is that 
a full \cft\ is already uniquely specified by this algebra; thus it was e.g.\ 
unclear whether the correlators constructed from the algebra of boundary fields 
in the manner described in \cite{tft1,tft5} coincide with those of the full 
\cft\ one started with. It is this issue that we address in the present paper.
We formulate a few universal conditions that should be met in every RCFT\,%
  \footnote{~We also have to make a technical assumption concerning the values
  of quantum dimensions. This condition might be stronger than necessary.}
and establish that under these conditions and for a given algebra of boundary 
fields the constraints on the system of correlators have a unique solution (see
theorem \ref{thm:unique}). Thus up to equivalence, the correlators 
must be the same as those obtained in the construction of \cite{tft1,tft5} 
from the algebra of boundary fields.  In other words, we are able to 
show that, under reasonable conditions, {\em every\/} consistent collection of 
RCFT correlators can be obtained by the methods of \cite{tft1,tft5}.

\medskip

Even for rational CFT, some major issues are obviously left unsettled by the 
approach of \cite{tft1,tft4,tft5,ffrs5}. While it efficiently identifies such 
quantities of a CFT which only depend on the topological and combinatorial 
data of the world sheet and the field insertions, in a complete picture the
conformal structure of the world sheet plays an important role and one must even
specify a concrete metric as a representative of its conformal equivalence
class. In particular the relation between chiral and full CFT is described
only at the level of topological surfaces, and the construction yields a
correlator just as an element of an abstract vector space of conformal blocks
and must be supplemented by a concrete description of the conformal blocks 
in terms of invariants in tensor products of modules over the chiral symmetry
algebra. (Note, however, that often the latter aspects are not of primary 
importance. For instance, a lot of interesting information about a CFT is 
contained in the coefficients of partition functions and in
the various types of operator product coefficients, and these 
can indeed be computed \cite{tft4} with our methods.) 
To alert the reader about this limitation, below we will refer to the surfaces 
we consider as {\em topological world sheets\/}. But this qualification must 
not be confused with the corresponding term for field theories. Our approach 
applies to all RCFTs, not only to two-dimensional topological field theories,
whose correlation functions are independent of the location of field insertions. 

To go beyond the combinatorial framework studied here, one has to promote
the geometric category of topological world sheets 
to a category of world sheets with metric and similarly for the relevant 
algebraic category of vector spaces, for the relevant functors between them and 
for natural transformations. Some ideas on how this can be achieved concretely 
are presented at the end of this paper. Confidence 
that this approach can be successful comes from the result of 
\cite{BAki} that the notions of a (\C-decorated) topological modular functor
and of a (\C-decorated) complex-analytic modular functor are equivalent.


\sect{Summary}

Let us briefly summarise the analysis of \cft\ pursued in sections 
\ref{sec:oc-sewing}\,--\,\ref{sec:unique-proof}.
We assume from the outset that we are given a definite
modular tensor category \C, and we make extensive use of the three-dimensional
\tft\ that is associated to \C. Note that for our calculations we do not have 
to assume that $\Cc$ is the category of representations of a suitable
chiral algebra (concretely, a conformal vertex algebra). However, if we 
want to interpret the quantities that describe correlators in our framework
as actual correlation functions of CFT on world sheets with metric, we do need 
an underlying chiral algebra $\Vc$ such that $\Cc\eq\RepVc$ and such that
in addition the 3-d TFT associated to \C\ correctly encodes the sewing and 
monodromy properties of the conformal blocks (compare section \ref{sec:wswm}).

\medskip

In section \ref{sec:worldsheet} we describe the relevant geometric category 
$\Wor$ whose objects are topological world sheets. Its morphisms do not only 
consist of homeomorphisms of topological world sheets, but we also introduce
sewings as morphisms; $\Wor$ is a symmetric monoidal category. 
In this paper we treat the boundary segments\,%
  \footnote{~In the terminology of section \ref{sec:worldsheet}, these are the
  `physical boundaries'.} 
of a world sheet as unlabeled. In more generality, one can assign different 
conformal boundary conditions to different connected components (or, in the 
presence of boundary fields, segments) of the boundary. In the category theoretic 
setting, boundary conditions are labeled by modules over the relevant Frobenius
algebra in $\Cc$, see \cite[sect.\,4]{tft1} and \cite[sect.\,4]{tft5}. Working 
with unlabeled boundaries corresponds to having selected one specific conformal
boundary condition which we then assign to all boundaries. 

In section \ref{sec:mtc-tft} we recall the 
definition of a modular tensor category and the way in which it gives rise to 
a 3-d TFT, i.e.\ to a monoidal functor $\tftC$ from a geometric category to 
the category $\Vect$ of finite-dimensional complex vector spaces. The 3-d TFT is
then used, in section \ref{sec:spaceblocks}, to construct a monoidal functor 
$\bl$ from $\Wor$ to $\Vect$. We also introduce a `trivial' functor $\triv{:}\ 
\Wor \To \Vect$, which assigns the ground field $\Cb$ to every object and 
$\id_\Cb$ to every morphism in $\Wor$. Given these two functors, we define in
section \ref{sec:form-sew} a collection $\cor$ of correlators as a monoidal 
natural transformation from $\triv$ to $\bl$. The properties of a monoidal 
natural transformation furnish a convenient way to encode the consistency 
conditions, or sewing constraints, that a collection of correlators must satisfy
(see section \ref{sec:wswm} for a discussion). Accordingly, we will say that 
$\cor$ {\em provides a solution $\sfc$ to the sewing constraints\/}. More
precisely, besides $\cor$ some other data need to be prescribed (see section 
\ref{sec:form-sew}), in particular the open and closed state spaces, which are 
objects $\Hop$ of \C\ and $\Hcl$ of the product category $\CxCb$, respectively.
Different solutions $\sfc$ and $\sfc'$ can describe CFTs that are physically
equivalent; a corresponding notion of equivalence of solutions to the sewing 
constraints is introduced in section \ref{sec:sew-equiv}.

\medskip

Section \ref{sec:fund-corr} recalls how sewing can be used to construct any 
world sheet from a small collection of fundamental world sheets. To apply
this idea to correlators one needs an operation of `projecting onto the closed
state vacuum'; this is studied in section  \ref{sec:projcl}.

The results of \cite{tft1,tft5} show in particular that any symmetric special 
Frobenius algebra $A$ in $\Cc$ gives rise to a solution $\sfc\eq\sfc(\Cc,A)$ to 
the sewing constraints. Together with some other background information
this is reviewed briefly in
\ref{sec:frob-to-sew}. Afterwards, in sections \ref{sec:sew-to-frob} and
\ref{sec:unique-thm}, we come to the main subject of this paper: we study how, 
conversely, a solution to the sewing constraints gives rise to a Frobenius 
algebra $A$ in \C. The ensuing uniqueness result is stated in theorem 
\ref{thm:unique}; it asserts that

\begin{quote}
{\em Every solution $\sfc$ to the sewing constraints is of the form $\sfc(\Cc,A)$,\\
with an (up to isomorphism) uniquely determined algebra $A$,\\
provided that the following conditions are fulfilled\/}:
\end{quote}

\noindent
(i)~~\,There is a unique `vacuum' state in $\Hcl$, in the sense that the vector
space $\Hom_{\CxCb}(\one {\times} \ol\one, \Hcl)$ is one-dimensional.
\\[2pt] 
(ii)~\,The correlator of a \disc\ with two boundary insertions is non-degenerate.
\\[2pt] 
(iii) The correlator of a sphere with two bulk insertions is non-degenerate.
\\[2pt] 
(iv) The quantum dimension of $\Hop$ is nonzero, and for each subobject
$U_i\Times\ol{U_j}$ of the full center of $A$ (as defined in section 
\ref{sec:frob-to-sew} below) the product $\dim(U_i)\hspace*{.5pt}\dim(U_j)$
of quantum dimensions is positive.

\medskip

The proof of this theorem is given in section \ref{sec:unique-proof}.
It shows in particular that a solution to the sewing constraints is determined 
up to equivalence 
by the correlators assigned to \disc s with one, two and three boundary 
insertions. The conditions (i), (ii) and (iii) are necessary; if any of them 
is removed, one can find counter examples, see remark \ref{rem:unique}\,(i). 
Condition (iv), on the other hand, appears to be merely a technical assumption 
used in our proof, and can possibly be relaxed, or dropped altogether.

\medskip

Let us conclude this summary with the following two remarks. First, as already
pointed out, even though in sections \ref{sec:oc-sewing}--\ref{sec:unique-proof}
we work exclusively with topological world sheets, we do not only describe
two-dimensional topological conformal field theories. The reason is that the
correlator $\cor_\Xr$ on a (topological) world sheet $\Xr$ is not itself a 
linear map between spaces of states, but rather it corresponds to a 
function on the moduli space of world sheets with metric (obtained as a
section in a bundle of multilinear maps over the moduli space).

Second, in the framework of local quantum field theory on $1{+}1$-dimensional
Minkowski space a result analogous to theorem \ref{thm:unique} has been given in
\cite{lore2}; see remark \ref{rem:unique}\,(vi) below.


\begin{table}[p]
\begin{center}
\begin{tabular}{|c|ll|ll|} \hline~&&&&\\[-.8em]
symbol & \mcc{quantity} & \mcc{introduced in}
                                  \\[-.9em]&&&& \\\hline&&&&\\[-.7em]
$\Wor$
  & \mcl{category of open/closed topological world sheets}
  & section \ref{sec:worldsheet}
  & p.\,\pageref{sec:worldsheet}
                                  \\[.3em]
$\Xr$, $\Yr$, ...
  & \mcl{world sheet (object of $\Wor$)}
  & definition \ref{def:ws}
  & p.\,\pageref{def:ws}
                                  \\[.3em]
$\Xhat$
  & \mcl{decorated double of a world sheet $\Xr$}
  & section \ref{sec:spaceblocks}
  & p.\,\pageref{def:Xhat}
                                  \\[.3em]
$\Xtil$
  & \mcl{surface used in the definition of a world sheet}
  & definition \ref{def:ws}
  & p.\,\pageref{def:ws}
                                  \\[.3em]
$b^\text{in}$, $b^\text{out}$
  & \mcl{set of in- resp.\ out-going boundary components}
  & definition \ref{def:ws}
  & p.\,\pageref{def:ws}
                                  \\[.3em]
$\sew$ 
  & \mcl{sewing of a world sheet}
  & definition \ref{def:oc-sewing}
  & p.\,\pageref{def:oc-sewing}
                                  \\[.3em]
$\xm \eq (\sew,f)$
  & \mcl{morphism of $\Wor$ (with $f$ a homeomorphism)}
  & definition \ref{def:ws-hom}
  & p.\,\pageref{def:ws-hom}
                                  \\[.3em]
$\fill_\sew(\Xr)$
  & \mcl{world sheet filled at $\sew$}
  & definition \ref{def:fill}
  & p.\,\pageref{def:fill}
                                  \\[.3em]
$\Mr_\Xr$
  & \mcl{connecting three-manifold}
  & equation \erf{eq:conmf-def}
  & p.\,\pageref{eq:conmf-def}
                       \\[-.8em]&&&& \\\hline \mcc{}&&&\\[-.5em]
\mcc{$\!\Xr_m,\Xr_\eta,\Xr_\Delta,\Xr_\eps,
\Xr_{Bb},\Xr_{B(3)},\Xr_{B\eta},\Xr_{B\eps}\!\!$}
  & {fundamental world sheets}
  & table\,\ref{table:fuwosh}\,/\,fig.\,\ref{fig:fund-world}\hspace*{-.7em}
  & p.\,\pageref{table:fuwosh} 
                       \\[.3em]
\mcc{$\Xr_p,\Xr_{Bp},\Xr_{B\Delta},\Xr_{Bm}\qquad$}
  & {some other simple world sheets}
  & table\,\ref{table:fuwosh}\,/\,fig.\,\ref{fig:addtl-world}\hspace*{-.7em}
  & p.\,\pageref{fig:addtl-world}
                       \\[-.6em]\mcc{}&&& \\\hline &&&&\\[-.7em]
$\C$
  & \mcl{a modular tensor category (here, the chiral sectors)}
  & section \ref{sec:mtc-tft}
  & p.\,\pageref{def:C}
                       \\[.3em]
$\tftC$
  & \mcl{TFT functor from geometric category $\mathcal{G}_\Cc$ to $\Vect$}
  & section \ref{sec:mtc-tft}
  & p.\,\pageref{def:tftC}
                       \\[.3em]
$\Hop$, $\Hcl$
  & \mcl{open\,/\,closed state spaces}
  & section \ref{sec:spaceblocks}
  & p.\,\pageref{def:HclHop}
                       \\[.3em]
$B_l$, $B_r$
  & \mcl{objects of \C\ such that $\Hcl$ is a retract of $\BlxBr$}
  & section \ref{sec:spaceblocks}
  & p.\,\pageref{sec:spaceblocks}
                       \\[.3em]
$\Omeka$
  & \mcl{auxiliary object in \C\ appearing in the description}
     &&\\&
  \mcl{$\qquad$~of $\Hcl$ as a retract (cf.\ lemma \ref{lem:Z(A)-retract})}
  & equation \erf{eq:omega-def}
  & p.\,\pageref{eq:omega-def}
                       \\[.3em]
$\wO$
  & \mcl{weighted sum of idempotents for subobjects of $\Omeka$}
  & equation \erf{eq:wO}
  & p.\,\pageref{eq:wO}
                       \\[.3em]
$H$
  & \mcl{object in $\C$ appearing in the description of TFT state}
      &&\\&
  \mcl{$\qquad$~spaces on surfaces with handles (cf.\ eq.\,\erf{eq:E-Hom-space})}
  & equation \erf{eq:omega-def}
  & p.\,\pageref{eq:omega-def}
                       \\[.3em]
$T_\Cc$
  & \mcl{canonical trivialising algebra (object in $\CxCb$)}
  & definition \ref{def:triv-alg}
  & p.\,\pageref{def:triv-alg}
                       \\[.3em]
$Z(A)$
  & \mcl{full center of a symmetric special}
     &&\\&
  \mcl{$\qquad$~Frobenius algebra $A$ in $\Cc$ (object in $\CxCb$)}
  & definition \ref{def:full-centre}
  & p.\,\pageref{def:full-centre}
                       \\[.3em]
$\varphi_\text{cl}^A$
  & \mcl{isomorphism from $\Hcl$ to $Z(A)$}
  & equation \erf{eq:phi_cl-def}
  & p.\,\pageref{eq:phi_cl-def}
                       \\[-.8em]&&&& \\\hline&&&&\\[-.9em]
$\triv$
& \mcl{monoidal functor $\Wor\To\Vect$
  with image $\Cb\,{\stackrel{\idsmall_\Cb}\longrightarrow}\,\Cb$}
  & definition \ref{def:triv}
  & p.\,\pageref{def:triv}
                       \\[.3em]
$\bl$
  & \mcl{monoidal functor $\Wor\To\Vect$}
  & definition \ref{def:bl}
  & p.\,\pageref{def:bl}
                       \\[.3em]
$\cor$
  & \mcl{monoidal natural transformation from $\triv$ to $\bl$}
  & section \ref{sec:form-sew}
  & p.\,\pageref{def:cor}
                       \\[.3em]
$\Alpha$
  & \mcl{natural isomorphism between functors of type $\bl$}
  & equation \erf{eq:alphaX-def}
  & p.\,\pageref{eq:alphaX-def}
                       \\[-.8em]&&&& \\\hline&&&&\\[-.7em]
$\sfc$
  & \mcl{solution to the sewing constraints}
  & definition \erf{def:sf}
  & p.\,\pageref{def:sf}
                       \\[.3em]
$\sfc(\Cc,A)$
  & \mcl{tuple $(\Cc,A,Z(A),A\otimes\Omeka,\Omeka,e_Z,r_Z,\cor_A)$}
     &&\\&
  \mcl{$\qquad$~furnishing a solution to the sewing constraints}
  & equation \erf{eq:sfc-def}
  & p.\,\pageref{eq:sfc-def}
                       \\[.3em]
$A_\sfc$
  & \mcl{algebra of open states associated to $\sfc$}
  & equation \erf{eq:Asfc-def}
  & p.\,\pageref{eq:Asfc-def}
                       \\[-.7em]&&&& \\\hline
\end{tabular} 
\caption{Symbols for basic quantities.}
\end{center}
\end{table}


\subsubsection*{List of symbols}

To achieve our goal we need to work with a variety of different structures. 
For the convenience of the reader, we collect some of them in table 1.


\sect{Open/closed sewing constraints}\label{sec:oc-sewing}

In this section we introduce the structures that we need for an algebraic
formulation of the sewing constraints. These are the category $\Wor$
of topological world sheets (section \ref{sec:worldsheet}),
   the three-dimensional topological field theory (3-d TFT) obtained from
   a modular tensor category (section \ref{sec:mtc-tft}), and a functor $\bl$ 
   from $\Wor$ to $\Vect$ which is constructed with the help of the 3-d TFT 
   (section \ref{sec:spaceblocks}). The notion of sewing constraints, and of 
   the equivalence of two solutions to these constraints, is discussed in 
sections \ref{sec:form-sew} and \ref{sec:sew-equiv}.


\subsection{Oriented open/closed topological world sheets}\label{sec:worldsheet}

\begin{figure}[bt]
  \begin{center} \begin{picture}(200,141)(0,0)
  \put(40,5){
     \put(0,0){\scalebox{.7}{\includegraphics{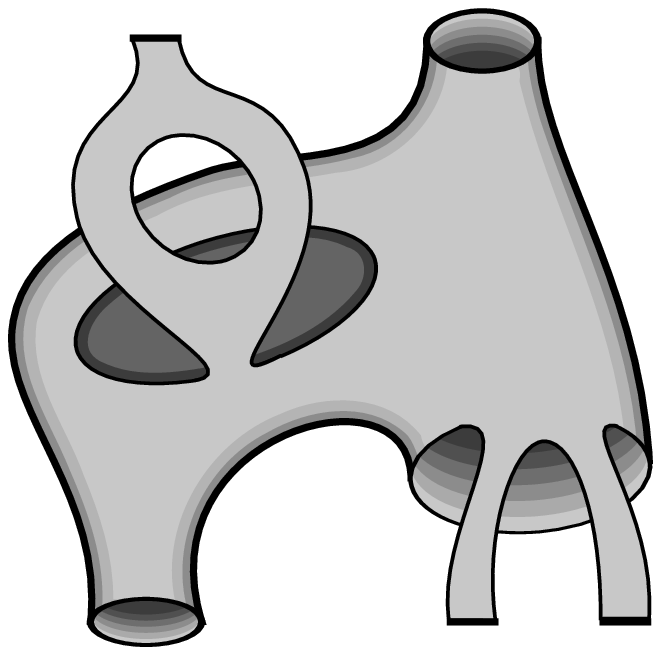}}}
     \put(20,-8){\sse c-in$_1$}
     \put(86,-3){\sse o-in$_1$}
     \put(117,-3){\sse o-in$_2$}
     \put(20,128){\sse o-out$_1$}
     \put(85,134){\sse c-out$_2$}
     \put(-90,110){\xymatrix{\txt{\small physical\\ \small boundary} \ar@/^/[rr] &&} }
     \put(120,25){ \xymatrix{&& \ar@/^/[ll]\txt{~\\\small physical\\ \small boundary} } }
     \put(90,80){ \xymatrix{&& \ar@/^/[ll] \txt{~\\orientation} } }
     }
  \end{picture} \end{center}
 \caption{A topological world sheet.} \label{fig:oc-world}
\end{figure}

We are concerned with CFT on oriented surfaces which may have empty or non-empty 
boundary. We call such surfaces oriented open/closed topological world sheets,
or just world sheets, for short, and refer to CFT on such surfaces as 
{\em open/closed CFT\/}. An example of a topological world sheet is displayed
in figure \ref{fig:oc-world}. Also recall 
from the introduction that when we want to describe correlators as actual 
functions, then we need to endow the world sheet in addition with a conformal 
structure and even a metric; this is discussed in section \ref{sec:wswm}.

As indicated in figure \ref{fig:oc-world}, a world sheet can have
five types of boundary components. Four of them signify the
presence of field insertions, while the fifth type describes a genuine physical 
boundary. These boundary types can be distinguished by their labeling: There 
are {\em in-going open state boundaries\/} (the intervals labeled o-in$_1$ and
o-in$_2$ in the example given in figure \ref{fig:oc-world}), {\em out-going 
open state boundaries\/} (o-out$_1$ in the example), {\em in-going closed state 
boundaries\/} (c-in$_1$ in the example) and {\em out-going closed state 
boundaries\/} (c-out$_1$ in the example), and finally 
{\em physical boundaries\/}, which are unlabeled. The open and closed state 
boundaries are parametrised by intervals and circles, respectively.
The physical boundaries are oriented, but not parametrised.

Geometrically the various boundary types can best be distinguished by 
describing the world sheet $\Xr$ as a quotient $\dot\Xr$ of a surface with 
boundary, $\Xtil$, by an orientation reversing involution $\iota$.
The surface shown in figure \ref{fig:oc-world} is $\dot\Xr$ rather than $\Xtil$.
With this description, a point $p$ on the boundary of $\dot\Xr$ belongs
to a physical boundary if its pre-image on $\Xtil$ is a fixed point of $\iota$,
and otherwise to an open state boundary if its pre-images lie on a single 
connected component of $\partial\Xtil$, and to a closed state boundary if it
has two pre-images on $\Xtil$ lying on two distinct connected components of 
$\partial\Xtil$. We denote the number of in-going open state boundaries of $\Xr$
by $|\text{o-in}|$, etc. An important operation on world sheets is 
{\em sewing\/} \cite{vafa0,sono2,lewe3,HUan}: one specifies a set 
$\sew$ of pairs, consisting of an out-going and an in-going state boundary of 
the same type. {}From $\sew$ one can obtain a new world sheet $\sew(\Xr)$ by 
sewing, that is, by identifying, for each pair in $\sew$, the two involved 
boundary components via the parametrisation of the state boundaries. In the 
example in figure \ref{fig:oc-world}, some possible sewings are
$\sew\eq\{ (\text{o-out}_1,\text{o-in}_1) \}$ and
$\sew\eq\{ (\text{c-out}_1,\text{c-in}_1), (\text{o-out}_1,\text{o-in}_2) \}$.

\medskip

Let us now describe these structures in a form amenable to our algebraic and
combinatoric framework. To this end we introduce a symmetric strict monoidal 
category $\Wor$ whose objects are oriented open/closed topological world 
sheets and whose morphisms are isomorphisms and sewings of such world sheets.

We denote by $S^1$ the unit circle $\{|z|\eq1\}$ in the complex plane, with 
counter-clockwise orientation. The map that assigns to a complex number its
complex conjugate is denoted by $C{:}\ \Cb \To \Cb$.
\\[-2em]

\dtl{Definition}{def:ws}
An {\em oriented open/closed topological world sheet\/}, or
{\em world sheet\/} for short, is a tuple  
  \be
  \Xr \,\equiv \big(\, \Xtil, \iota_\Xr, \Varphi_\Xr,
  b^\text{in}_\Xr, b^\text{out}_\Xr, \text{or}_\Xr \,\big)
  \ee
consisting of:
\\[.3em]
\nxt An oriented compact two-dimensional topological manifold $\Xtil$.
The (possibly empty) boundary $\partial\Xtil$ of
$\Xtil$ is oriented by the inward-pointing normal.
\\[.3em]
\nxt A continuous orientation-reversing involution
  \be
  \iota_\Xr{:}\quad \Xtil \to \Xtil \,.
  \ee
\\[.3em]
\nxt A continuous orientation-preserving map
which parametrises all boundary components of $\Xtil$, i.e.\ a map
  \be
  \Varphi_\Xr{:}\quad \partial \Xtil \to S^1
  \ee
that is an isomorphism when restricted to a connected component of 
$\partial \Xtil$, and which treats boundary components that are mapped to each 
other by $\iota_\Xr$ in a compatible manner, i.e.\ intertwines the involutions 
on $\Xtil$ and $\Cb$, $\Varphi_\Xr\Circ\iota_\Xr \eq C\Circ\Varphi_\Xr$.
\\[.3em]
\nxt A partition of the set $\pi_0(\partial \Xtil)$ of connected components 
of $\partial\Xtil$ into two subsets $b^\text{in}_\Xr$ and $b^\text{out}_\Xr$
(i.e.\ $b^\text{in}_\Xr \,{\cap}\, b^\text{out}_\Xr \eq \emptyset$ and
$b^\text{in}_\Xr \,{\cup}\, b^\text{out}_\Xr \eq \pi_0(\partial \Xtil)$).
The subsets $b^\text{in}_\Xr$ and $b^\text{out}_\Xr$ are required to be fixed
(as sets, not necessarily element-wise) under the involution ${\iota_\Xr}_*$ 
on $\pi_0(\partial \Xtil)$ that is induced by $\iota_\Xr$.    
\\[.3em]
\nxt Denoting by $\tilde\pi_\Xr {:}\ \Xtil \To \dot\Xr$ the canonical projection
to the quotient surface $\dot\Xr \,{:=}\, \Xtil / \langle \iota_\Xr \rangle$, 
$\text{or}_\Xr$ is a global section of
the bundle $\Xtil\,{\overset{\tilde\pi_\Xr}\longrightarrow}\,\dot\Xr$,
i.e.\ $\text{or}_\Xr {:}\ \dot\Xr \To \Xtil$ is a continuous
function such that $\tilde\pi_\Xr \Circ \text{or}_\Xr \eq \id_{\dot\Xr}$.
In particular, a global section exists.
We also demand that for a connected component $c$ of $\partial\Xtil$, 
$\delta_\Xr(\text{im} \, \text{or}_\Xr \,{\cap}\, c)$ is either $\emptyset$, or $S^1$, 
or the upper half circle $\{\mathrm e^{i\theta} \,|\, 0 \,{\le}\, \theta \,{\le}\, \pi \}$.

\dtl{Remark}{rem:regions}
(i) Since $\Xtil$ is compact, the number $|\pi_0(\partial \Xtil)|$ of connected 
components of $\partial\Xtil$ is finite. Also, the existence of a global section
$\text{or}_\Xr{:}\ \dot\Xr \To \Xtil$ implies that $\dot\Xr$ is orientable, and 
in fact is provided with an orientation by demanding
$\text{or}_\Xr$ to be orientation-preserving.
\\[.3em]
(ii) 
As mentioned at the beginning of this section, the boundary of the 
quotient surface $\dot\Xr$ can be divided into segments each of which is of 
one of five types. A point $p$ on $\partial\dot\Xr$ lies on a {\em physical 
boundary\/} iff $p$ has a single pre-image under $\tilde\pi_\Xr$, which is
hence a fixed point of $\iota_\Xr$.
The point $p$ lies on a {\em state boundary\/} if $p$ has two pre-images under 
$\tilde\pi$. If both pre-images lie on the same connected component of 
$\partial\Xtil$, then $p$ lies on an {\em open\/} state boundary, otherwise it
lies on a {\em closed\/} state boundary. (Note that an open state boundary is 
a parametrised open interval on $\dot\Xr$.) Let $a$ be a boundary
component that contains a pre-image of $p$. If $a \iN b^\text{in}$,
then the state boundary containing $p$ is {\em in-going\/}, otherwise
it is {\em out-going\/}. Altogether we thus have five types:
A region of $\partial\dot\Xr$ can be a physical boundary, or an
in/out-going open, or an in/out-going closed state boundary.

\medskip

World sheets are the {\em objects\/} of the category $\Wor$ we wish to
define. For {\em morphisms\/} we need the notion of sewing.

\dtl{Definition}{def:oc-sewing}
Let $\Xr \eq \big( \Xtil, \iota, \Varphi, b^\text{in},
b^\text{out}, \text{or} \big)$ be a world sheet.
\\[.3em]
(i)~\,{\em Sewing data for\/} $\Xr$, or a {\em sewing of\/} $\Xr$, 
is a (possibly empty) subset $\sew$
   of $b^\text{out} \Times b^\text{in}$ such that if $(a,b) \iN \sew$ then\\
-- $\sew$ does not contain any other pair of the form $(a,\cdot)$ or 
   $(\cdot,b)$,\\
-- also $(\iota_*(a),\iota_*(b)) \iN \sew$,\\
-- the boundary component $a$ has non-empty intersection with the image of 
   $\,\text{or}{:}\ \dot\Xr\To\Xtil$ iff the boundary component $b$
   does (i.e.\ $\sew$ preserves the orientation).
\\[.3em]
(ii)~For a sewing $\sew$ of $\Xr$, the {\em sewn world sheet\/} $\sew(\Xr)$ is 
the tuple $\sew(\Xr) \,{\equiv}\, \big( \Xtil',\iota',\Varphi',{b^\text{in}}',
{b^\text{out}}',\text{or}' \big)$ that is obtained as follows. 
For $a \iN \pi_0(\partial \Xtil)$ denote by
  \be
  \Varphi_a := \delta\big|_{\partial \Xtil^{(a)}}
  \ee
the restriction of the boundary parametrisation $\Varphi$ to the connected 
component $a$ of $\partial\Xtil$; $\Varphi_a$ is an isomorphism. Then we set 
$\Xtil' \,{:=}\, \Xtil / {\sim}$, where $\Varphi_a^{-1}(z) \,{\sim}\, 
\Varphi_b^{-1}\Circ C(-z)$ for all $(a,b) \iN \sew$ and $z \iN S^1$. Next, 
denote by $\pi_{\sew,\Xr}$ the projection from $\Xtil$ to $\Xtil'$ that takes a 
point of $\Xtil$ to its equivalence class in $\Xtil'$. Then $\iota'{:}\ \Xtil' 
\To \Xtil'$ is the unique involution
such that $\iota'\Circ\pi_{\sew,\Xr} \eq \pi_{\sew,\Xr}\Circ\iota$. Further, 
$\Varphi'$ is the restriction of $\Varphi$ to $\partial\Xtil'$,
${b^\text{out}}'\eq \{ a \iN b^\text{out}| (a,\cdot) \,{\notin}\, \sew \}$,
${b^\text{in}}' \eq \{ b \iN b^\text{in} | (\cdot,b) \,{\notin}\, \sew \}$,
and $\text{or}'$ is the unique continuous section of
$\Xtil'\,{\overset{\tilde\pi_{\Xr'}}{\longrightarrow}}\,{\dot\Xr}'$ such that 
the image of $\,\text{or}'$ coincides with the image of
$\pi_{\sew,\Xr} \Circ \text{or}$.

\medskip

One can verify that the procedure in (ii) does indeed define a world sheet.

\dtl{Definition}{def:ws-hom}
Let $\Xr$ and $\Yr$ be two world sheets.
\\[2pt]
(i) A {\em homeomorphism of world sheets\/} is a homeomorphism $f{:}\ \Xtil 
\To \widetilde\Yr$ that is compatible with all chosen structures on $\Xtil$, 
i.e.\ with orientation, involution and boundary parametrisation. That is, $f$ 
satisfies
  \be
  f \circ \iota_\Xr = \iota_\Yr \circ f \,,\quad~
  \Varphi_\Yr \circ f = \Varphi_\Xr \,,\quad~
  f_* b^\text{in/out}_\Xr = b^\text{in/out}_\Yr 
  \ee
(where $f_*{:}\ \pi_0(\partial\Xtil) \To \pi_0(\partial\widetilde\Yr)$
is the isomorphism induced by $f$), and the image of 
$f\Circ\text{or}_\Xr$ coincides with the image of $\,\text{or}_\Yr$.
\\[2pt]
(ii) A {\em morphism\/} $\xm {:}\ \Xr \To \Yr$
is a pair $\xm \eq (\sew,f)$ where $\sew$ are sewing data for $\Xr$ and $f{:}\ 
\widetilde{\sew(\Xr)} \To \widetilde\Yr$ is a homeomorphism of world sheets.
\\[2pt]
(iii) The set of all morphisms from $\Xr$ to $\Yr$ is denoted
by $\Hom(\Xr,\Yr)$.

\medskip

Given two morphisms $\xm \eq (\sew,f){:}\ \Xr \To \Yr$ and $\xm'\eq (\sew',g){:}
\ \Yr \To \Zr$, the composition $\xm'\Circ\xm$ is defined as follows. The union 
$\sew'' \eq \sew \,{\cup}\, (f \Circ\pi_{\sew,\Xr})_*^{\,-1}(\sew')$ 
is again a sewing of $\Xr$. Furthermore there exists a unique isomorphism
$h{:}\ \widetilde{\sew''(\Xr)} \To \widetilde{\Zr}$ such that the diagram
  \be
  \xymatrix@C=2em{
  \Xtil \ar[d]^{\pi_{\sew'',\Xr}} \ar[r]^{\pi_{\sew,\Xr}} &
  \widetilde{\sew(\Xr)} \ar[r]^{f} &
  \widetilde{\Yr} \ar[r]^{\pi_{\sew',\Yr}} & \widetilde{\sew'(\Yr)} \ar[r]^{g}
  & \widetilde{\Zr} \\
  \widetilde{\sew''(\Xr)} \ar[urrrr]_{h} }
  \ee
commutes. We define the composition
$\circ\,{:}\ \Hom(\Yr,\Zr) \Times \Hom(\Xr,\Yr) \To \Hom(\Xr,\Zr)$ as
  \be
  (\sew',g) \circ (\sew,f) = (\sew'',h) \,.
  \ee
One verifies that the composition is associative. The identity morphism on 
$\Xr$ is the pair $\id_\Xr \eq (\emptyset,\id_{\Xtil})$.

Finally, we define a monoidal structure on $\Wor$ by taking the tensor product 
to be the disjoint union, both on world sheets and on morphisms,
and the unit object to be the empty set.
We also define the isomorphism $c_{\Xr,\Yr}{:}\ \Xr \,{\sqcup}\, \Yr \To
\Yr \,{\sqcup}\, \Xr$ to be the homeomorphism that exchanges the two
factors of the disjoint union.
In this way, $\Wor$ becomes a symmetric strict monoidal category.

\dt{Remark}
(i)~~The set of morphisms from $\one$ (the empty set) to any world sheet $\Xr$ 
with $\Xtil \,{\neq}\, \emptyset$ is empty.  Thus there does not exist a duality
on $\Wor$, nor is there an initial or a final object in $\Wor$.
\\[.3em]
(ii)\,~What we refer to as physical boundaries of $\dot\Xr$ are called 
`boundary sector boundaries' in \cite{laza}, 
`free boundaries' in \cite{moor10,sega13},   
`coloured boundaries' in \cite{lapf},        
and `constrained boundaries' in \cite{moSe}. 
\\[.3em]
(iii)~$\Wor$ is different from the category of open/closed two-dimensional 
cobordisms considered in the context of two-dimensional open/closed 
topological field theory in \cite{laza,moor10,sega13,lapf,moSe}. There, objects 
are disjoint unions of circles and intervals, and morphisms are equivalence 
classes of cobordisms between these unions. One can also consider 
two-dimensional open/closed cobordisms as a 2-category as in \cite{till3,bacR}.
Then objects are defined as just mentioned, 1-morphisms are surfaces embedded 
in $\Rb^3$ which have the union of circles and intervals as boundary, and 
2-morphisms are homeomorphisms between these surfaces. The 1- and 2-morphisms 
in this definition correspond to the objects and some of the morphisms in 
$\Wor$, but they do not include the sewing operation.

\medskip

{}From section \ref{sec:spaceblocks} on we will, when drawing a world sheet 
$\Xr$, usually only draw the surface $\dot\Xr$, give the orientation on 
$\dot\Xr$, and indicate the decomposition of $\partial\dot\Xr$ into segments
as well as the type of each segment (see remark \ref{rem:regions}\,(ii)). As an 
example, consider the surface $\Xtil$ given by a sphere with six small equally 
spaced holes along a great circle,
  \eqpic{pic-fjfrs2_45} {190} {73} {
  \put(45,0)     {\Includeourbeautifulpicture 45{}}
  \put(0,61)     {$ \Xtil~= $}
  \put(39,55)    {$ \imath $}
  \put(154,158)  {$ \imath' $}
  \put(114,18)   {\small$ in $}
  \put(99,50)    {\small$ in $}
  \put(105,105)  {\small$ in $}
  \put(152,120)  {\small$ out $}
  \put(168,90)   {\small$ out $}
  \put(165,38)   {\small$ out $}
  \put(116,120)  {$ E' $}
  \put(135,46)   {$ E $}
  } 
In the figure it is also indicated how $\pi_0(\partial\Xtil)$ is partitioned 
into $b^\text{in}$ and $b^\text{out}$. In addition two great circles $E$ 
and $E'$ are drawn. Denote by
$\iota$ the reflection with respect to the plane intersecting $\Xtil$ at $E$ and
$\iota'$ the reflection for the plane intersecting at $E'$. We obtain two world 
sheets $\Xr$ and $\Xr'$ which only differ in their involution and orientation,
  \be
  \Xr = (\Xtil,\iota,\Varphi,b^\text{in},b^\text{out},\text{or})
  \qquad \text{and} \qquad
  \Xr' = (\Xtil,\iota',\Varphi,b^\text{in},b^\text{out},\text{or}')\,.
  \labl{eq:ws-example-1}
The orientation $\,\text{or}$ is fixed by requiring its image
in $\Xtil$ to be the half-sphere above $E$ (say), together with $E$,
and for $\text{or}'$ one can take the half-sphere in front of $E'$.
The quotients $\dot\Xr$ and $\dot\Xr'$ for these two world sheets
then look as follows.
  \eqpic{pic-fjfrs2_46} {380} {42} {
  \put(45,0)     {\Includeournicelargepicture 46a}
  \put(0,43)     {$ \dot\Xr~= $}
  \put(265,0)    {\Includeournicelargepicture 46b}
  \put(218,43)   {$ \dot\Xr'~= $}
  \put(40,43)    {\small$ in $}
  \put(64,43)    {\small$ in $}
  \put(259,43)   {\small$ in $}
  \put(277,79)   {\small$ in $}
  \put(277,7)    {\small$ in $}
  \put(104,43)   {\small$ out $}
  \put(127,43)   {\small$ out $}
  \put(350,43)   {\small$ out $}
  \put(329,80)   {\small$ out $}
  \put(330,6.6)  {\small$ out $}
  \put(72,74)    {\tiny{$2$}}
  \put(79,65)    {\tiny{$1$}}
  \put(317,63)   {\tiny{$2$}}
  \put(325,54)   {\tiny{$1$}}
          } 
Note that $\dot\Xr$ and $\dot\Xr'$ have different topology.


\subsection{Modular tensor categories and three-dimensional
  topological field theory}\label{sec:mtc-tft}

The starting point of the algebraic formulation of the sewing
constraints is a {\em modular tensor category\/} $\Cc$. By this we mean
a strict monoidal category $\Cc$ such that\label{def:C}
\\[.3em]
(i)~~\,The tensor unit is simple.
\\[.3em]
(ii)~\,\,$\Cc$ is abelian, $\Cb$-linear and semisimple.
\\[.3em]
(iii)~\,$\Cc$ is ribbon:\,%
    \footnote{~%
    Besides the qualifier `ribbon' \cite{retu2}, which emphasises the fact that
    (see e.g.\ chapter XIV.5.1 of \cite{KAss}) the category of ribbons is
    universal for this class of categories, also the
    terms `tortile' \cite{joSt6} and `balanced rigid braided' are in use.}
  There are families $\{c_{U,V}\}$ of braiding, $\{\theta_U\}$ of
  twist, and $\{d_U,b_U\}$ of evaluation and coevaluation morphisms
  satisfying the usual properties.
\\[.3em]
(iv)~$\Cc$ is Artinian (or `finite'), i.e.\ the number of isomorphism
  classes of simple objects is finite.
\\[.3em]
(v)~\,The braiding is maximally non-degenerate: the numerical matrix $s$ with
  entries
  \be
    s_{i,j} \,{:=}\, (d_{U_j}^{}\oti\tilde d_{U_i}^{}) \circ
    [\id_{U_j^\vee}\oti(c_{U_i,U_j}^{}{\circ}\,c_{U_j,U_i}^{})\oti\id_{U_i^\vee}]
    \circ (\tilde b_{U_j}^{}\oti b_{U_i}^{})\,
  \labl{eq:s-mat-def}
  is invertible.

\medskip\noindent
Here we denote by $\{U_i\,|\,i\iN\Ic\}$ a (finite) set of representatives
of isomorphism classes of simple objects; we also take $U_0\,{:=}\,\one$ as the
representative for the class of the tensor unit.
The properties we demand of a modular tensor category are slightly
stronger than in the original definition in \cite{TUra}.

It is worth mentioning that every ribbon category is {\em sovereign\/}, i.e.\
besides the right duality given by $\{d_U,b_U\}$ there is
also a left duality (with evaluation and coevaluation morphisms to be denoted
by $\{\tilde d_U,\tilde b_U\}$), which coincides with the left duality in the
sense that ${{}^\vee\!}U\eq U^\vee$ and ${{}^\vee\!}\!f\eq f^\vee$.

\medskip

We also make use of the following notions. An {\em idempotent} is an 
endomorphism $p$ such that $p \Circ p \eq p$. A {\em retract} of an object $W$ 
is a triple $(V,e,r)$ with $e \iN \Hom(V,W)$, $r \iN \Hom(W,V)$ and $r \Circ e 
\eq \id_V$. Note that $e \Circ r$ is an idempotent in $\mathrm{End}(W)$. Because 
of property (ii) above, a modular tensor category is idempotent-complete, 
i.e.\ every idempotent is split and thus gives rise to a retract.

The {\em dual category\/} $\ol\Cc$ of a monoidal category $(\Cc, \otimes)$ is
the monoidal category $(\Cc^{\rm opp}, \otimes)$.
More concretely, when marking quantities in $\ol\Cc$ by an overline, we have
  \be\begin{array}{ll}
  {\rm Objects:}  &  \quad
    \Obj(\ol\Cc) = \Obj(\Cc)\,, \ {\rm i.e.} \ \
    \ol U \in \Obj(\ol\Cc) \ \ {\rm iff}\ \ U \in \Obj(\Cc)\,, \\[5pt]
  {\rm Morphisms:} &  \quad
    \ol\Hom(\ol U,\ol V) = \Hom(V,U)\,, \\[5pt]
  {\rm Composition:} &  \quad
    \ol f \bar\circ \ol g = \ol{g{\circ}f}\,,   \\[5pt]
  {\rm Tensor\ product:} &  \quad
    \ol U \bar\oti \ol V = \ol{U{\otimes}V}\,, \quad
    \ol f \bar\oti \ol g = \ol{f{\otimes}g}\,, \\[5pt]
  {\rm Tensor\ unit:} &  \quad
    \bar\one = \one \,.
  \end{array}\ee
If $\Cc$ is in addition ribbon, then we can turn $\ol\Cc$ into a ribbon 
category by taking ${\ol c}_{\ol U,\ol V} \,{:=}\, \ol{(c_{U,V})^{-1}}$ and
$\ol\theta_{\ol U} \,{:=}\, \ol{(\theta_U^{-1})}$ for braiding and twist,
and ${\ol b}_{\ol U} \,{:=}\, \ol{(\tilde d_U)}$, etc., for the dualities.
More details can be found e.g.\ in section 6.2 of \cite{ffrs}.

Alternatively, as in \cite[sect.\,7]{muge9} one can define a category
$\widetilde\Cc$ identical to $\Cc$ as a monoidal category, but
with braiding and twist replaced by their inverses. As $\Cc$ is ribbon,
we have a duality compatible with braiding and twist, and it turns out
that $\ol\Cc$ and $\widetilde\Cc$ are equivalent as ribbon categories.
For our purposes it is more convenient to work with $\ol\Cc$.

Let $\CxCb$ be the product of $\Cc$ and $\overline{\Cc}$ in the sense of 
enriched (over $\Vect$) categories, i.e.\ the modular tensor category 
obtained by idempotent completion of the category whose objects are pairs of 
objects of $\Cc$ and $\overline{\Cc}$ and whose morphism spaces are tensor 
products over $\Cb$ of the morphism spaces of $\Cc$ and $\overline{\Cc}$ 
(see \cite[Definition 6.1]{ffrs} for more details).

\medskip

Next we briefly state our conventions for the 3-d TFT that we will use; for 
more details see e.g.\ \cite{TUra,BAki,KArt} or section 2 of \cite{tft1}.

Given a modular tensor category $\Cc$, the construction of \cite{TUra} allows 
one to construct a 3-d TFT, that is, a monoidal functor $\tftC$ from a 
geometric category $\mathcal{G}_\Cc$ to $\Vect$.\label{def:tftC} The 
geometric category is defined as follows. An object $\mathrm E$ of 
$\mathcal{G}_\Cc$ is an {\em extended surface\/}; an oriented, closed surface 
with a finite number of marked arcs labeled by pairs $(U,\epsilon)$, where 
$U\iN \mathcal{O}bj(\Cc)$ and $\epsilon\in\{+,-\}$, and with a choice of 
Lagrangian subspace $\lambda\,{\subset}\, H^1(\mathrm E,\Rb)$. Following 
\cite{stte}, we define a morphism $a{:}\ \mathrm E\To \mathrm F$ to be one of 
two types: 
\\
(i) a homeomorphism of extended surfaces (a homeomorphism of the underlying 
surfaces preserving orientation, marked arcs and Lagrangian subspaces) 
\\
(ii) a triple $(\Mr, n, h)$ where $\Mr$ is a cobordism of extended surfaces,
$h{:}\ \partial\Mr\To\bar{\mathrm E}\sqcup\Fr$ is a homeomorphism of extended 
surfaces, and $n\iN\mathbb{Z}$ is a weight which is needed (see 
\cite[sect.\,IV.9]{TUra}) to make $\tftC$ anomaly-free. The cobordism $\Mr$ can
contain ribbons, which are labeled by objects of $\Cc$ and coupons, which are 
labeled by morphisms of $\Cc$. Ribbons end on coupons or on the arcs of 
$\mathrm E$ and $\mathrm F$. We denote by $h^-$ and $h^+$ the restrictions of 
$h$ to the in-going component $\partial_-\Mr$ of $\partial\mathrm{M}$ (the 
pre-image of $\bar{\mathrm E}$ under $h$) and the out-going component 
$\partial_+\mathrm{M}$ (the pre-image of $\mathrm F$). 

Two cobordisms $(\Mr,n,h)$ and $(\Mr',n,h')$ from ${\rm E}$ to ${\rm F}$
are {\em equivalent\/} iff there exists a homeomorphism 
$\varphi{:}\ \Mr \To \Mr'$ taking ribbons and coupons of $\Mr$ to identically 
labeled ribbons and coupons of $\Mr'$ and obeying $h \eq h' \Circ \varphi$. 
The functor $\tftC$ is constant on equivalence classes of cobordisms.

Composition of two morphisms is defined as 
follows: For $f{:}\ \mathrm E\To \mathrm E'$ and $g{:}\ \mathrm E'\To \mathrm F$
both homeomorphisms, the composition is simply the composition $g\Circ f{:}\ 
\mathrm E\To \mathrm F$ of maps. Morphisms $(\mathrm{M}_1,n_1,h_1){:}\ 
\mathrm E\To \mathrm E'$ and $(\Mr_2,n_2,h_2){:}\ \mathrm E'\To\Fr$ are composed
to $(\mathrm{M},n,h){:}\ \mathrm E\To \Fr$, where $\mathrm{M}$ is the cobordism 
obtained by identifying points on $\partial_+\Mr_1$ with points on 
$\partial_-\Mr_2$ using the homeomorphism $(h_2^-)^{-1}\circ h_1^+$. The 
homeomorphism $h{:}\ \partial\mathrm{M}\To \bar {\mathrm E}\sqcup \mathrm F$ is 
defined by $h|_{\partial_-\mathrm{M}}\,{:=}\, h_1^-$, $h|_{\partial_+\mathrm{M}}
\,{:=}\, h_2^+$, and the integer $n$ is computed from the two morphisms 
$\mathrm E\To \mathrm E'$ and $\mathrm E'\To\Fr$ according to an algorithm 
described in \cite[sect.\,IV.9]{TUra}. Composition of a homeomorphism 
$f{:}\ \mathrm E\To\mathrm E'$ and a cobordism $(\mathrm{M},n,g){:}\ \mathrm E'
\To \mathrm F$ is the cobordism $(\mathrm{M},n,h){:}\ \mathrm E\To \mathrm F$,
where $h{:}\ \partial\mathrm{M}\To \bar {\mathrm E}\,{\sqcup}\,\mathrm F$
is defined as $h|_{\partial_-\mathrm{M}}\,{:=}\, f^{-1}\Circ g^-$, 
$h|_{\partial^+\mathrm{M}} \,{:=}\, g^+$. The category $\mathcal{G}_\Cc$ is a
strict monoidal category with monoidal structure given by disjoint union,
and the empty set (interpreted as an extended surface) as the tensor unit.

\medskip

Given a modular tensor category $\Cc$ with label set $\Ic$ for
representatives of the simple objects, consider the objects
  \be
  \Omeka := \bigoplus_{k\in\Ic} U_k \qquad \text{and} \qquad
  H := \bigoplus_{k\in\Ic} U_k\otimes U_k^\vee
  \labl{eq:omega-def}
in $\Cc$. Note that we can choose a nonzero epimorphism $r_H$ from 
$\Omeka \oti \Omeka^\vee \eq \bigoplus_{i,j\in\Ic} U_i\oti U_j^\vee$ to $H$.
The dimension of the category $\Cc$ is defined to be that of the object $H$,
  \be
  \mathrm{Dim}(\Cc) = \dim(H) = \sum_{k\in\Ic} \dim(U_k)^2 \,.
  \ee
The objects $H$ and $\Omeka$ are useful to describe the state spaces
of the 3-d TFT constructed from $\Cc$: Let $\mathrm E$
be a connected extended surface of genus $g$ with marked points
$\{ (V_i,\eps_i) \,|\, i\eq1,...\,,m \}$ where $\eps_i\iN\{\pm1\}$. By 
construction \cite[sect.\,IV.2.1]{TUra}, the state space $\Htft(\mathrm E)$
is isomorphic to
  \be
  \Hom\big(\bigotimes_{i=1}^m V_i^{\eps_i} \oti H^{\otimes g} ,\one\big) \,,
  \labl{eq:E-Hom-space}
where $V_i^+ \eq V_i$ and $V_i^- \eq V_i^\vee$. An isomorphism
can be given by choosing a handle body $\hb$ with $\partial \hb \eq \mathrm E$,
inserting a coupon labeled by an element in \erf{eq:E-Hom-space} such that
the $V_i$-ribbons starting at the boundary arcs are joined to the ingoing side
of the coupon. For each $H$-ribbon attached to the coupon insert the restriction 
morphism $r_H$ from $\Omeka \oti \Omeka^\vee$ to $H$ and a $\Omeka$-ribbon 
starting and ending at this restriction morphism, 
so that one $\Omeka$-ribbon passes through each of the handles of $\hb$. 
For example, if $\mathrm E$ is a genus two surface with marked arcs labeled by
$(U,+)$, $(V,-)$ and $(W,+)$, then we have $f \iN \Hom(U \oti V^\vee \oti W
\oti H \oti H,\one)$ and the relevant handle body is
  \eqpic{pic_fjfrs2_59} {210} {45} {
  \put(41,10)    {\Includeourbeautifulpicture 59{}}
  \put(0,52)     {$ \hb~= $}
  \put(52,33)    {\tiny{$(U,+)$} }
  \put(76,27)    {\tiny{$(V,-)$} }
  \put(99,23)    {\tiny{$(W,+)$} }
  \put(125,30)   {\tiny{$K$} }
  \put(176,36)   {\tiny{$K$} }
  \put(182,50.5) {\tiny{$r_H$} }
  \put(137,42)   {\tiny{$r_H$} }
  \put(177,63)   {\tiny{$H$} }
  \put(143,52)   {\tiny{$H$} }
  \put(108,43)   {\tiny{$W$} }
  \put(85,42)    {\tiny{$V$} }
  \put(77,53)    {\tiny{$U$} }
  \put(128,70.5) {\tiny{$f$} }
  }

We call the cobordism from $\emptyset$ to $\mathrm E$ obtained in this way
a {\em handle body for\/} $\mathrm E$ and denote it by $\hb(f)$, 
where $f$ is the element of \erf{eq:E-Hom-space} labeling the coupon. Then
  \be
  f \longmapsto \tftC(\hb(f))
  \labl{eq:handle-iso}
is an isomorphism from \erf{eq:E-Hom-space} to the space of linear maps from 
$\Cb$ to $\Htft(\mathrm E)$, which we may identify with $\Htft(\mathrm E)$.
For non-connected $\mathrm E$ one defines $\Htft(\mathrm E)$
as the tensor product of the state spaces of its connected components.

\medskip

Later on we will need the morphism $\wO \iN \Hom(\Omeka,\Omeka)$ given by
  \be
  \wO := \frac{1}{\text{Dim}(\Cc)}\sum_{k\in\Ic} \dim(U_k)\, P_k \,,
  \labl{eq:wO}
where $P_k \iN \Hom(\Omeka,\Omeka)$ is the idempotent projecting onto
the subobject $U_k$ of $\Omeka$. Note that
  \be
  \mathrm{tr}(\wO)
   = \frac{1}{\text{Dim}(\Cc)} \sum_{k\in\Ic} \dim(U_k)^2 = 1 \,.
  \labl{eq:Omeka-trS}
Let $V$ be an object of $\Cc$ and let $e_{k\alpha} \iN \Hom(U_k,V)$ and 
$r_{k\alpha} \iN \Hom(V,U_k)$ be embedding and restriction morphisms of the 
various subobjects $U_k$, so that we have $\sum_{k,\alpha} e_{k\alpha} 
\Circ r_{k\alpha} \eq \id_V$. The following identity holds:
  \eqpic{eq:omega-project} {190} {44} {
  \put(15,10)    {\Includeournicelargepicture 20a}
  \put(135,0)    {\Includeournicelargepicture 20b}
  \put(78,49)    {$ =~\dsty\sum_\alpha $}
  \put(-4,57)    {\footnotesize{$ w_K$} }
  \put(25,78)    {\tiny{$ K $} }
  \put(138.5,61) {\small$e_{0\alpha}$}
  \put(138.5,39) {\small$r_{0\alpha}$}
  \put(142.3,-10){\sse$V$}
  \put(143,109)  {\sse$V$}
  \put(35.7,0)   {\sse$V$}
  \put(36.4,102) {\sse$V$}
  }
To see this, note that (by the properties of the matrix $s$ for a modular 
tensor category)
  \bea \begin{picture}(420,57)(5,50)
  \put(107,-10)  {\Includeournicelargepicture 21b}
  \put(290,0)    {\Includeournicelargepicture 21a}
  \put(0,46)     {$ \dsty\frac1{\text{Dim}(\Cc)}\sum_l\text{dim}(U_l) $}
  \put(155,46)   {$ =~~\dsty\frac1{\text{Dim}(\Cc)}\sum_l s_{0,l}\,
                    \frac{s_{k,l}}{s_{0,k}} $}
  \put(331,46)   {$ =~~\delta_{k,0} \,e_{k\alpha} \Circ r_{k\alpha}\,, $}
  \put(119,36)   {\tiny$ U_k $}
  \put(127,54)   {\tiny$ U_l $}
  \put(302,40)   {\tiny$ U_k $}
  \put(110,69)   {\small$e_{k\alpha}$}
  \put(110,26)   {\small$r_{k\alpha}$}
  \put(293,69)   {\small$e_{k\alpha}$}
  \put(293,26)   {\small$r_{k\alpha}$}
  \put(119,5)    {\tiny$ V $}
  \put(119,90)   {\tiny$ V $}
  \put(302,5)    {\tiny$ V $}
  \put(302,90)   {\tiny$ V $}
  \epicture29 \labl{pic-fjfrs2_21}
which implies that when expanding $\id_V$ into a sum over the identity morphisms
for the simple subobjects $U_k$ of $V$, only $U_0 \eq \one$ gives
a nonzero contribution, so that one arrives at \erf{eq:omega-project}.


\subsection{Assigning the space of blocks to a world sheet}%
\label{sec:spaceblocks}

The sewing constraints will be formulated as a natural transformation between 
two symmetric monoidal functors $\triv$ and $\bl$ from $\Wor$ to the category 
$\Vect$ of finite-dimensional complex vector spaces. The first one is 
introduced in
\\[-2.3em]

\dtl{Definition}{def:triv}
The functor $\triv$ from $\Wor$ to $\Vect$ is given by setting
  \be
  \triv(\Xr) := \Cb \qquad{\rm and}\qquad \triv(\xm) := \id_\Cb
  \labl{eq:triv-def}
for objects $\Xr$ and morphisms $\xm$ of $\Wor$, respectively.

\bigskip

The second functor, $\bl \,{\equiv}\, \bl(\Cc,\Hop,\Hcl,B_l,B_r,e,r)$ is 
obtained as follows. We first assign to a world sheet $\Xr$ an extended 
surface $\Xhat$, called the decorated double of $\Xr$; to $\Xhat$ we can 
apply the 3-d TFT functor $\Htft$ obtained from a modular tensor 
category $\Cc$; finally we select a suitable subspace of $\Htft(\Xhat)$. 
Analogous steps must be performed for morphisms. The precise description 
of these manipulations will take up most of the rest of this section.
 
We will call the vector space that $\bl$ assigns to a world sheet $\Xr$ the 
{\em space of blocks for\/} $\Xr$. It depends on the following data.
\\[.2em]
\nxt A modular tensor category $\Cc$.
\\[.2em]
\nxt A nonzero object $\Hop$ of $\Cc$, called the {\em open state space\/},
 and a nonzero object $\Hcl$ of $\CxCb$, called the {\em closed state space\/}.
\label{def:HclHop}
\\[.2em]
\nxt 
Auxiliary objects $B_l$ and $B_r$ of $\Cc$, together with morphisms $e \iN 
\Hom_{\Cc \boxtimes \ol\Cc}(\Hcl,\BlxBr)$ and $r \iN \Hom_{\CxCb}(\BlxBr,\Hcl)$
such that $(\Hcl,e,r)$ is a retract of $\BlxBr$.
\\[.3em]
At the end of this section we will show that different 
realisations of $\Hcl$ as a retract lead to equivalent functors $\bl$.

\medskip

As a start, 
from a world sheet $\Xr$ we construct an extended surface $\Xhat \;{\equiv}\;
\Xhat(\Hop,B_l,B_r)$,\label{def:Xhat} 
which we call the {\em decorated double of\/} $\Xr$. It is obtained by gluing
a standard \disc\ with a marked arc to each boundary component of $\Xtil$:
Let $\vec D$ be the unit \disc\ $\{|z|\,{\le}\,1\} \,{\subset}\, \Cb$ with
a small arc embedded on the real axis, centered at $0$ and pointing towards
$+1$. The orientation of $\vec D$ is that induced by $\Cb$. Then we set
  \be
  \Xhat := \Xtil \sqcup \big( \pi_0(\partial\Xtil) \Times \vec D \big) /{\sim}
  \ee
where the equivalence relation divided out specifies the gluing 
in terms of the boundary parametrisation according to
  \be
  (a,z) \sim \Varphi_a^{-1} \Circ C(-z) \qquad {\rm for}\quad
  a \iN \pi_0(\partial\Xtil),\, z \iN \partial \vec D \,.
  \ee
(Here the complex conjugation $C$ is needed for $\Xhat$ to be oriented.)
Further, for $a \iN \pi_0(\partial\Xtil)$
the arc on the \disc\ $\{a\} \Times \vec D$ is marked by $(U_a,\eps_a)$,
where $U_a \iN \{\Hop,B_l,B_r\}$ and $\eps_a \iN \{\pm\}$ are chosen
as follows.
\\[.2em]
\nxt If $\iota_*(a) \eq a$, then $U_a \eq \Hop$. If in addition
$a\iN b^\text{in}$, then $\eps_a\eq{+}\,$,\, otherwise $\eps_a\eq{-}\,$.
\\[.2em]
\nxt If $\iota_*(a) \,{\neq}\, a$ and the boundary component $a$ lies in the 
image of the orientation $\,\text{or}$\,, then $U_a\eq B_l$. If in addition 
$a\iN b^\text{in}$, then $\eps_a\eq{+}\,$,\, otherwise $\eps_a\eq{-}\,$.
\\[.2em]
\nxt If $\iota_*(a) \,{\neq}\, a$ and the boundary component $a$ does not lie in
the image of the orientation $\,\text{or}$\,, then $U_a\eq B_r$. If in addition 
$a\iN b^\text{in}$, then $\eps_a\eq{-}\,$,\, otherwise $\eps_a\eq{+}\,$.

Note that the involution $\iota{:}\ \Xtil \To \Xtil$ can be extended
to an involution $\hat\iota{:}\ \Xhat \To \Xhat$ by taking it to be
complex conjugation on each of the \disc s $\vec D$ glued to $\Xtil$.
Finally, to turn $\Xhat$ into an extended surface we need to
specify a Lagrangian subspace $\lambda \,{\subset}\, H^1(\Xhat,\Rb)$.
To do this we start by taking the {\em connecting manifold\/}
  \be
  \Mr_{\Xr} = \Xhat \Times [-1,1] /{\sim}
  \qquad \text{where for all}~ x\iN\Xhat ~,~~ (x,t) \sim (\hat\iota(x),-t) \,,
  \labl{eq:conmf-def}
which has the property that $\partial \textrm{M}_{\Xr} \eq \Xhat$. Then 
$\lambda$ is the kernel of the resulting homomorphism 
$H^1(\Xhat,\Rb) \To H^1(\textrm{M}_{\Xr},\Rb)$.
We refer to appendix B.1 of \cite{tft5} for more details.

\medskip

As an example, consider the world sheet $\Xr$ in \erf{eq:ws-example-1}. In 
this case the decorated double is given by a sphere with six marked arcs,
  \eqpic{pic-fjfrs2_47} {230} {45} {
  \put(115,0)    {\Includeournicelargepicture 47{}}
  \put(0,50)     {$ \Xhat(\Hop,B_l,B_r)~= $}
  \put(146,10)   {\tiny{$(B_r{,}-)$} }
  \put(133,36)   {\tiny{$(\Hop{,}+)$} }
  \put(136,69)   {\tiny{$(B_l{,}+)$} }
  \put(175,30)   {\tiny{$(B_r{,}+)$} }
  \put(175,68)   {\tiny{$(\Hop{,}-)$} }
  \put(169,91)   {\tiny{$(B_l{,}-)$} }
  }

\medskip

The 3-d TFT assigns to the extended surface $\Xhat\,{\equiv}\,\Xhat(\Hop,B_l,
B_r)$ a complex vector space $\Htft(\Xhat)$. In order to define the action of 
the functor $\bl$ on objects of $\Wor$, one needs to reduce the auxiliary object
$\BlxBr$ to its retract $\Hcl$ in $\CxCb$ (this will also show that the precise 
choice of objects $B_l$ and $B_r$ is immaterial). To this end we first introduce
a certain linear map between the vector spaces assigned to decorated doubles. 
More specifically, given a world sheet $\Xr$, a choice of objects $\Hop,B_l,B_r,
\Hop',B_l',B_r'$ and morphisms $o^\text{in} \iN \Hom_\Cc(\Hop',\Hop)$ and
$o^\text{out} \iN \Hom_\Cc(\Hop,\Hop')$, as well as
$c^\text{in} \iN \Hom_{\CxCb}(B_l'\times\ol B_r',\BlxBr)$ and
$c^\text{out} \iN \Hom_{\CxCb}(\BlxBr,B_l'\times\ol B_r')$,
we will introduce a linear map
  \be
  F_\Xr(o^\text{in},o^\text{out},c^\text{in},c^\text{out})
  :\quad \Htft\big( \Xhat(\Hop,B_l,B_r)\big) \To
  \Htft\big(\Xhat(\Hop',B_l',B_r') \big) \,.
  \labl{eq:F_X-def}
The slightly tedious definition proceeds as follows. Since the morphism spaces 
of $\CxCb$ are given in terms of tensor products, we can write
  \be
  c^\text{in} = \sum_{\alpha \in I^\text{in}}
  c^\text{in}_{l,\alpha} \oti c^\text{in}_{r,\alpha} \qquad \text{and} \qquad
  c^\text{out} = \sum_{\beta \in I^\text{out}}
  c^\text{out}_{l,\beta} \oti c^\text{out}_{r,\beta}
  \labl{eq:c-decomp}
with suitable morphisms $c^\text{in}_{l,\alpha} \iN \Hom_\Cc(B_l',B_l)$,
$c^\text{in}_{r,\alpha} \iN \Hom_\Cc(B_r,B_r')$,
$c^\text{out}_{l,\beta} \iN \Hom_\Cc(B_l,B_l')$ and
$c^\text{out}_{r,\beta} 
        $\linebreak[0]$
\iN \Hom_\Cc(B_r',B_r)$,
and index sets $I^\text{in}$ and $I^\text{out}$. 
Denote by $S^\text{in}$ all in-going closed state boundaries of $\Xr$, i.e.\ 
$S^\text{in} \eq \{ a\iN b^\text{in} \,|\, \iota_*(a)\,{\ne}\,a \}$, and 
similarly $S^\text{out} \eq \{ a\iN b^\text{out} \,|\, \iota_*(a)\,{\ne}\,a \}$.
We say that a map $\alpha{:}\ S^\text{in} \To I^\text{in}$ or
$\alpha{:}\ S^\text{out} \To I^\text{out}$ is
$\iota$-{\em invariant\/} iff $\alpha \Circ \iota \eq \alpha$.
Given two $\iota$-invariant maps $\alpha{:}\ S^\text{in} \To I^\text{in}$ and
$\beta{:}\ S^\text{out} \To I^\text{out}$, we construct a cobordism
  \be
  N_\Xr(\alpha,\beta)
  :\quad \Xhat(\Hop,B_l,B_r) \To \Xhat(\Hop',B_l',B_r')
  \labl{eq:NX-def}
as follows. Start from the cylinder $\Xhat(\Hop,B_l,B_r) \Times [0,1]$. On each 
vertical ribbon insert a coupon. Relabel the half of the vertical ribbon between
the coupon and the out-going boundary component $\Xhat(\Hop,B_l,B_r) \Times \{1\}$ 
from $\Hop,B_l,B_r$ to $\Hop',B_l',B_r'$, respectively. The coupon
attached to a ribbon starting on the \disc\ $\{a\}{\times}\vec D
\,{\subset}\, \Xhat(\Hop,B_l,B_r)$ for some $a \iN \pi_0(\partial\Xtil)$
is labeled by $o^\text{in}$, $c^\text{in}_{l,\alpha(a)}$ or 
$c^\text{in}_{r,\alpha(a)}$ if $a \iN b^\text{in}$, and
$o^\text{out}$, $c^\text{out}_{l,\beta(a)}$ or $c^\text{out}_{r,\beta(a)}$
if $a \iN b^\text{out}$. Which of the three morphisms one must choose is 
determined in an obvious manner by the labels of the
ribbons attached to the coupon.

\medskip

For the world sheet $\Xr$ from example \erf{eq:ws-example-1}, 
the cobordism $N_\Xr(\alpha,\beta)$ looks as follows.
  \eqpic{pic-fjfrs2_48} {265} {64} {
  \put(85,0)    {\Includeourbeautifulpicture 48{}}
  \put(0,69)    {$ N_\Xr(\alpha,\beta)~= $}
  \put(129,14)  {\tiny$ B_r'$}
  \put(142,37)  {\tiny$ B_r$}
  \put(114,37)  {\tiny$ c_{r,\alpha}^\text{in}$}
  \put(93,78)   {\tiny$ \Hop'$}
  \put(118.5,78){\tiny$ \Hop$}
  \put(105,60)  {\tiny$ o^\text{in}$}
  \put(130,125) {\tiny$ B_l'$}
  \put(142,104) {\tiny$ B_l$}
  \put(116,105) {\tiny$ c_{l,\alpha}^\text{in}$}
  \put(189,117) {\tiny$ B_l'$}
  \put(176,95)  {\tiny$ B_l$}
  \put(159,112) {\tiny$ c_{l,\beta}^\text{out}$}
  \put(186,78)  {\tiny$ \Hop$}
  \put(210,78)  {\tiny$ \Hop'$}
  \put(194,60)  {\tiny$ o^\text{out}$}
  \put(191,21)  {\tiny$ B_r'$}
  \put(177,44)  {\tiny$ B_r$}
  \put(159,25)  {\tiny$ c_{r,\beta}^\text{out}$}
  \put(249,30)  {\small$ \widehat{X}\Times \{0\}$}
  \put(244,133) {\small$ \widehat{X}\Times \{1\}$}
  }

\medskip

The linear map $F_\Xr(o^\text{in},o^\text{out},c^\text{in},c^\text{out})
{:}\ \Htft\big( \Xhat(\Hop,B_l,B_r)\big) \To
\Htft\big(\Xhat(\Hop',B_l',B_r') \big)$ is given by
  \be
  F_\Xr(o^\text{in},o^\text{out},c^\text{in},c^\text{out})
  = \sum_{\alpha,\beta}
  \tftC\big( N_\Xr( \alpha,\beta )\big) \,,
  \labl{eq:FX-def}
where the sum is over all $\iota$-invariant functions $\alpha {:}\ S^\text{in} 
\To I^\text{in}$ and $\beta {:}\ S^\text{out} \To I^\text{out}$. This definition
is independent of the choice of decomposition \erf{eq:c-decomp} because the 
functor $\tftC$ is multilinear in the labels of the coupons.

Suppose we are in addition given objects $\Hop''$, $B_l''$, $B_r''$
and morphisms $p^\text{in} \iN \Hom_\Cc(\Hop'',\Hop')$,
$p^\text{out} \iN \Hom_\Cc(\Hop',\Hop'')$ as well as
$d^\text{in} \iN \Hom_{\CxCb}(B_l''\Times\ol B_r'',B_l'\Times\ol B_r')$ and
$d^\text{out} \iN \Hom_{\CxCb}(B_l'\Times\ol B_r',
  $\linebreak[0]$
B_l''\Times\ol B_r'')$.
Using the definition of $F_\Xr$ and functoriality of $\tftC$ one can verify that
  \be
  F_\Xr(p^\text{in}\!,p^\text{out}\!,d^\text{in}\!,d^\text{out}) \,
  F_\Xr(o^\text{in}\!,o^\text{out}\!,c^\text{in}\!,c^\text{out})
  = F_\Xr(
    o^\text{in} {\circ}\, p^\text{in}\!,
    p^\text{out}{\circ}\, o^\text{out}\!,
    c^\text{in} {\circ}\, d^\text{in}\!,
    d^\text{out}{\circ}\, c^\text{out} ) \,.
  \labl{eq:FX-comp}
We have now gathered all the ingredients for defining the functor $\bl$ on 
objects of $\Wor$. Denote by $P_\Xr\eq P_\Xr(\Hop,\Hcl,B_l,B_r,e,r)$ the 
endomorphism of $\Htft(\Xhat(\Hop,B_l,B_r))$ that is given by
  \be
  P_{\Xr} := F_\Xr(\id_{\Hop},\id_{\Hop},e \Circ r,e \Circ r)
  \labl{eq:PX-def}
with morphisms $e$ and $r$ such that $(\Hcl,e,r)$ is a retract of $\BlxBr$.
Equation \erf{eq:FX-comp} immediately implies that
$P_\Xr P_\Xr \eq P_\Xr$, i.e.\ $P_\Xr$ is an idempotent.

Now we define, for a world sheet $\Xr$,
  \be
  \bl(\Xr) := \text{Im}(P_{\Xr}) \subseteq \Htft( \Xhat ) \,,
  \labl{eq:bl(X)-def}
where we abbreviate $\bl \,{\equiv}\, \bl(\Cc,\Hop,\Hcl,B_l,B_r,e,r)$,
$P_\Xr\eq P_\Xr(\Hop,\Hcl,B_l,B_r,e,r)$ as well as
$\Xhat \,{\equiv}\, \Xhat(\Hop,B_l,B_r)$.

\medskip

Next we turn to the definition of $\bl(\xm)$ for a morphism
$\xm \eq (\sew,f) \iN \Hom(\Xr,\Yr)$ of $\Wor$. First note that we can extend 
the isomorphism $f{:}\ \widetilde{\sew(\Xr)} \To \widetilde Y$
to an isomorphism $\hat f {:}\ \widehat{\sew(\Xr)} \To \widehat{Y}$ by taking 
it to be the identity map on the \disc s $\vec D$ which are glued to the 
boundary components of $\widetilde{\sew(\Xr)}$ and $\widetilde Y$. To $\hat f$ 
the 3-d TFT assigns an isomorphism $\tftC(\hat f){:}\, \Htft(\widehat{\sew(\Xr)}) 
\To \Htft(\widehat Y)$. Next we construct a morphism
$\hat\sew{:}\, \Xhat \To \widehat{\sew(\Xr)}$ as a cobordism. It is given by 
the cylinder over $\Xhat$ modulo an equivalence relation,
  \be
  \hat\sew := \Xhat \times [0,1] / {\sim} \,.
  \labl{eq:shat-cobord}
The equivalence relation identifies certain points on the boundary $\Xhat\Times 
\{1\}$ of $\Xhat \Times [0,1]$. Namely, for each pair $(a,b) \iN \sew$ and 
for all $z \iN \vec D$ we identify the points $(a,z,1) \iN \{a\}\Times\vec D 
\Times\{1\}$ and $(b,C(-z),1) \iN \{b\}\Times\vec D\Times\{1\}$. In terms of the 
morphisms $\hat f$ and $\hat\sew$ we now define, for $\xm \eq (\sew,f)$,
  \be
  \bl(\xm) := \tftC(\hat f) \circ \tftC(\hat\sew) \big|_{\bl(\Xr)} \,,
  \labl{eq:bl(m)-def}
i.e.\ the restriction of the linear map $\tftC(\hat f) \Circ \tftC(\hat\sew)$ 
to the subspace $\bl(\Xr)$ of $\Htft(\Xhat)$. As it stands, $\bl(\xm)$ is a 
linear map from $\bl(\Xr)$ to $\Htft(\widehat Y)$. We now must verify that
the image of $\bl(\xm)$ is indeed contained in $\bl(\Yr)$. This follows from
  \be
  P_\Yr \circ \bl(\xm) \circ P_\Xr = \bl(\xm) \circ P_\Xr \,,
  \ee
which can again be checked by substituting the definitions.
Note that, on the other hand, $P_\Yr \Circ \bl(\xm) \Circ P_\Xr$ is in general 
not equal to $P_\Yr \Circ \bl(\xm)$.

The discussion above is summarised in the
\\[-2.5em]

\dtl{Definition}{def:bl} 
The {\em block functor\/}
  \be
  \bl \equiv \bl(\Cc,\Hop,\Hcl,B_l,B_r,e,r): \quad \Wor \to \Vect
  \ee
is the assignment 
\erf{eq:bl(X)-def} on objects and \erf{eq:bl(m)-def} on morphisms of $\Wor$. 

\medskip

That $\bl$ is indeed a functor is established in 
\\[-2.5em]

\dtl{Proposition}{prop:bl-funct}
The mapping $\bl {:}\ \Wor \To \Vect$ is a symmetric monoidal functor.

\medskip\noindent
Proof:\\
We must show that $\bl(\id_\Xr) \eq \id_{\bl(\Xr)}$ and $\bl(\xm \Circ \xm') 
\eq \bl(\xm) \Circ \bl(\xm')$ (functoriality), that $\bl(\emptyset) \eq \Cb$,
$\bl(\Xr \,{\sqcup}\, \Yr) \eq \bl(\Xr) \oti \bl(\Yr)$
and $\bl(\xm \,{\sqcup}\, \xm') \eq \bl(\xm) \oti \bl(\xm')$ (monoidal),
and finally that $\bl(c_{\Xr,\Yr}) \eq c_{\bl(\Xr),\bl(\Yr)}$ (symmetric).
Here, $c_{U,V} \iN \Hom(U \oti V,V \oti U)$ is the isomorphism
in $\Vect$ that exchanges the two factors in a tensor product.
\\[2pt]
Functoriality and symmetry follow immediately from the definition
\erf{eq:bl(m)-def} and functoriality of $\tftC$. The same holds for the
monoidal property on morphisms. To verify that
$\bl$ is monoidal also on objects one uses in addition
that the projector \erf{eq:PX-def}, in terms of which $\bl$ is defined,
satisfies $P_{\Xr \,{\sqcup}\, \Yr} \eq P_\Xr \oti P_\Yr$. This latter
property is not difficult to see upon substituting the explicit definition 
\erf{eq:FX-def} of $F_\Xr$ in terms of cobordisms.
\qed

\dt{Remark}
The definition of $\bl$ is closely related to that of a two-dimensional
modular functor, see 
\cite{sega8} as well as \cite{Mose}, \cite[chapter 5]{BAki},
\cite[chapter V]{TUra} or \cite{anue3}. The main difference is that $\bl$ 
starts from a different category, namely one in which the two-dimensional 
surfaces are in addition equipped with an involution, and in which the boundaries
of the surface are not labeled by objects of some decoration category.

\medskip

Next we show that any two functors $\bl$ that are constructed in the manner 
described above from isomorphic objects $\Hop$ and $\Hcl$ are equivalent as 
symmetric monoidal functors. We abbreviate $\bl\eq \bl(\Cc,\Hop,\Hcl,B_l,
B_r,e,r)$ and $\bl'\eq \bl(\Cc,\Hop',\Hcl',B_l',B_r',e',r')$. Let further
$\varphi_\text{op} \iN \Hom_\Cc(\Hop',\Hop)$ and $\varphi_\text{cl} 
\iN \Hom_{\CxCb}(\Hcl',\Hcl)$ be isomorphisms. Define linear maps 
  \be
  \beta_\Xr(\varphi_\text{op},\varphi_\text{cl}){:}\quad \Htft\big(\Xhat
  (\Hop,B_l,B_r)\big) \to \Htft\big(\Xhat(\Hop',B_l',B_r')\big)
  \ee
by setting
  \be
  \beta_\Xr(\varphi_\text{op},\varphi_\text{cl})
  := F_\Xr( \varphi_\text{op}, \varphi_\text{op}^{-1},
  e \circ \varphi_\text{cl} \circ r',
  e' \circ \varphi_\text{cl}^{-1} \circ r ) \,.
  \ee
We now show that $\beta_\Xr$ restricts to an isomorphism from $\bl(\Xr)$ to 
$\bl(\Xr')$. Note that, with the abbreviations 
$P_\Xr \eq P_\Xr(\Hop,\Hcl,B_l,B_r,e,r)$
and $P_\Xr' \eq P_\Xr(\Hop',\Hcl',B_l',B_r',e',r')$, we have
  \beaa
  P_\Xr' \circ \beta_\Xr \!\!\!&
  = F_\Xr(\id_{\Hop'},\id_{\Hop'},e' \Circ r',e' \Circ r') \,
  F_\Xr( \varphi_\text{op}, \varphi_\text{op}^{-1},
  e \Circ \varphi_\text{cl} \Circ r', e'\Circ \varphi_\text{cl}^{-1}\Circ r )
  \enL &
  = F_\Xr( \varphi_\text{op}, \varphi_\text{op}^{-1},
  e \Circ \varphi_\text{cl} \Circ r', e'\Circ \varphi_\text{cl}^{-1}\Circ r )
  \enL &
  = F_\Xr( \varphi_\text{op}, \varphi_\text{op}^{-1},
  e \Circ \varphi_\text{cl} \Circ r', e'\Circ \varphi_\text{cl}^{-1}\Circ r ) \,
  F_\Xr(\id_{\Hop},\id_{\Hop},e \Circ r,e \Circ r)
  = \beta_\Xr \circ P_\Xr \,,
  \end{array}
  \ee
so that $\beta_\Xr$ maps $\bl(\Xr)$ to $\bl'(\Xr)$. We can thus define a linear 
map $\Alpha_\Xr {:}\ \bl(\Xr) \To \bl'(\Xr)$ by restricting $\beta_\Xr$,
  \be
  \Alpha_\Xr(\varphi_\text{op},\varphi_\text{cl})
  := F_\Xr( \varphi_\text{op}, \varphi_\text{op}^{-1},
  e \circ \varphi_\text{cl} \circ r',
  e' \circ \varphi_\text{cl}^{-1} \circ r ) \big|_{\bl(\Xr)} \,.
  \labl{eq:alphaX-def}
~\\[-2.5em]

\dtl{Proposition}{prop:nat-iso}
Let $\bl\eq \bl(\Cc,\Hop,\Hcl,B_l,B_r,e,r)$ and
$\bl'\eq \bl(\Cc,\Hop',\Hcl',B_l',B_r',e',r')$. For any two isomorphisms
$\varphi_\text{op} \iN \Hom_\Cc(\Hop',\Hop)$ and
$\varphi_\text{cl} \iN \Hom_{\CxCb}(\Hcl',\Hcl)$, the family $\{ \Alpha_\Xr
(\varphi_\text{op},\varphi_\text{cl}) \}$ of linear maps \erf{eq:alphaX-def} 
is a monoidal natural isomorphism from $\bl$ to $\bl'$.

\dtl{Remark}{rem:nat-iso}
In order to keep the notation simple we consider in proposition 
\ref{prop:nat-iso} only the case when $\bl$ and $\bl'$ involve the same 
modular tensor category $\Cc$. One can also define a monoidal natural 
isomorphism from $\bl$ to $\bl'$ if in $\bl'$ one allows a modular tensor 
category $\Cc'$ equivalent (as a braided monoidal category) to $\Cc$, and 
inserts the equivalence functor at the appropriate places.

\medskip

The proof of proposition \ref{prop:nat-iso} is based on two lemmas. For 
world sheets $\Xr$ and $\Yr$, consider first a homeomorphism
$f {:}\ \Xtil \To \widetilde\Yr$ of world sheets. By gluing \disc s with 
appropriately labeled arcs, from $f$ and the data in proposition 
\ref{prop:nat-iso} we obtain two morphisms of extended surfaces,
  \bea
  \hat f\,:\quad \Xhat(\Hop,B_l,B_r) \longrightarrow
  \widehat\Yr(\Hop,B_l,B_r)
  \qquad \text{and} \enL
  \hat f' :\quad \Xhat(\Hop',B_l',B_r') \longrightarrow
  \widehat\Yr(\Hop',B_l',B_r') \,.
  \end{array}
  \ee

\dtl{Lemma}{lem:nat-iso-1}
We have
  \bea
  F_\Yr(\varphi_\text{op},\varphi_\text{op}^{-1},e\Circ\varphi_\text{cl}\Circ r'
       , e' \Circ \varphi_\text{cl}^{-1} \Circ r ) \circ \tftC(\hat f)
  \\{}\\[-.7em]\hspace*{10.9em}
  = \tftC(\hat f') \circ F_\Xr(\varphi_\text{op},\varphi_\text{op}^{-1},
    e\Circ\varphi_\text{cl}\Circ r', e'\Circ\varphi_\text{cl}^{-1}\Circ r ) \,.
  \end{array}
  \ee
%
Proof:\\
The statement follows because the two cobordisms
$N_\Yr(\alpha,\beta) \Circ \hat f$ and $\hat f' \Circ N_\Xr(\alpha,\beta)$,
with $N$ as in \erf{eq:NX-def}, are in the same equivalence class of cobordisms.
\qed

On the other hand, given a sewing $\sew$ of a world sheet $\Xr$, we have
\\[-2.2em]

\dtl{Lemma}{lem:nat-iso-2}
With the arguments of $F_\Xr$ and $F_{\sew(\Xr)}$
the same as in lemma \ref{lem:nat-iso-1}, we have
  \be
  F_{\sew(\Xr)} \circ \tftC(\hat \sew) \circ P_\Xr
  = \tftC(\hat\sew) \circ F_\Xr \,.
  \labl{FstsP=tsF}
%
Proof:\\
The claim follows by substituting the various definitions in terms of 
cobordisms. The additional idempotent accounts for the projector resulting 
from composing $e \Circ \varphi_\text{cl} \Circ r'$ and
$e'\Circ \varphi_\text{cl}^{-1} \Circ r$ at sewings of closed state boundaries 
in $\tftC(\hat\sew) \Circ F_\Xr$. For example, consider the world sheet $\Xr$ 
in \erf{eq:ws-example-1} and a sewing $\sew$ which consists of sewing the two 
open state boundaries and the two closed state boundaries (the set $\sew$ thus 
consists of three pairs). Then $\sew(\Xr)$ has no state boundaries (i.e.\ 
$\widetilde{\sew(\Xr)}$ has empty boundary). Expand the morphism $p\,{:=}\,
e\Circ r$ as $p \eq \sum_\alpha p_{l,\alpha}\oti p_{r,\alpha}$. Substituting 
the definitions, we find that the left hand side of \erf{FstsP=tsF} is given by
  \eqpic{eq:nat-iso-2-aux1} {355} {65} {
  \put(160,0)    {\Includeourbeautifulpicture 49{}}
  \put(0,69)     {$ F_{\sew(\Xr)}\circ\tftC(\hat\sew)\circ P_\Xr~~=~~
                    \dsty\sum_{\alpha,\beta} $}
  \put(231,19)   {\tiny$ B_r$ }
  \put(243,38)   {\tiny$ B_r$ }
  \put(212,35)   {\tiny$ p_{r,\alpha}$ }
  \put(201,78)   {\tiny$ \Hop $}
  \put(231,123)  {\tiny$ B_l$ }
  \put(242,104)  {\tiny$ B_l$ }
  \put(216,103)  {\tiny$ p_{l,\alpha}$ }
  \put(287,114)  {\tiny$ B_l$ }
  \put(276,95)   {\tiny$ B_l$ }
  \put(259,113)  {\tiny$ p_{l,\beta}$ }
  \put(294,78)   {\tiny$ \Hop $}
  \put(288,25)   {\tiny$ B_r$ }
  \put(278,43)   {\tiny$ B_r$ }
  \put(259,26)   {\tiny$ p_{r,\beta}$ }
  \put(332,129)  {\small$ A'$ }
  \put(355,78)   {\small$ B'$ }
  \put(331,11)   {\small$ C'$ }
  \put(176,12)   {\small$ C$ }
  \put(151,82)   {\small$ B$ }
  \put(174,128)  {\small$ A$ }
  }
In this figure, the \disc\ $A$ has to be identified with the \disc\ $A'$,
as well as $B$ with $B'$, and $C$ with $C'$. The application of $\tftC$ to the 
cobordism is understood. Note that since $\widetilde{\sew(\Xr)}$ has empty 
boundary, $F_{\sew(\Xr)}$ is just the identity on $\bl(\sew(\Xr))$. For the 
right hand side of the \erf{FstsP=tsF}, write $c \eq e \Circ \varphi_\text{cl} 
\Circ r'$, $d \eq e' \Circ \varphi_\text{cl}^{-1} \Circ r$ and expand
$c \eq \sum_\alpha c_{l,\alpha} \oti c_{r,\alpha}$,
$d \eq \sum_\alpha d_{l,\alpha} \oti d_{r,\alpha}$.
Inserting the definitions, one finds
  \eqpic{pic-fjfrs2_50} {320} {63} {
  \put(125,0)    {\Includeourbeautifulpicture 50{}}
  \put(0,69)     {$ \tftC(\hat\sew)\circ F_\Xr~~=~~\dsty\sum_{\alpha,\beta} $}
  \put(196,17)   {\tiny$ B_r'$ }
  \put(207,37)   {\tiny$ B_r$ }
  \put(179,34)   {\tiny$ c_{r,\alpha}$ }
  \put(162,78)   {\tiny$ \Hop' $}
  \put(183,77.3) {\tiny$ \Hop $}
  \put(169,62)   {\tiny$ \varphi_\text{op}$ }
  \put(196,123)  {\tiny$ B_l'$ }
  \put(208,104)  {\tiny$ B_l$ }
  \put(183,103)  {\tiny$ c_{l,\alpha}$ }
  \put(252,114)  {\tiny$ B_l'$ }
  \put(241,95)   {\tiny$ B_l$ }
  \put(226,113.5){\tiny$ d_{l,\beta}$ }
  \put(251,77.3) {\tiny$ \Hop $}
  \put(276,77.3) {\tiny$ \Hop' $}
  \put(260,61)   {\tiny$ \varphi^{-1}_\text{op}$ }
  \put(253,25)   {\tiny$ B_r'$ }
  \put(242,44)   {\tiny$ B_r$ }
  \put(230,23)   {\tiny$ d_{r,\beta}$ }
  \put(295,129)  {\small$ A' $}
  \put(319,77)   {\small$ B' $}
  \put(295,11)   {\small$ C' $}
  \put(141,12)   {\small$ C $}
  \put(115,81)   {\small$ B $}
  \put(140,128)  {\small$ A $}
  }
Here the identifications are as in \erf{eq:nat-iso-2-aux1}.
Taking the morphisms $d_{l/r}$ and $\varphi_\text{op}^{-1}$ through
the identifications, one obtains
\bea
  \tftC(\hat\sew) \circ F_\Xr(\varphi_\text{op},\varphi_\text{op}^{-1},c,d)
  \enL \hspace*{5em}
  = \tftC(\hat\sew)\circ F_\Xr(\varphi_\text{op}^{-1},\id_{\Hop},d,\id_{\BlxBr})
  \circ F_\Xr(\varphi_\text{op},\id_{\Hop},c,\id_{\BlxBr})
  \enL \hspace*{5em}
  = \tftC(\hat\sew) \circ F_\Xr(\varphi_\text{op} \Circ \varphi_\text{op}^{-1},
    \id_{\Hop},c \Circ d,\id_{\BlxBr}) \,,
  \end{array}\ee
where the last step uses \erf{eq:FX-comp}. Now $c \Circ d \eq e \Circ 
\varphi_\text{cl} \Circ r' \Circ e' \Circ \varphi_\text{cl}^{-1} \Circ r 
\eq e \Circ r \eq p$. Replacing $p$ by $p \circ p$ and redoing the above 
steps in opposite order (without inserting $\varphi_\text{op}$),
one indeed arrives at \erf{eq:nat-iso-2-aux1}.
\qed

\medskip\noindent
{\bf Proof of proposition \ref{prop:nat-iso}:}\\[2pt]
To see that each of the linear maps $\Alpha_\Xr(\varphi_\text{op},
\varphi_\text{cl})$ is an isomorphism, one verifies that it has
$\Alpha_\Xr(\varphi_\text{op}^{-1},\varphi_\text{cl}^{-1})$ as a two-sided 
inverse. That $\Alpha_\Xr(\varphi_\text{op}^{-1},\varphi_\text{cl}^{-1})$ 
is a left-inverse follows directly by using the rule \erf{eq:FX-comp}:
  \bea
  \Alpha_\Xr(\varphi_\text{op}^{-1},\varphi_\text{cl}^{-1}) \,
  \Alpha_\Xr(\varphi_\text{op},\varphi_\text{cl})
  \enL \hspace*{2em}
  = F_\Xr( \varphi_\text{op}^{-1}, \varphi_\text{op},
  e' \Circ \varphi_\text{cl}^{-1} \Circ r, e \Circ \varphi_\text{cl} \Circ r' )
  \, F_\Xr( \varphi_\text{op}, \varphi_\text{op}^{-1},
  e \Circ \varphi_\text{cl} \Circ r', e' \Circ \varphi_\text{cl}^{-1} \Circ r )
  \enL \hspace*{2em}
  = F_\Xr(\id_{\Hop}, \id_{\Hop}, e \circ r, e \circ r )
  = P_\Xr  = \id_{\bl(\Xr)} \,.
  \end{array}
  \ee
The right-inverse property follows similarly.
\\
To see that $\Alpha \,{\equiv}\, \Alpha(\varphi_\text{op},\varphi_\text{cl})$
defines a natural transformation, we must check that for each morphism
$\xm {:}\, \Xr \To \Yr$ between two world sheets, the square
  \be
  \xymatrix@C=2.9em{
  \bl(\Xr) \ar[d]^{\Alpha_\Xr} \ar[r]^{\bl(\xm)} & \bl(\Yr) \ar[d]^{\Alpha_\Yr} \\
  \bl'(\Xr) \ar[r]^{\bl'(\xm)} & \bl'(\Yr)  }
  \ee
commutes. This follows from substituting definition \erf{eq:bl(m)-def} of 
$\bl(\xm)$ and \erf{eq:alphaX-def} of $\Alpha$, and applying lemmas 
\ref{lem:nat-iso-1} and \ref{lem:nat-iso-2}. Finally, the property that the 
natural transformation $\Alpha$ is monoidal amounts to the statement that
  \be
  \xymatrix{
  \bl(\Xr{\sqcup}\Yr) \ar[d]^{\Alpha_{\Xr\sqcup\Yr}} \ar[r]^{\cong~~~} &
  \bl(\Xr) \oti \bl(\Yr) \ar[d]^{\Alpha_\Xr \oti \Alpha_\Yr}   \\
  \bl'(\Xr{\sqcup}\Yr) \ar[r]^{\cong~~~} & \bl'(\Xr) \oti \bl'(\Yr)  }
  \labl{eq:bl-comm-diag}
commutes. That this is indeed satisfied is a direct consequence of the fact 
that $\tftC$ is a monoidal functor, and the isomorphisms in 
\erf{eq:bl-comm-diag} follow from isomorphisms $\tftC(\Mr\,{\sqcup}\, 
\mathrm{N}) \overset{\cong}{\longrightarrow}\tftC(\Mr) \oti \tftC(\mathrm{N})$ 
which form part of the data specifying a monoidal functor.
\qed


\subsection{Correlators and sewing constraints}\label{sec:form-sew}

With the help of the concepts introduced in sections \ref{sec:worldsheet} and 
\ref{sec:spaceblocks}, we can finally formulate the central notion of our 
investigation, namely what we call a `solution to the sewing constraints,' 
or synonymously, a `consistent collection of correlators.'
\\[-2.2em]

\dtl{Definition}{def:sf}
For given data $\Cc,\Hop,\Hcl,B_l,B_r,e,r$, a {\em solution to the sewing 
constraints\/}, or {\em consistent collection of correlators\/} is given by a
monoidal natural transformation $\cor$\label{def:cor} from $\triv$ to
the block functor $\bl\,{\equiv}\,\bl(\Cc,\Hop,\Hcl,B_l,B_r,e,r)$.
\\
We also refer to the tuple
  \be
  \sfc = (\Cc,\Hop,\Hcl,B_l,B_r,e,r,\cor)
  \ee
as a solution to the sewing constraints and
call $\cor$ the {\em collection of correlators\/}.

\medskip

Given a solution $\sfc$ we will denote the data $e$ and $r$ also by
$e_\sfc$ and $r_\sfc$, and write 
  \be 
  p_\sfc := e_\sfc \circ r_\sfc \,;
  \labl{eq:pS-intro}
$p_\sfc$ is an idempotent.
	
\medskip

Let us disentangle the meaning of this definition. 
\\[2pt] \nxt
First of all, as a natural 
transformation, $\cor$ assigns to every world sheet $\Xr$ a linear map
$\cor_\Xr{:}\ \triv(\Xr) \To \bl(\Xr)$; we call $\cor_\Xr{:}\ \Cb \To \bl(\Xr)$ 
the {\em correlator of the world sheet\/} $\Xr$. 
\\[2pt] \nxt
Next, by definition of a natural transformation, the diagram 
  \be
  \xymatrix@C=3.6em{
  \Cb= \hspace*{-5.3em} & \triv(\Xr)\ar[d]^{\cor_\Xr}\ar[r]^{~~~\triv(\xm)~~~} &
    \triv(\Yr) \ar[d]^{\cor_\Yr} & \hspace*{-5.3em} =\Cb  \\
  & \bl(\Xr) \ar[r]^{~~~\bl(\xm)~~~} & \bl(\Yr)  }
  \ee
commutes for every morphism $\xm{:}\ \Xr \To \Yr$ of world sheets.
Since $\triv(\xm) \eq\id_\Cb$, commutativity of the diagram means that
  \be
  \cor_\Yr = \bl(\xm) \circ \cor_\Xr \,.
  \labl{eq:cor-natxfer-prop}
\nxt
The relation \erf{eq:cor-natxfer-prop} expresses the {\em covariance of the 
correlators\/} under arbitrary morphisms of $\Wor$, i.e.\ both homeomorphisms 
and sewings. It includes in particular the usual covariance property, namely 
when $\xm \eq (\emptyset,f)$ for two world sheets $\Xr$ and $\Yr$ and a 
homeomorphism $f{:}\ \Xtil \To \widetilde\Yr$, i.e.\ the case that there is no 
sewing. In this case transporting $\cor_\Xr{:}\ \Cb \To \bl(\Xr)$ from 
$\bl(\Xr)$ to $\bl(\Yr)$ using the linear map $\bl\big((\emptyset,f)\big)$ 
results in $\cor_\Yr$.
\\[2pt] \nxt
Similarly, given a world sheet $\Xr$ and sewing data $\sew$ for $\Xr$,
we can apply \erf{eq:cor-natxfer-prop} to the morphism
$\xm \eq (\sew,\id_{\widetilde{\sew(\Xr)}}) {:}\ \Xr \To \sew(\Xr)$. It states 
that the correlator on $\Xr$ and on the sewn world sheet $\sew(\Xr)$ are 
related by the linear map
  \be
  \bl\big((\sew,\id_{\widetilde{\sew(\Xr)}}) \big){:}\quad
  \bl(\Xr)\To\bl(\sew(\Xr))
  \ee
between the spaces of blocks for the world sheet and the sewn word sheet. This 
expresses {\em consistency of the correlators with sewing\/}
and thereby justifies our terminology.
\\[2pt] \nxt
Finally, that $\cor$ is monoidal implies that\,%
  \footnote{~In writing this equality it is understood that one has to apply the
  natural isomorphism $\tftC(-\,{\sqcup}\,-) \overset{\cong}{\longrightarrow}
  \tftC(-) \oti \tftC(-)$ to the left hand side. Here and below
  we do not spell out this isomorphism explicitly. }
  \be
  \cor_{\Xr \sqcup \Yr} = \cor_\Xr \oti \cor_\Yr \,,
  \labl{eq:cor-tensor-prop}
i.e.\ the correlator evaluated on a disconnected world sheet 
$\Xr \,{\sqcup}\, \Yr$ is just the tensor product of the correlators 
evaluated on the individual world sheets $\Xr$ and $\Yr$.


\subsection{Equivalence of solutions to the sewing constraints}
\label{sec:sew-equiv}

It is not difficult to convince oneself that different tuples $\sfc$ and $\sfc'$
may describe CFTs that one wants to consider as `equal' on physical grounds.
In other words, we need to introduce a suitable equivalence relation. The notion
of equivalence must be broad enough to accommodate the following.

First, a solution to the sewing constraints can be obtained from a symmetric 
special 
Frobenius algebra; we recall this construction in section \ref{sec:frob-to-sew}.
Furthermore, as shown in \cite{tft1,ffrs5}, correlators obtained from Morita 
equivalent algebras differ only by constants related to the Euler character of 
the world sheet
(provided the boundary conditions are related as described in \cite{tft1,ffrs5}).

Next, $B_l$ and $B_r$ are only auxiliary data. Accordingly, two solutions $\sfc$
and $\sfc'$ which only differ in the way $\Hcl$ is realised as a retract (of 
$\BlxBr$ or of $B_l' \times \ol{B}_r'$, respectively) should be equivalent.
In other words, if two functors $\bl$ and $\bl'$ are related by 
$\bl' \eq \Alpha(\varphi_\text{op},\varphi_\text{cl}) \Circ \bl $
(see propostion \ref{prop:nat-iso}), then the two solutions 
$\cor{:}\ \triv\To\bl$ and $\cor'{:}\ \triv\To\bl'$ should be equivalent.

Moreover, working with fields rather than states, as is possible owing to the 
field-state correspondence in CFT (and is natural from the point of view of 
statistical mechanics), one should regard two CFTs as equivalent if upon a 
suitable isomorphism between the spaces of fields all expectation values 
(correlators normalised such that the identity field has expectation value one) 
agree. This leaves the freedom to modify the correlators by a multiplicative
constant, as such a constant cancels when passing to expectation values. Thus 
two solutions $\sfc$ and $\sfc'$ are to be regarded as equivalent if they only 
differ in the 
assignment of correlators in such a way that $\corP(\Xr) \eq f(\Xr)\,\cor(\Xr)$
for some function $f$ that assigns a nonzero constant to every world sheet 
$\Xr$. Consistency with sewing then requires $f(\Xr)$ to be of the form 
  \be
  f(\Xr) = \gamma^{2\chi(\Xr)}
  \labl{f.gamma}
for some $\gamma \iN \Cb^\times$, with $\chi(\Xr)$ the Euler character of $\Xr$,
which by
  \be
  \chi(\Xr) := \tfrac12\, \chi(\Xtil)
  \labl{eq:euler}
is defined through the one of $\Xtil$. For connected $\Xtil$, the latter is 
given by $\chi(\Xtil) \eq 2-2g(\Xtil)-b(\Xtil)$, with $g(\Xtil)$ the genus of 
$\Xtil$ (or rather, of the surface with empty boundary obtained by closing all 
holes of $\Xtil$ with \disc s), and $b$ the number of boundary components of 
$\Xtil$. In terms of the quotient surface $\dot\Xr$, we can write
  \be
  \chi(\Xr) = 2-2g(\dot\Xr) - b(\dot\Xr) - \tfrac12 |o| - |c| \,,
  \ee
where $g(\dot\Xr)$ is the genus of $\dot\Xr$, $b(\dot\Xr)$ the number of
connected components of $\partial\dot\Xr$, $|c|$ the number of
closed state boundaries of $\Xr$ and $|o|$ the number of open state
boundaries of $\Xr$. For example, for $\Xr$ a \disc\ with
$m$ open state boundaries one has $\chi(\Xr) \eq 1\,{ -}\, \tfrac12 m$.

\medskip

The following observations are relevant when formalising the notion of 
equivalence.
\\[-2.2em]

\dtl{Lemma}{lem:Euler-nat}
For any $\gamma \iN \Cb^\times$, the assignment
$\Xr \,{\mapsto}\, G^\gamma_\Xr \,{:=}\, \gamma^{\chi(\Xtil)} \id_{\bl(\Xr)}$
defines a monoidal natural self-equi\-va\-len\-ce $G^\gamma$ of
$\bl(\Cc,\Hop,\Hcl,B_l,B_r,e,r)$.

\medskip\noindent
Proof:\\
That $G^\gamma$ is a natural transformation amounts to verifying that for every
morphism $\xm {:}\ \Xr \To \Yr$ we have $\bl(\xm)\Circ G^\gamma_\Xr \eq 
G^\gamma_\Yr\Circ \bl(\xm)$. One checks that both for $\xm \eq (\emptyset,f)$ 
and for $\xm \eq (\sew,\id_{\widetilde{\sew(\Xr)}})$ one has 
$\chi(\Xtil) \eq \chi(\widetilde\Yr)$, and thus this remains true in the 
general case which is a composition of the two. The monoidal structure 
is just the additivity of the Euler character with respect to disjoint union.
\qed

Now let $\sfc$ and $\sfc'$ be two solutions to the sewing constraints.
As in proposition \ref{prop:nat-iso} we only consider the situation that $\sfc$ 
and $\sfc'$ involve the same modular tensor category $\Cc$. (This is again
just for simplicity of presentation, compare remark \ref{rem:nat-iso}.)
Thus $\sfc \eq (\Cc,\Hop,\Hcl,B_l,B_r,e,r,\cor)$ and
$\sfc'\eq (\Cc,\Hop',\Hcl',B_l',B_r',e',r',\corP)$.
\\[-2.2em]

\dtl{Lemma}{lem:compare}
Given two solutions $\sfc$ and $\sfc'$ as above, and given two isomorphisms
$\varphi_\text{op} \iN \Hom(\Hop',\Hop)$ and
$\varphi_\text{cl} \iN \Hom(\Hcl',\Hcl)$, abbreviate
$\Alpha \,{\equiv}\, \Alpha(\varphi_\text{op},\varphi_\text{cl})$.
Suppose that $\corP_{\Yr_\mu} \eq \gamma^{2\chi(\Yr_\mu)} \Alpha_{\Yr_\mu} 
{\circ}\, \cor_{\Yr_\mu}$ for some $\gamma\iN\Cb^\times$ and for world sheets 
$\Yr_\mu$, $\mu\iN\{1,2,...\,,n\}$.
Let $\Xr$ be a world sheet for which there exists a morphism
$\xm{:}\ \Yr_{\mu_1} \,{\sqcup}\, \cdots \,{\sqcup}\, \Yr_{\mu_n} \To \Xr$.
Then also $\corP_\Xr \eq \gamma^{2\chi(\Xr)} \Alpha_\Xr \Circ \cor_\Xr$.

\medskip\noindent
Proof:\\
Abbreviating also $\bl \,{\equiv}\, \bl(\Cc$, $\Hop$, $\Hcl$, $B_l,B_r,e,r)$
and $\bl' \,{\equiv}\, \bl(\Cc,\Hop',\Hcl',B_l',B_r',e',r')$, we have
  \beaa
  \corP_\Xr \!\!\!&
  \overset{(1)}{=}
  \bl'(\xm) \circ \corP_{\Yr_{\mu_1} \sqcup \cdots \sqcup \Yr_{\mu_n}}
  \overset{(2)}{=}
  \bl'(\xm) \circ \big( \corP_{\Yr_{\mu_1}} \Oti\, \cdots \oti
  \corP_{\Yr_{\mu_n}} \big)
  \enL&
  \overset{(3)}{=}
  \gamma^{2\chi(\Yr_{\mu_1})+\cdots +2\chi(\Yr_{\mu_n})}\, \bl'(\xm) \circ
  \big( \Alpha_{\Yr_{\mu_1}} \Oti\, \cdots \oti \Alpha_{\Yr_{\mu_n}} \big) \circ
  \big( \cor_{\Yr_{\mu_1}} \Oti\, \cdots \oti \cor_{\Yr_{\mu_n}} \big)
  \enL&
  \overset{(4)}{=}
  \gamma^{2\chi(\Xr)} \, \bl'(\xm) \circ
  \Alpha_{\Yr_{\mu_1} \sqcup \cdots \sqcup \Yr_{\mu_n}} {\circ}\,
  \cor_{\Yr_{\mu_1} \sqcup \cdots \sqcup \Yr_{\mu_n}}
  \enL&
  \overset{(5)}{=}
  \gamma^{2\chi(\Xr)} \, \Alpha_\Xr \circ \bl(\xm) \circ
  \cor_{\Yr_{\mu_1} \sqcup \cdots \sqcup \Yr_{\mu_n}}
  \,\overset{(6)}{=}\,
  \gamma^{2\chi(\Xr)} \, \Alpha_\Xr \circ \cor_\Xr \,.
  \end{array}
  \ee
Steps (1) and (6) are examples of the identity \erf{eq:cor-natxfer-prop}, 
i.e.\ naturality of $\cor$ and $\corP$; step (2) holds because $\corP$ is 
monoidal, step (3) holds by the assumption of the lemma, step (4) combines
monoidality of $\Alpha$ and $\cor$ with additivity of 
the Euler character, and finally step (5) is naturality of $\Alpha$.
\qed

Combining all these considerations we are led to the following notion of
equivalence.
\\[-2.2em]

\dtl{Definition}{def:sol-equiv}
Two solutions $\sfc \eq (\Cc,\Hop,\Hcl,B_l,B_r,e,r,\cor)$ and
$\sfc'\eq (\Cc,\Hop',\Hcl',B_l',B_r',e',r',\corP)$ to the sewing constraints 
that are based on the same category $\Cc$ are called {\em equivalent\/}
iff there exists a $\gamma\iN\Cb^\times$ and isomorphisms
$\varphi_\text{op} \iN \Hom(\Hop',\Hop)$ and
$\varphi_\text{cl} \iN \Hom(\Hcl',\Hcl)$ such that the identity 
  \be
  \corP = G^\gamma \circ \Alpha(\varphi_\text{op},\varphi_\text{cl}) \circ \cor
  \,.
  \labl{eq:sol-equiv}
between natural transformation holds.


\sect{Frobenius algebras and solutions to the sewing constraints}

Solutions to the sewing constraints are intimately related with Frobenius 
algebras in the category $\Cc$ that enters the formulation of the sewing 
constraints. From any symmetric special Frobenius algebra in $\Cc$ one can 
construct a solution to the sewing constraints; this result of \cite{tft1,tft5} 
will be recalled in section \ref{sec:frob-to-sew}.
In section \ref{sec:sew-to-frob} we will show that, conversely, any solution 
$\sfc$ gives rise to a symmetric Frobenius algebra in $\Cc$. Under suitable 
assumptions on $\sfc$, this algebra is also special. We can then state, in 
section \ref{sec:unique-thm}, our main result, namely that the procedures of 
constructing correlators from a symmetric special Frobenius algebra and of 
determining an algebra from a solution to the sewing constraints are inverse 
to each other. In the next two sections we start by collecting some notations 
and tools that we will need, in particular the fundamental world sheets from 
which all world sheets can be obtained via sewing (section \ref{sec:fund-corr})
and the notion of projecting onto the closed state vacuum 
(section \ref{sec:projcl}).


\subsection{Fundamental correlators}\label{sec:fund-corr}

\begin{table}[tb] 
\begin{center} 
   \hspace*{1.7em} \begin{tabular}{|c|l|} \hline~&\\[-.6em]
$\Xr_m$
   & ~a \disc\ with two in-going and one out-going open state boundaries~
   \\[.3em]
$\Xr_\eta$
   & ~a \disc\ with one out-going open state boundary
   \\[.3em]
$\Xr_\Delta$
   & ~a \disc\ with one in-going and two out-going open state boundaries 
   \\[.3em]
$\Xr_\eps$
   & ~a \disc\ with one in-going open state boundary
   \\[-.9em]& \\\cline{1-1}&\\[-.7em]
\raisebox{-.5em}{$\Xr_{Bb}$}
   & ~a \disc\ with one in- and one out-going open state boundary 
   \\[-.2em]& \hspace*{11.9em}and one in-going closed state boundary
   \\[-.9em]& \\\cline{1-1}&\\[-.7em]
$\Xr_{B(3)}$
   & ~a sphere with three in-going closed state boundaries 
   \\[.3em]
$\Xr_{B(1)} \equiv \Xr_{B\eps}$
   & ~a sphere with one in-going closed state boundary
   \\[.3em]
$\Xr_{oo}$
   & ~a sphere with two out-going closed state boundaries
   \\[-.7em]& \\\hline
\multicolumn2c{~}
   \\\hline~&\\[-.6em]
$\Xr_p$
   & ~a \disc\ with one in- and one out-going open state boundary
   \\[.3em]
$\Xr_{B\eta}$
   & ~a sphere with one out-going closed state boundary
   \\[.3em]
$\Xr_{Bp}$
   & ~a sphere with one in- and one out-going closed state boundary
   \\[.3em]
$\Xr_{B\Delta}$
   & ~a sphere with one in- and two out-going closed state boundaries
   \\[.3em]
$\Xr_{Bm}$
   & ~a sphere with two in- and one out-going closed state boundaries
   \\[-.7em]& \\\hline
\end{tabular} 
   \caption{Fundamental and other simple world sheets, as listed
   in figures \ref{fig:fund-world} and \ref{fig:addtl-world}, respectively}
\label{table:fuwosh} 
\end{center}
\end{table}

  \begin{figure}[tb]
  \begin{picture}(190,325)
    \put(0,195){
  \put(30,0)     {\Includeournicelargepicture 09a \put(49,-18){$\Xr_m$} }
  \put(190,30)   {\Includeournicelargepicture 09b \put(13,-18){$\Xr_\eta$} }
  \put(270,0)    {\Includeournicelargepicture 09c \put(49,-18){$\Xr_\Delta$} }
  \put(420,30)   {\Includeournicelargepicture 09d \put(13,-18){$\Xr_\eps$} }
    }
    \put(0,25){
  \put(1,0)      {\Includeournicelargepicture 14a \put(34,-18){$\Xr_{Bb}$} }
  \put(115,10)   {\Includeournicemediumpicture 14g\put(76,-18){$\Xr_{B(3)}$} }
  \put(312,40)   {\Includeournicelargepicture 14h \put(42,-18){$\Xr_{oo}$} }
  \put(424,10)   {\Includeournicelargepicture 14e \put(0,-18) {$\Xr_{B\eps}
                                                          \equiv\Xr_{B(1)}$} }
    }
  \end{picture}
  \caption{List of fundamental world sheets. Any world sheet $\Xr$ can be 
  decomposed into surfaces in this list by repeatedly cutting along intervals 
  or circles. In each of these world sheet pictures the bottom boundaries 
  are in-going and the top boundaries out-going, while the closed state
  boundary drawn in the middle of $\Xr_{Bb}$ is in-going.}
  \label{fig:fund-world} \end{figure}
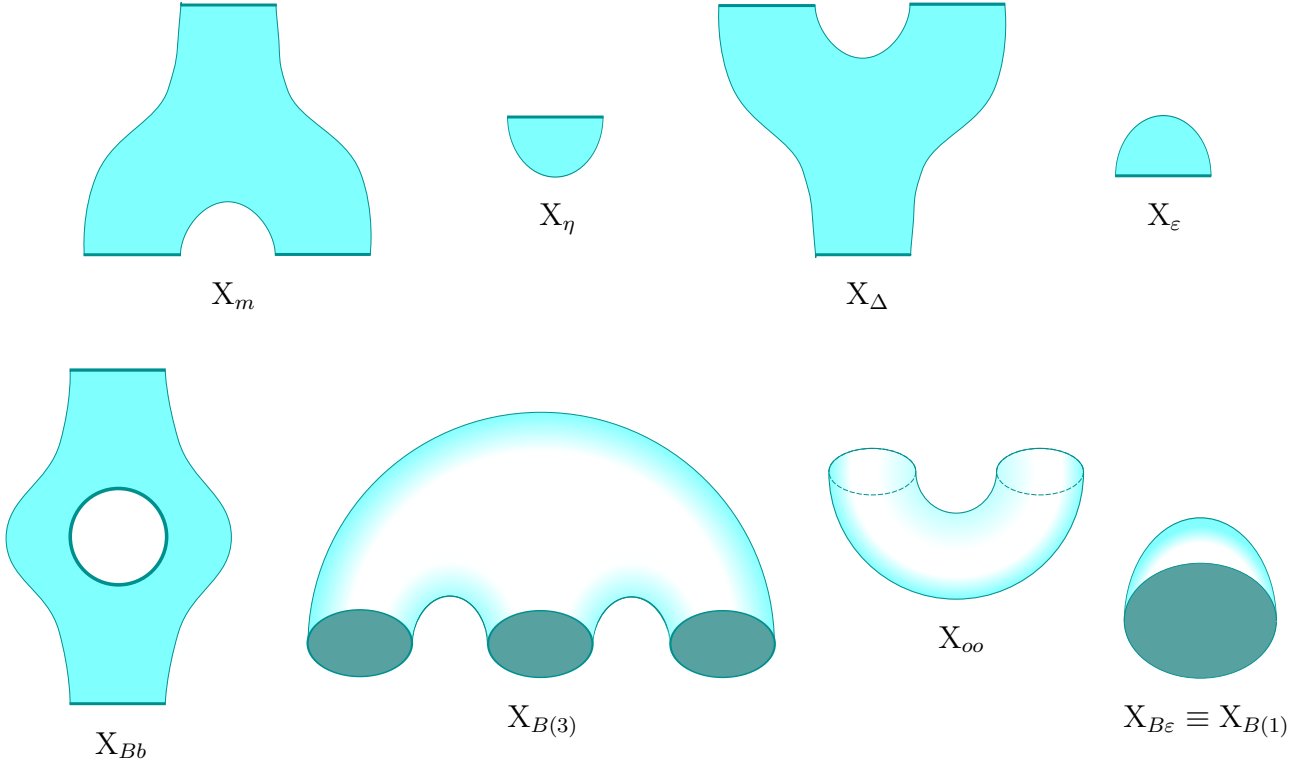

Every world sheet $\Xr$ can be obtained by applying sewing to a small collection
of fundamental world sheets \cite{lewe,prss3,lapf,kaPe}. In terms of the category
$\Wor$ this means that for every world sheet $\Xr$ there is a (non-unique) 
morphism
  \be
  \xm:\quad \bigsqcup_{\alpha\in C_{\Xr}} \Yr_{\!\alpha} \longrightarrow \Xr \,,
  \ee
where $C_{\Xr}$ is a finite index set and each of the world sheets 
$\Yr_{\!\alpha}$ is one of the world sheets that are displayed in figure 
\ref{fig:fund-world}. We will refer to them as {\em fundamental\/} world sheets;
the 
symbols $m,\eta,\Delta,\eps$ refer to the morphisms in formula \erf{pic_navf_c} 
below, while ``$B$'' stands for bulk. Of course, one may also use other 
sets of fundamental world sheets. For instance, one could replace $\Xr_{B\eps}$ 
in figure \ref{fig:fund-world} by $\Xr_{B\eta}$ in figure \ref{fig:addtl-world}. 

  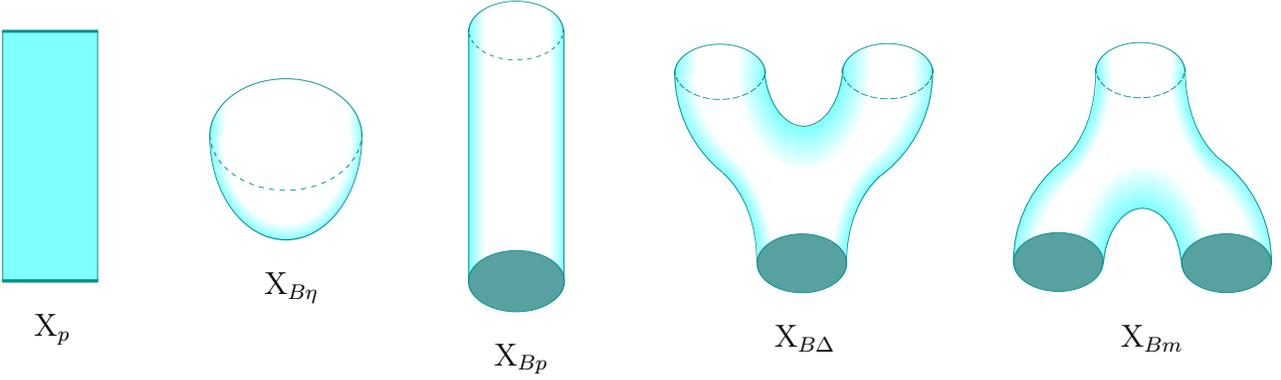
\begin{figure}[tb]
  \begin{picture}(190,165)
    \put(0,25){
  \put(2,11)     {\Includeournicelargepicture 09e \put(13,-20){$\Xr_p$} }
  \put(81,27)    {\Includeournicelargepicture 14c \put(21,-20){$\Xr_{B\eta}$} }
  \put(179,0)    {\Includeournicelargepicture 14f \put(10,-20){$\Xr_{Bp}$} }
  \put(257,7)    {\Includeournicelargepicture 14d \put(38,-20){$\Xr_{B\Delta}$}}
  \put(385,7)    {\Includeournicelargepicture 14b \put(41,-20){$\Xr_{Bm}$} }
    }
  \end{picture}
  \caption{Some other simple world sheets, included for convenience.} 
  \label{fig:addtl-world} \end{figure}

We also display, in figure 
\ref{fig:addtl-world}, five other simple world sheets, namely $\Xr_{B\eta}$,
the projectors $\Xr_p$ and $\Xr_{Bp}$ which will be used below, e.g.\ to 
formulate the conditions in the uniqueness theorem \ref{thm:unique}, and the 
`pairs of pants' $\Xr_{Bm}$ and $\Xr_{B\Delta}$ which have been used as 
fundamental world sheets elsewhere in the literature. These are only shown 
to point out clearly that also these particular world sheets can be obtained 
by gluing world sheets from figure \ref{fig:fund-world}. For convenience, 
all these world sheets are also collected in table \ref{table:fuwosh}.

\smallskip

By invoking lemma \ref{lem:compare} in the special case $\corP\eq\cor$ (and 
hence $\gamma\eq1$), it follows that a collection $\cor$ of correlators is 
uniquely determined on all of $\Wor$ already 
by the finite subset $\{\cor(\Xr)\}$ for the fundamental world sheets $\Xr$.

\medskip

  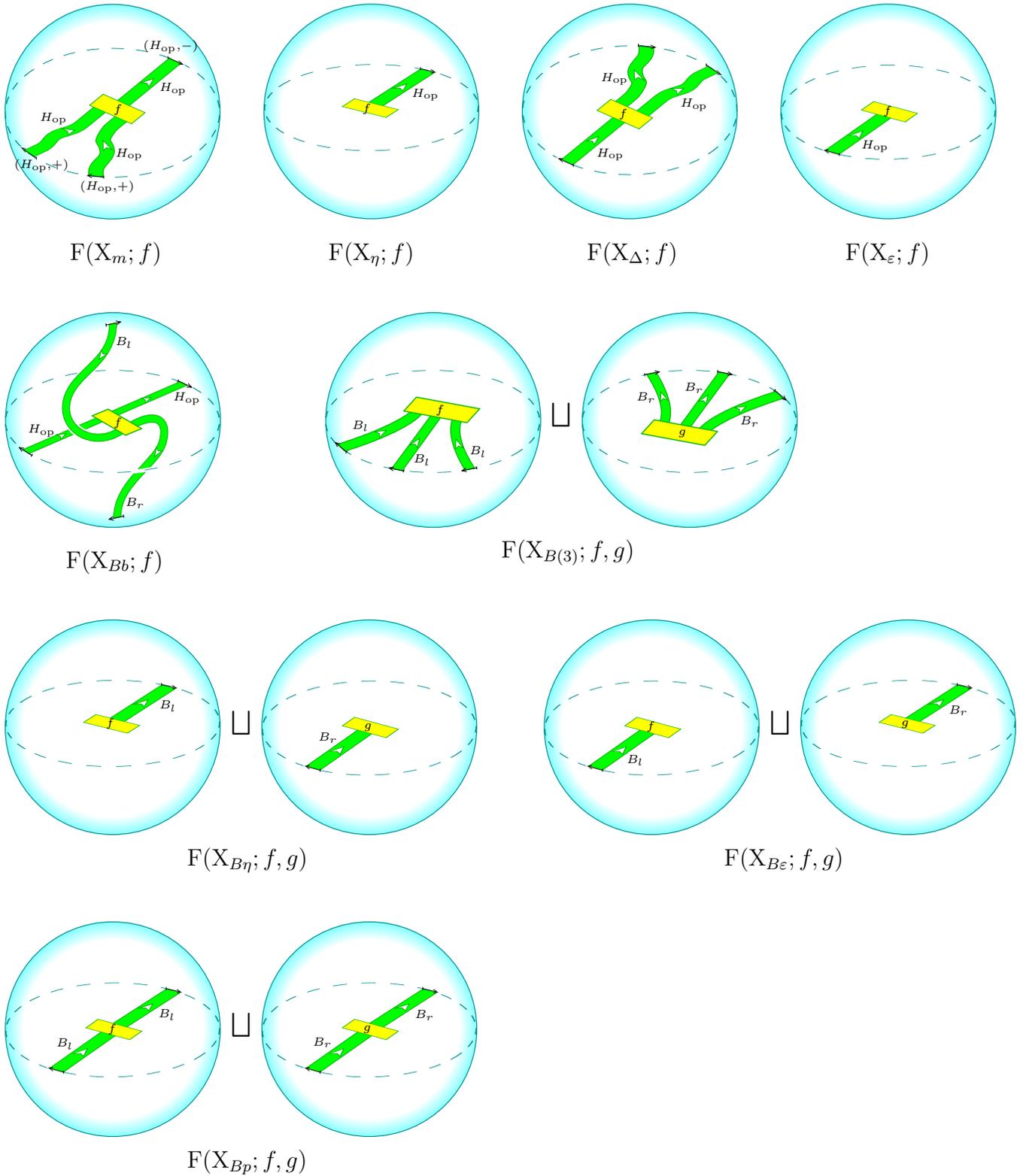
\begin{figure}[p]
  \begin{picture}(190,620)
    \put(0,490){
  \put(0,0)    {\Includeourbeautifulpicture 10{}\put(33,-20){$\Fr(\Xr_{m};f)$}}
  \put(18,50)     {\tiny$ \Hop $}
  \put(56,32)     {\tiny$ \Hop $}
  \put(78,65)     {\tiny$ \Hop $}
  \put(70,86)     {\tiny$ (\Hop{,}-) $}
  \put(5,26)      {\tiny$ (\Hop{,}+) $}
  \put(38,15.5)   {\tiny$ (\Hop{,}+) $}
  \put(55,54)     {\tiny$ f $}
  \put(130,0)  {\Includeourbeautifulpicture 11a \put(33,-20){$\Fr(\Xr_{\eta};f)$}}
  \put(205,62)    {\tiny$ \Hop $}
  \put(180,54.4)  {\tiny$ f $}  
  \put(260,0)  {\Includeourbeautifulpicture 11b \put(33,-20){$\Fr(\Xr_{\Delta};f)$}}
  \put(300,70)    {\tiny$ \Hop $}
  \put(298,32)    {\tiny$ \Hop $}
  \put(340,57)    {\tiny$ \Hop $}
  \put(309,52)    {\tiny$ f $}
  \put(390,0)  {\Includeourbeautifulpicture 11c \put(33,-20){$\Fr(\Xr_{\eps};f)$}}
  \put(431,36)    {\tiny$ \Hop $}
  \put(442,53.2)  {\tiny$ f $}
    }
    \put(0,335){
  \put(0,0)    {\Includeourbeautifulpicture 15a \put(31,-20){$\Fr(\Xr_{Bb};f)$}}
  \put(85,64)     {\tiny$ \Hop $}
  \put(12,48)     {\tiny$ \Hop $}
  \put(61,11)     {\tiny$ B_r $}
  \put(56,92)     {\tiny$ B_l $}
  \put(55,51.5)   {\tiny$ f $}
  \put(161,0)  {\Includeourbeautifulpicture 11x \put(114,54){$\bigsqcup$}}
  \put(175,50)    {\tiny$ B_l $}
  \put(205.8,33)  {\tiny$ B_l $}
  \put(234,37)    {\tiny$ B_l $}
  \put(217,57.7)  {\tiny$ f $}
  \put(290,0)  {\Includeourbeautifulpicture 11y \put(-40,-15){$\Fr(\Xr_{B(3)};f,g)$}}
  \put(318,64)    {\tiny$ B_r $}
  \put(342,69)    {\tiny$ B_r $}
  \put(370,52)    {\tiny$ B_r $}
  \put(339,47)    {\tiny$ g $}
    }
    \put(0,180){
  \put(0,0)    {\Includeourbeautifulpicture 11a \put(114,54){$\bigsqcup$}}
  \put(129,0)  {\Includeourbeautifulpicture 11c \put(-37,-15){$\Fr(\Xr_{B\eta};f,g)$}}
  \put(271,0)  {\Includeourbeautifulpicture 11c \put(114,54){$\bigsqcup$}}
  \put(400,0)  {\Includeourbeautifulpicture 11a \put(-38,-15){$\Fr(\Xr_{B\eps};f,g)$}}
  \put(51,55)     {\tiny$ f $}
  \put(78,64)     {\tiny$ B_l $}
  \put(158,47.8)  {\tiny$ B_r $}
  \put(181,54)    {\tiny$ g $}
  \put(313,36.7)  {\tiny$ B_l $}
  \put(323,53.5)  {\tiny$ f $}
  \put(475,61.5)  {\tiny$ B_r $}
  \put(451,55.5)  {\tiny$ g $}
    }
    \put(0,28){
  \put(0,0)    {\Includeourbeautifulpicture 11s \put(114,54){$\bigsqcup$}}
  \put(129,0)  {\Includeourbeautifulpicture 11s \put(-37,-15){$\Fr(\Xr_{Bp};f,g)$}}
  \put(0,0){
     \setlength{\unitlength}{.38pt}\put(-73,-361){
     \put(210,501)   {\tiny$ f $}
     \put(142,482)   {\tiny$ B_l $}
     \put(277,520)   {\tiny$ B_l $}
     }\setlength{\unitlength}{1pt}}
  \put(129,0){
     \setlength{\unitlength}{.38pt}\put(-73,-361){
     \put(209,502.5) {\tiny$ g $}
     \put(142,483)   {\tiny$ B_r $}
     \put(277,519)   {\tiny$ B_r $}
     }\setlength{\unitlength}{1pt}}
    }
  \end{picture}
  \caption{Cobordisms $\Fr(\Xr;f){:}\ \emptyset \To \widehat \Xr$ for each of 
  the fundamental world sheets in figure \ref{fig:fund-world}
  and for $\Xr_{Bp}$ from figure \ref{fig:addtl-world}.
}
  \label{fig:fund-cobord}
  \end{figure}

The correlators assigned to fundamental world sheets can be related to specific
morphisms of $\Cc$ with the help of the suitable cobordisms; these cobordisms
are listed in figure \ref{fig:fund-cobord}. Consider for example the world sheet
$\Xr_m$. The decorated double of $\Xr_m$ is a sphere with two arcs marked by 
$(\Hop,+)$ and one arc marked by $(\Hop,-)$. According to \erf{eq:E-Hom-space}, 
the space $\bl(\Xr_m)\eq \Htft(\widehat\Xr_m)$ is isomorphic to 
$\Hom(\Hop\oti\Hop\oti\Hop^\vee,\one) \,{\cong}\, \Hom(\Hop\oti\Hop,\Hop)$. 
An isomorphism is provided by considering the cobordism 
$\Fr(\Xr_{m};f)\,{\equiv}\,\Fr(\Xr_m;\Cc,\Hop;f){:}\ \emptyset\To\widehat \Xr_m$
shown as the first picture of figure \ref{fig:fund-cobord},
where $f$ is an element of $\Hom(\Hop\oti\Hop,\Hop)$:\,\footnote{~%
   Strictly speaking, $\tftC(\Fr(\Xr_m;f))$ is a linear map
   $\Cb \To \Htft(\widehat \Xr_m)$. One obtains an element of $\Htft(\widehat\Xr_m)$
   by evaluating this linear map on $1 \iN \Cb$.
   It is understood that this is done implicitly where necessary.}
  \be
  \Psi_m:\quad \Hom(\Hop \otimes \Hop,\Hop)\To \Htft(\widehat \Xr_m) \,,
  \qquad f \stackrel{\Psi_m}\longmapsto \tftC(\Fr(\Xr_m;f)) \,.
  \labl{eq:Xm-iso}
Analogously, given the cobordisms in figure \ref{fig:fund-cobord}, we define
\\[.3em]
\nxt for $\Xr\iN \{\Xr_\eta, \Xr_\Delta, \Xr_\eps, \Xr_p\}$ cobordisms 
$\,\Fr(\Xr;f) \,{\equiv}\, \Fr(\Xr;\Cc,\Hop;f) {:}\ \emptyset \To \widehat \Xr$;
\\[.3em]
\nxt for $\Xr_{Bb}$ a cobordism $\Fr(\Xr_{Bb};f) \,{\equiv}\, 
\Fr(\Xr_{Bb};\Cc,\Hop,B_l,B_r;f) {:}\ \emptyset \To \widehat \Xr_{Bb}$;
\\[.3em]
\nxt for $\Xr \iN \{\Xr_{B(1)}, \Xr_{oo}, \Xr_{B(3)}, \Xr_{B\eta}, \Xr_{Bp} \}$ 
cobordisms $\Fr(\Xr;f,g) \,{\equiv}\, \Fr(\Xr;\Cc,B_l,B_r;f,g){:}\
\emptyset \To \widehat \Xr$.
\\[.5em]
As in \erf{eq:Xm-iso}, applying the 3-d TFT to these cobordisms yields 
linear isomorphisms between certain
morphism spaces of $\Cc$ and $\Htft(\Xhat)$. For example, \be
  \tftC(\Fr(\Xr_{Bb}; \cdot )).1 :\quad \Hom(\Hop\oti B_l,\Hop \oti B_r)
  \overset{\cong}{\longrightarrow} \Htft(\widehat\Xr_{Bb}) \,.
  \ee 
However, when closed state boundaries are present on $\Xr$, according to the 
projection in prescription \erf{eq:bl(X)-def} we do in general no longer 
have $\bl(\Xr) \eq \Htft(\Xhat)$, but only $\bl(\Xr)\,{\subset}\,\Htft(\Xhat)$.


\subsection{Projecting onto the closed state vacuum}\label{sec:projcl}

In this section we define the operation of {\em projecting onto the closed 
state vacuum\/}. It can be thought of as `pinching' a circle on the world sheet,
i.e.\ replacing an annulus-shaped subset of the surface $\dot\Xr$ by two 
half-spheres. This procedure can be applied in all CFTs for which the closed 
state vacuum is unique, i.e.\ when
  \be
  \dim_\Cb \Hom_{\Cc \boxtimes \ol\Cc}(\one {\times} \ol\one, \Hcl) = 1 \,;
  \ee
it will be crucial when proving the uniqueness theorem in section
\ref{sec:unique-proof} below.

\medskip

Let $\sfc \eq (\Cc,\Hop,\Hcl,B_l,B_r,e,r,\cor)$ be a solution to the sewing 
constraints. We first analyse the correlators of the closed world sheets 
$\Xr_{B\eta}$ and $\Xr_{B\eps}$ in figure \ref{fig:fund-world}. The correlator
for the first of them is a linear map
$\cor_{\Xr_{B\eta}}{:}\ \Cb \To \Htft(\widehat\Xr_{B\eta})$, where
  \be
  \Htft(\widehat\Xr_{B\eta}) \cong
  \Hom(\one,B_l) \otic \Hom(B_r,\one)
  = \Hom_{\Cc \boxtimes \ol\Cc}(\one{\times}\ol\one, \BlxBr) \,.
  \ee
An isomorphism between the first two spaces is obtained via the cobordism
$\Fr(\Xr_{B\eta}; \,\cdot\,,\,\cdot\,)$ in figure \ref{fig:fund-cobord}, while 
the equality of the second and third expressions holds by definition of the 
morphism spaces of $\CxCb$, see section \ref{sec:mtc-tft}. Thus there exists a 
unique $u_{B\eta} \iN \Hom(\one,B_l) \otic \Hom(B_r,\one)$ such that 
when written as $u_{B\eta} \eq \sum_\alpha u_\alpha' \oti u_\alpha''$ with
$u_\alpha'\iN\Hom(\one,B_l)$ and $u_\alpha''\iN\Hom(B_r,\one)$ one has
  \be
  \cor_{\Xr_{B\eta}} = \sum_\alpha
  \tftC\big( \Fr( \Xr_{B\eta} ; u_\alpha' , u_\alpha'' ) \big) \,.
  \labl{eq:XB-eta-u} 
Similarly, via the cobordism $\Fr(\Xr_{B\eps}; \,\cdot\,,\,\cdot\,)$ the 
correlator $\cor_{\Xr_{B\eps}}$ corresponds to an element $u_{B\eps} 
\iN \Hom(B_l,\one) \otic \Hom(\one,B_r)$.

Given any world sheet $\Xr$, one can embed a parametrised little \disc\ $D$ 
into $\dot\Xr$ 
and write $\dot\Xr$ as the union of $D$ and $\dot\Xr{\setminus}D$. Similarly
one can find a world sheet $\Yr$ such that there exists a morphism $\xm {:}\ 
\Yr\,{\sqcup}\,\Xr_{B\eta}\To \Xr$. Writing $\xm \eq (\sew,f)$, this gives
  \beaa
  \cor_\Xr \!\!\!&
   = \bl(\xm) \circ \cor_{\Yr \sqcup \Xr_{B\eta}}
   = \bl(\xm) \circ \big( \cor_{\Yr} \oti \cor_{\Xr_{B\eta}} \big)
   \enL&\displaystyle
   = \sum_\alpha \tftC(f) \circ \tftC(\hat\sew) \circ \big( \cor_{\Yr} \oti
   \tftC\big( \Fr( \Xr_{B\eta} ; u_\alpha' , u_\alpha'' ) \big) \big)
   \enL&\displaystyle
   = \sum_\alpha \tftC(f) \circ \tftC\big(\hat\sew
     \circ (\id_\Yr \sqcup \Fr( \Xr_{B\eta} ; u_\alpha' , u_\alpha'' )) \big)
     \circ \cor_{\Yr}  \,.
  \end{array}
  \labl{eq:cut-disc} 
In a neighbourhood of $\Xr_{B\eta}$, the cobordism appearing in the last line 
looks as
  \eqpic{pic-fjfrs2_19} {320} {84} {
      \put(0,7){
  \put(165,96)   {\Includeourbeautifulpicture 19a}
  \put(165,-11)  {\Includeourbeautifulpicture 19b}
  \put(0,85)     {$ \hat\sew\circ(\id_\Yr
                    \sqcup\Fr(\Xr_{B\eta};u_\alpha',u_\alpha''))~= $}
  \put(252.5,135){\sse $B_l$}
  \put(252.5,59) {\sse $B_r$}
  \put(242,153)  {\sse $u_\alpha'$}
  \put(242,28)   {\sse $u_\alpha''$}
  } }
In this picture, the top and bottom surfaces are part of
$\widehat{\sew(\Yr\,{\sqcup}\,\Xr_{B\eta})} \,{\cong}\, \widehat\Xr$, and
the two inner boundaries (on which the $B_l$ and $B_r$ ribbons end/start) 
are part of $\widehat\Yr$.

The above discussion motivates us to formulate
\\[-2.3em]

\dtl{Lemma}{lem:semisphere-nonzero} 
If there exists at least one world sheet $\Xr$ with $\cor_\Xr \,{\ne}\, 0$, then 
the morphisms $u_{B\eta}$ and $u_{B\eps}$ are nonzero.

\medskip\noindent
Proof:\\
Let $\Xr$ be a world sheet such that $\cor_\Xr\,{\ne}\,0$ and let $\Yr$ be a
world sheet for which there exists a morphism $\xm{:}\ \Yr\,{\sqcup}\,\Xr_{B\eta}
\To \Xr$. Then the right hand side of \erf{eq:cut-disc} must be nonzero. 
For this to be the case it is necessary that  $\sum_\alpha u_\alpha' \oti
u_\alpha'' \,{\ne}\, 0$. That also $u_{B\eps}\,{\ne}\,0$ can be seen similarly. 
\qed

\medskip

By a {\em purely closed sewing\/} of a world sheet $\Xr$ we mean sewing data 
$\sew$ for $\Xr$ such that for all pairs $(a,b) \iN \sew$ we have 
$(a,b) \,{\ne}\, (\iota_*(a),\iota_*(b))$. Given a purely closed sewing, we 
define a cobordism $\mathrm{M}^\text{vac}_{\sew,\Xr}{:}\ \widehat{\sew(\Xr)} \To 
\widehat{\sew(\Xr)}$ as follows. Start from the cylinder 
$\widehat{\sew(\Xr)}\Times[0,1]$. For each pair $(a,b)\iN \sew$ define the 
circle $C_{(a,b)} \,{:=}\, \pi_{\sew,\Xr}(C_a) \eq \pi_{\sew,\Xr}(C_b)$ 
embedded in $\widetilde{\sew(\Xr)} \,{\subset}\, \widehat{\sew(\Xr)}$, where 
$C_a$ and $C_b$ are the boundary components of $\partial\Xtil$ corresponding to 
$a, b \iN \pi_0(\partial\Xtil)$. On each annulus $C_{(a,b)} \Times [0,1] 
\,{\subset}\, \mathrm{M}^\text{vac}_{\sew,\Xr}$ insert a coupon labeled by the 
morphism $\wO$ defined in \erf{eq:wO} and an annulus-shaped $\Omeka$-ribbon 
starting and ending on this coupon, in such a way that the core of the 
$\Omeka$-ribbon lies on $C_{(a,b)} \Times \{\tfrac12\}$. In
a neighbourhood of such an annulus, $\Mr^{\text{vac}}_{\sew,\Xr}$ looks as 
  \eqpicc{pic-fjfrs2_16} {420} {70} {
  \put(15,23)    {\Includeournicesmallpicture 16a}
  \put(240,-6)   {\Includeournicesmallpicture 16b}
  \put(-22,78)   {$ \sew(\Xr)\!=$ }
  \put(167,74)   {$ \longrightarrow \ \Mr^{\text{vac}}_{\sew,\Xr}~=$}
  \put(75,75)    {\small$ C_{(a,b)} $}
  \put(118,120)  {\small$ L $}
  \put(118,35)   {\small$ L' $}
      \put(7,0){
  \put(295,95)   {\sse$ \Omeka $}
  \put(322,133)  {\sse \begin{turn}{-80}$w_\Omeka$\end{turn}}
  \put(295,8)    {\sse$ \Omeka $}
  \put(322,46)   {\sse \begin{turn}{-80}$w_\Omeka$\end{turn}}
  \put(445,133)  {\small$ A $}
  \put(378,80)   {\small$ A' $}
  \put(445,46)   {\small$ B $}
  \put(378,-9)   {\small$ B' $}
      }
  }
Here the lines $L$ and $L'$ are to be identified. Likewise, the faces $A$ and 
$A'$, as well as $B$ and $B'$, must be
identified. Now define a linear map $P^\text{vac}_{\sew,\Xr}$ as
  \be
  P^\text{vac}_{\sew,\Xr} := \tftC\big(\mathrm{M}^{\text{vac}}_{\sew,\Xr}\big)
                          :\quad \bl(\sew(\Xr)) \To \bl(\sew(\Xr)) 
  \,.
  \labl{eq:Pvac-def} 
It is not difficult to check that $P^\text{vac}_{\sew,\Xr}$, which is initially 
a linear map from $\Htft(\widehat{\sew(\Xr)})$ to itself, indeed restricts to an 
endomorphism of $\bl(\sew(\Xr))$.  In fact, given two solutions $\sfc$, $\sfc'$ 
to the sewing constraints and denoting by $P^\text{vac}_{\sew,\Xr}$ the linear 
map \erf{eq:Pvac-def} with ribbons labeled by the data in $\sfc$ and by 
$P^\text{vac}_{\sew,\Xr}\phantom{|}\!'$ the corresponding map with ribbons 
labeled by the data in $\sfc'$, it is straightforward to verify by 
substituting the explicit form of the cobordisms that
  \be
  F_\Xr(o^\text{in},o^\text{out},c^\text{in},c^\text{out})\,
  P^\text{vac}_{\sew,\Xr}
  = P^\text{vac}_{\sew,\Xr}\phantom{|}\!'\,
  F_\Xr(o^\text{in},o^\text{out},c^\text{in},c^\text{out}) \,,
  \ee
where $F_\Xr$ is the linear map defined in \erf{eq:FX-def}. In particular, by 
the definition of $\Alpha$ in \erf{eq:alphaX-def}, for isomorphisms 
$\varphi_\text{op} \iN \Hom(\Hop',\Hop)$ and
$\varphi_\text{cl} \iN \Hom(\Hcl',\Hcl)$ we have
  \be
  \Alpha(\varphi_\text{op},\varphi_\text{cl})_\Xr^{} \, P^\text{vac}_{\sew,\Xr}
  = P^\text{vac}_{\sew,\Xr}\phantom{|}\!' \,
  \Alpha(\varphi_\text{op},\varphi_\text{cl})_\Xr^{} \,.
  \labl{eq:Pvac-alpha-comm}
Furthermore, $P^\text{vac}_{\sew,\Xr}$ has the following property.
\\[-2.2em]

\dtl{Lemma}{lem:Pvac-disc} 
Let $\Xr \eq \Yr \,{\sqcup}\, \Xr_{B\eta}$ for some world sheet $\Yr$. Let $a 
\iN \pi_0(\partial\Xtil_{B\eta})$ and $b \iN \pi_0(\partial\widetilde\Yr)$ be 
such that $\sew \eq \{(a,b),(\iota_*(a),\iota_*(b) \}$ are sewing data for 
$\Xr$. Then
  \be
  P^\text{vac}_{\sew,\Xr} \circ \cor_{\sew(\Xr)} = \cor_{\sew(\Xr)} \,.
  \ee
Proof:\\
In a neighbourhood of the \disc\ $\Xr_{B\eta}$ the cobordism $\Mr
^{\text{vac}}_{\sew,\Xr}$ constructed above takes the following simple form:
  \eqpicc{pic-fjfrs2_22} {420} {63} {
  \put(-22,22)   {\Includeournicesmallpicture 22a}
  \put(256,-15)  {\Includeournicesmallpicture 22b}
  \put(167,70)   {$ \longrightarrow ~~~ \Mr^{\text{vac}}_{\sew,\Xr}~=$}
  \put(45,80)    {\sse \rm{\disc} $ \Xr_{B\eta}^{} $}
  \put(313,-3)   {\sse $\Omeka$}
  \put(336,13)   {\sse\begin{turn}{-90}$w_\Omeka$\end{turn}}
  \put(313,87)   {\sse $\Omeka$}
  \put(336,102)  {\sse\begin{turn}{-90}$w_\Omeka$\end{turn}}
  }
By \erf{eq:Omeka-trS}, the two annulus-shaped $\Omeka$-ribbons can be omitted 
without changing $\tftC(\mathrm{M}^{\text{vac}}_{\sew,\Xr})$. The resulting 
cobordism is just the cylinder over $\widehat{\sew(\Xr)}$, and hence 
$P^\text{vac}_{\sew,\Xr} \eq 
\tftC(\mathrm{M}^{\text{vac}}_{\sew,\Xr}) \eq \id_{\Htft(\widehat{\sew(\Xr)})}$. 
\qed

\medskip

\dtl{Definition}{def:fill} 
Let $\Xr \eq \big( \Xtil,\iota,\Varphi,b^\text{in},b^\text{out},\text{or} \big)$
be a world sheet and $\sew$ be a purely closed sewing of $\Xr$. The world sheet
  \be
  \fill_\sew(\Xr) := \big( \Xtil', \iota', \Varphi', {b^\text{in}}',
  {b^\text{out}}', \text{or}' \big)
  \ee 
(the world sheet {\em filled at\/} $\sew$) is defined by gluing unmarked \disc s
to all boundary components of $\Xtil$ that are listed in $\sew$:
  \be
  \Xtil' := \big( \Xtil\,{\sqcup}\,(\sew{\times}D)\,{\sqcup}\,(\sew{\times}D)
  \big) /{\sim} \,,
  \ee 
where $D \eq \{|z| \,{\le}\, 1\} \,{\subset\,} \Cb$ is the unit \disc. Denoting 
by $((a,b),z)_{\!k}$, $k\eq1,2$, elements of the first and second copy
of $\sew \Times D$, the identification is, for all $(a,b) \iN \sew$ and all
$z \iN \partial D$, given by $((a,b),z)_1 \,{\sim}\, \Varphi_a^{-1} 
\Circ C(-z)$ and $((a,b),z)_2 \,{\sim}\, \Varphi_b^{-1} \Circ C(-z)$.
The involution $\iota'$ is defined to equal $\iota$ on $\Xtil$ and as 
$\iota'\big(((a,b),z)_k\big) \eq ((\iota_*(a),\iota_*(b)),C(-z))_k$ on the 
\disc s $D$. $\Varphi'$ is the restriction of $\Varphi$ to 
$\partial\Xtil'$, and ${b^\text{in}}'$ and ${b^\text{out}}'$ are the 
restrictions of ${b^\text{in}}$ and ${b^\text{out}}$, respectively, to 
$\pi_0(\partial\Xtil')$. Finally, $\,\text{or}'$ is defined to be the unique 
continuous extension of $\,\text{or}$ to $\Xtil'/\langle \iota'\rangle$.

\bigskip

For $(a,b)\iN\sew$, in a neighbourhood of the circles $C_a$, $C_b$, the sewed
world sheet $\sew(\Xr)$ and the filled world sheet $\fill_\sew(\Xr)$ look as
follows (as usual we draw the quotient surface)
  \eqpicc{pic_fjfrs2_17} {420} {59} {
  \put(24,7)     {\Includeournicelargepicture 17a}
  \put(300,7)    {\Includeournicelargepicture 17b}
  \put(-21,68)   {$ \sew(\Xr)= $}
  \put(246,68)   {$ \fill_\sew(\Xr)= $}
  \put(103,43)   {\small$ C_{(a,b)} $}
  \put(335,50)   {\small$ C_a $}
  \put(419,50)   {\small$ C_b $}
  }

We proceed by defining, for a world sheet $\Xr$ and a purely closed sewing 
$\sew$, a linear map $E^{\text{vac}}_{\sew,\Xr}$ (the symbol $E$ reminds of 
`embedding') 
from $\bl(\fill_\sew(\Xr))$ to $\bl(\sew(\Xr))$. Consider the cobordism
  \be
  \Mr_\sew := \widehat{\fill_\sew(\Xr)} \Times [0,1] \,/{\sim} \,,
  \labl{Msew}
where the equivalence relation `$\sim$' identifies $\big((a,b),z\big)_1 \Times 
\{1\}$ with $\big((a,b),C(-z)\big)_2 \Times \{1\}$  
for all points $\big((a,b),z\big)$ in $\sew \Times D$. For $(a,b)\iN\sew$, in 
a neighbourhood of the circle $C_{(a,b)}$, $\Mr_\sew$ looks as follows.
  \eqpic{pic-fjfrs2_18} {290} {146} {
  \put(0,150)    {$ \Mr_\sew~= $}
  \put(70,0)     {\Includeourbeautifulpicture 18{}}
  \put(70,0){
     \setlength{\unitlength}{.38pt}\put(0,0){
  \put(-15,721)   {$ D $}
  \put(-36,351)   {$ D' $}
  \put(-16, 80)   {$ D $}
  \put(605,561)   {$ A $}
  \put(605,192)   {$ B $}
     }\setlength{\unitlength}{1pt}}
  }
Here the two regions marked $D$ are to be identified. In fact, $D$ and $D'$
are the \disc s on the boundary of $\widehat{\fill_\sew(\Xr)} \Times [0,1]$
that get identified in \erf{Msew}. The part of the boundary of $\Mr_\sew$
marked $A$ is the part of the decorated double $\widehat{\fill_\sew(\Xr)}$
that corresponds to a neighbourhood of $C_a$ in \erf{pic_fjfrs2_17},
and similarly for $B$ and $C_b$.
    
One verifies that with these identifications, $\Mr_\sew$ is a cobordism 
from $\widehat{\fill_\sew(\Xr)}$ to $\widehat{\sew(\Xr)}$. We then set
  \be
  E^{\text{vac}}_{\sew,\Xr}
  :\quad \Htft(\widehat{\fill_\sew(\Xr)}) \longrightarrow \Htft(\widehat{\sew(\Xr)})
  \,, \qquad E^{\text{vac}}_{\sew,\Xr} := \tftC(\Mr_\sew) \,.
  \labl{eq:Evac-def} 
It is again not difficult to check that $E^{\text{vac}}_{\sew,\Xr}$ restricts 
to a linear map from $\bl(\fill_\sew(\Xr))$ to $\bl(\sew(\Xr))$. Also, following 
the same reasoning that led to \erf{eq:Pvac-alpha-comm}, one shows that
  \be
  \Alpha(\varphi_\text{op},\varphi_\text{cl})_{\sew(\Xr)}^{} \,
  E^{\text{vac}}_{\sew,\Xr}
  = E^{\text{vac}}_{\sew,\Xr}\phantom{|}\!' \,
  \Alpha(\varphi_\text{op},\varphi_\text{cl})_{\fill_\sew(\Xr)}^{} \,,
  \labl{eq:alpha-past-Evac}
where again the ribbons in the cobordism representing $E^{\text{vac}}
_{\sew,\Xr}$ are labeled by the data in a solution $\sfc$, and those for 
$E^{\text{vac}}_{\sew,\Xr}\phantom{|}\!'$ by the data in a solution $\sfc'$.

\medskip

The following property of $E^{\text{vac}}_{\sew,\Xr}$ will be needed below.
\\[-2.4em]

\dtl{Lemma}{lem:Evac-inject} 
The linear map $E^{\text{vac}}_{\sew,\Xr}{:}\ \bl(\fill_\sew(\Xr)) \To 
\bl(\sew(\Xr))$ is injective.

\medskip\noindent
Proof:\\
Let $\widehat{\sew(\Xr)}\eq\bigsqcup_{\beta} \mathrm{L}_\beta$ be the 
decomposition of the decorated double $\widehat{\sew(\Xr)}$ into connected 
components. Figure \erf{pic_fjfrs2_17} illustrates that one can easily find 
examples where $\widehat{\fill_\sew(\Xr)}$ has more connected components 
than $\sew(\Xr)$. Let us write
$\mathrm{K}^{(\beta)}\eq\bigsqcup_{\alpha} \mathrm{K}^{(\beta)}_\alpha$
for the decomposition of the part $\mathrm{K}^{(\beta)} \,{\subseteq}\, 
\widehat{\fill_\sew(\Xr)}$ that corresponds to
the single component $\mathrm{L}_\beta$ in $\widehat{\sew(\Xr)}$. More 
precisely, $\mathrm{K}^{(\beta)}_\alpha$ are those connected components of 
$\widehat{\fill_\sew(\Xr)}$ for which 
$\pi_{\sew,X}\big( \mathrm{K}^{(\beta)}_\alpha{\cap}\widetilde \Xr \big)$
has nonzero intersection with $L_\beta\,{\cap}\,\widetilde{\sew(\Xr)}$.
\\[2pt]
By construction, the cobordism $\Mr_\sew$ in \erf{Msew} then decomposes as 
$\Mr_\sew\eq\bigsqcup_\beta \Mr_\sew^{(\beta)}$, where $\Mr_\sew^{(\beta)}$
is a cobordism from $\mathrm{K}^{(\beta)}$ to $\mathrm{L}_\beta$. To prove the 
lemma, it is enough to show that all $\tftC(\Mr_\sew^{(\beta)})$ are injective.
Below we will consider one fixed value of $\beta$, so let us abbreviate 
$\mathrm{L} \,{\equiv}\, \mathrm{L}_\beta$, 
$\mathrm{K}_\alpha \,{\equiv}\, \mathrm{K}^{(\beta)}_\alpha$,
$\Mr \,{\equiv}\, \Mr^{(\beta)}$, and set $E \eq \tftC(\Mr)$.
\\[2pt]
Let $\mathrm{K} \eq \bigsqcup_\alpha \mathrm{K}_\alpha$ and let
$(V_i,\eps_i)$, $i\eq1,2,...\,,m$, be the labels of the marked arcs of 
$\mathrm K$. Let $\hb_\alpha(\,\cdot\,)$
be a handle body for $\mathrm K_\alpha$ (as in formula \erf{eq:handle-iso}). 
Then $\tftC(\hb_\alpha(\,\cdot\,))$ defines an isomorphism
  \be
  \Hom\Big({\textstyle \bigotimes_i^{(\alpha)}} V_i^{\eps_i} \otimes 
  H^{\otimes g_\alpha},\one\Big) \stackrel\cong\longrightarrow 
  \Htft(\mathrm K_\alpha) \,,
  \ee 
where the tensor product $\bigotimes_i^{(\alpha)}$ extends over all marked arcs
$(V_i,\eps_i)$ that lie in $\mathrm K_\alpha$ and $g_\alpha$ is the genus of 
$\mathrm K_\alpha$. A 
handle body $\hb'$ for $\mathrm L$, on the other hand, provides an isomorphism
  \be
  \tftC(\hb'(\,\cdot\,)) :\quad
  \Hom\Big({\textstyle \bigotimes_i} V_i^{\eps_i} \otimes
   H^{\otimes g_L},\one\Big) \stackrel\cong\longrightarrow \Htft(\mathrm L) \,,
  \labl{eq:Evac-inject-aux1} 
where the tensor product is over {\em all\/} marked arcs $(V_i,\eps_i)$ and 
$g_{\mathrm L}$ is the genus of $\mathrm L$. 
\\
A crucial observation is now that one can choose $\hb'(\,\cdot\,)$ to
be given, as a three-manifold, by $\Mr\Circ\bigsqcup_\alpha
\hb_\alpha(\,\cdot\,)$ and choose the ribbons in $\hb'$ so that
  \be
  \tftC\big( \Mr \circ (\mbox{$\bigsqcup_\alpha$}
  \hb_\alpha(f_\alpha)) \big)
  = \tftC\big( \hb'((\mbox{$\bigotimes_{\!\alpha}$} f_\alpha)
  \otimes (\tilde d_\one)^{\otimes n}) \big)\,,
  \ee
where 
$\tilde d_\one \iN \Hom(\one \oti \one^\vee, \one) \eq \Hom(\one,\one)$ is 
the duality morphism, and $n \eq g_{\mathrm K}\,{-}\sum_\alpha g_\alpha$
is the number of additional handles arising in the gluing process. By 
construction, in $\Mr\Circ (\bigsqcup_\alpha\! \hb_\alpha(f_\alpha))$ 
there are no ribbons running through these additional handles, and so one 
obtains the duality $\tilde d_\one$, interpreted (via the restriction of 
$H$ to $\one\oti\one$) as a morphism in $\Hom(H,\one)$, for each such handle.
\\[2pt]
Every vector $v \iN \Htft(\mathrm K)$ can be written as $v \eq \sum_i 
\tftC\big( \bigsqcup_\alpha\! \hb_\alpha(f_\alpha^{(i)}) \big)$ for
appropriate morphisms $f_\alpha^{(i)}$. Thus, invoking also the definition 
\erf{Msew} of $\Mr_\sew$, for $E \eq \tftC(\Mr)$ we obtain
  \be
  E(v)
  = \sum_i \tftC\big( \Mr \Circ (\mbox{$\bigsqcup_\alpha$}
  \hb_\alpha(f_\alpha^{(i)})) \big)
  = \tftC\big( \hb'((\mbox{$\sum_i\bigotimes_{\!\alpha}$} f_\alpha^{(i)})
  \oti (\tilde d_\one)^{\otimes n}) \big) \,.
  \ee 
Since the latter map is just the isomorphism \erf{eq:Evac-inject-aux1}, it
follows that if we have $E(v) \eq 0$, then also 
$(\sum_i\bigotimes_\alpha f_\alpha^{(i)}) \oti (\tilde d_\one)^{\otimes n} \eq0$, 
which in turn implies $v \eq 0$. Hence $E$ is injective.
\qed

\medskip

In the sequel we abbreviate by $S^2$ the following world sheet: 
$\widetilde S^2 \eq S^2 \,{\sqcup}\, (-S^2)$ with $S^2$ the two-sphere, 
$\iota$ is the permutation of the two components of $\widetilde S^2$, 
so that $\widetilde S^2 / \langle \iota \rangle$ is again a two-sphere, and 
$\,\text{or}$ is the identification of $S^2$ with the first factor. The 
correlator $\cor_{S^2}$ is an 
element of $\Htft(S^2\,{\sqcup}\,(-S^2))$. Denoting by $B^3$
the unit three ball, there is thus a constant $\Lambda_\sfc \iN \Cb$ such that
  \be
  \cor_{S^2} = \Lambda_\sfc \, \tftC( B^3 \sqcup (-B^3) ) \,.
  \labl{eq:LamS-def}
With these ingredients, we are in a position to state
\\[-2.3em]

\dtl{Proposition}{prop:proj-vac} Let $\sfc$ be a solution to the sewing 
constraints such that $\dim_\Cb \Hom_{\Cc \boxtimes \ol\Cc}(\one{\times}\ol\one, 
\Hcl) \eq 1$ and such that there is at least one nonzero correlator. Then 
the constant $\Lambda_\sfc$ in \erf{eq:LamS-def} is nonzero, and for every 
world sheet $\Xr$ and every purely closed sewing $\sew$ of $\Xr$ we have
  \be
  P^\text{vac}_{\sew,\Xr} \circ \cor_{\sew(\Xr)}
  = \Lambda_\sfc^{\,-|\sew|/2}\,
  E^\text{vac}_{\sew,\Xr} \circ \cor_{\fill_\sew(\Xr)} \,,
  \labl{eq:proj-vac} 
where $|\sew|$ is the number of pairs in $\sew$.

\medskip\noindent
Proof:\\
The left hand side $L$ of \erf{eq:proj-vac} can be written as
  \be
  L = P^\text{vac}_{\sew,\Xr} \circ \bl\big((\sew,\id)\big) \circ \cor_{\Xr}
  = \tftC(\Mr_\sew) \circ \cor_{\Xr} \,,
  \ee 
where the cobordism $\Mr_\sew{:}\, \Xhat \To \widehat{\sew(\Xr)}$ 
coincides with the cobordism $\hat\sew$ defined in \erf{eq:shat-cobord} 
everywhere except in the annuli $C_{(a,b)} \Times [0,1]$ created by the sewing 
$(a,b) \iN \sew$, where there are additional $\Omeka$-ribbons from 
$P^\text{vac}_{\sew,\Xr}$.  Specifically, in a neighbourhood of 
one of the annuli $C_{(a,b)} \Times [0,1]$, $\Mr_\sew$ looks as follows.
  \eqpic{pic-fjfrs2_23} {420} {142} {
  \put(53,6)     {\Includeourbeautifulpicture 23a}
  \put(256,6)    {\Includeourbeautifulpicture 23b}
  \put(-20,156)  {$ \tftC(\Mr_\sew)~= $}
  \put(211,156)  {$ =~\dsty\sum_{\alpha,\beta} $}
  \put(370,7)    {$ =~\dsty\sum_{\alpha,\beta} \tftC(\Mr_{\alpha,\beta}) $}
  \put(124,273)  {\sse$ B_l$}
  \put(124,140)  {\sse$ B_r$}
  \put(136.5,152){\sse$ \Omeka$}
  \put(101,143)  {\sse$ w_{\!\Omeka}^{} $}
  \put(124,16)   {\sse$ B_l$}
  \put(136.5,27) {\sse$ \Omeka$}
  \put(101,18)   {\sse$ w_{\!\Omeka}^{} $}
  \put(327,275)  {\sse$ B_l$}
  \put(327,164)  {\sse$ B_r$}
  \put(327,139)  {\sse$ B_r$}
  \put(327,37)   {\sse$ B_l$}
  \put(307.1,159){\sse$ \tilde r_\beta$}
  \put(307.1,147){\sse$ \tilde e_\beta$}
  \put(307.1,32) {\sse$ r_\alpha$}
  \put(307.1,21) {\sse$ e_\alpha$}
  }
where it is understood that $\tftC(\,\cdot\,)$ is applied to each cobordism 
shown in the picture; the $(\one,e_\alpha,r_\alpha)$ label a basis for the 
different ways to realise $\one$ as a retract of $B_l$, 
$(\one,\tilde e_\alpha,\tilde r_\alpha)$ the ways to realise $\one$ as a 
retract of $B_r$, and we used \erf{eq:omega-project} twice.
Since $\cor_\Xr{:}\ \Cb \To \bl(\Xr)$, we can write
  \be
  L = \tftC(\Mr_\sew) \circ \cor_\Xr
  = \sum_{\alpha,\beta} \tftC(\Mr_{\alpha,\beta}) \circ P_\Xr \circ \cor_\Xr
  \labl{eq:proj-vac-aux1} 
with $P_\Xr$ the projector introduced in \erf{eq:PX-def}. 
Since by assumption there is a nonzero correlator, according to
lemma \ref{lem:semisphere-nonzero} the two morphisms $u_{B\eta}$ and 
$u_{B\eps}$ are both nonzero. Since $\dimc \Hom_{\Cc \boxtimes \ol\Cc}
(\one {\times} \ol\one, \Hcl) \eq 1$, there exist numbers
$\lambda_{\alpha\beta}, \tilde \lambda_{\alpha\beta} \iN \Cb$ such that
  \bea
  p_\sfc \circ (e_\alpha \Times \tilde r_\beta)
  = \lambda_{\alpha\beta}\, u_{B\eta} \in
  \Hom_{\Cc \boxtimes \ol\Cc}(\one {\times} \ol\one, \BlxBr)
  \qquad \text{and} \enL
    (r_\alpha \Times \tilde e_\beta) \circ p_\sfc
  = \tilde \lambda_{\alpha\beta}\, u_{B\eps} \in
  \Hom_{\Cc \boxtimes \ol\Cc}(\BlxBr, \one {\times} \ol\one)
  \end{array} \ee 
with $p_\sfc$ the idempotent in $\Hom(\BlxBr,\BlxBr)$ introduced in 
(\ref{eq:pS-intro}).
This allows us to replace $e_\alpha, \tilde e_\beta, r_\alpha, \tilde r_\beta$ 
in each of the terms $\tftC(\Mr_{\alpha,\beta}) \Circ P_{\Xr}$ in the sum 
\erf{eq:proj-vac-aux1} by the morphisms occurring in the decompositions 
$u_{B\eta} \eq \sum_\gamma u_\gamma' \oti u_\gamma''$ and $u_{B\eps} \eq \sum_
\delta n_\delta' \oti n_\delta''$, up to the constants $\lambda_{\alpha\beta}$ 
and $\tilde \lambda_{\alpha\beta}$. We can then use \erf{eq:cut-disc} 
and the corresponding identity for $u_{B\eps}$ to conclude that
  \be
  L = \lambda^{|\sew|/2} \,
  E^\text{vac}_{\sew,\Xr} \circ \cor_{\fill_\sew(\Xr)} \,,
  \labl{eq:proj-vac-aux2} 
where $|\sew|$ is the number of pairs in $\sew$ and $\lambda \eq 
\sum_{\alpha,\beta} \lambda_{\alpha\beta} \tilde \lambda_{\alpha\beta}$. The 
constant $\lambda$ is independent of $\Xr$. In particular, 
\erf{eq:proj-vac-aux2} must hold if we take $\Xr$, $\Yr$ and $\sew$ as in lemma
\ref{lem:Pvac-disc}. Then by lemma \ref{lem:Pvac-disc} we have 
$L \eq \cor_{\sew(\Xr)}$, so that in this case \erf{eq:proj-vac-aux2} becomes
  \be
  \cor_{\sew(\Xr)} = \lambda \,
  E^\text{vac}_{\sew,\Xr} \circ \cor_{\fill_\sew(\Xr)} \,.
  \labl{eq:proj-vac-aux3} 
To establish \erf{eq:proj-vac} it remains to show that $\lambda \eq \Lambda
_\sfc^{\,-1}$. Denote by $R$ the right hand side of \erf{eq:proj-vac-aux3}. 
Since $\Xr \eq \Yr 
\,{\sqcup}\, \Xr_{B\eta}$, the world sheet $\fill_\sew(\Xr)$ is isomorphic to 
the union of $\sew(\Xr)$ and a copy of $S^2$. Inserting the explicit form 
\erf{eq:Evac-def} for $E^\text{vac}_{\sew,\Xr}$ and substituting 
$\cor_{S^2} \eq \Lambda_\sfc\, \tftC(B^3\,{\sqcup}\,(-B^3))$, one finds that
  \be
  R = \lambda \, \Lambda_\sfc \, \cor_{\sew(\Xr)} \,.
  \ee 
Comparing this result with \erf{eq:proj-vac-aux3} and recalling that we may 
choose $\sew(\Xr)$ to be a world sheet with $\cor_{\sew(\Xr)} \,{\neq}\, 0$, it 
follows that $R \,{\neq}\, 0$ (and hence in particular $\Lambda_\sfc\,{\ne}\,0$)
and that $\lambda \eq \Lambda_\sfc^{\,-1}$. Thus $L$ in \erf{eq:proj-vac-aux2} 
is indeed equal to the right hand side of \erf{eq:proj-vac}. 
\qed

\dt{Remark}
Equation \erf{eq:proj-vac} is the analogue of the operation on world sheets with
metric of taking the limit in which a cylindrical neighbourhood of the image of 
$\sew$ in $\sew(\Xr)$ gets infinitely long, such that only the closed state 
vacuum can ``propagate along the 
cylinder.'' Proposition \ref{prop:proj-vac} also demonstrates that if there is 
at least one nonzero correlator and if $\dimc \Hom_{\Cc\boxtimes\ol\Cc}(\one
{\times}\ol\one, \Hcl) \eq 1$, then automatically $\cor_{S^2} \,{\ne}\, 0$.


\subsection{From Frobenius algebras to a solution
  to the sewing constraints}\label{sec:frob-to-sew}

In \cite{tft1,tft4,tft5}, a solution to the sewing constraints was explicitly
constructed using a symmetric special Frobenius algebra in the category $\Cc$. 
The version of this construction presented here, borrowed from \cite{rffs}, 
differs from the one in \cite{tft1,tft4,tft5} in the respect that we consider 
only a single boundary condition, and that individual boundary and bulk fields
are combined into algebra objects in $\Cc$ and $\CxCb$, respectively.

A symmetric special Frobenius algebra in $\Cc$ is a quintuple 
$(A,m,\eta,\Delta,\varepsilon)$, where $A\iN\Obj(\Cc)$, and $m,\eta,\Delta$ 
and $\varepsilon$ are the multiplication, unit, comultiplication and counit
morphisms. These morphisms can be visualised graphically as follows:
  \eqpic{pic_navf_c} {360} {16} {
     \put(0,0){
  \put(10,0)     {\Includeourbeautifulpicture 62a}
  \put(-29,20)   {$ m~= $}
  \put(6.8,-8.8) {\sse$ A $}
  \put(22.5,46.5){\sse$ A $}
  \put(36.8,-8.8){\sse$ A $}
     } \put(120,6){
  \put(10,0)     {\Includeourbeautifulpicture 62b}
  \put(-25,14)   {$ \eta~= $}
  \put(9.5,32.2) {\sse$ A $}
     } \put(220,0){
  \put(10,0)     {\Includeourbeautifulpicture 62c}
  \put(-29,20)   {$ \Delta~= $}
  \put(7.5,46.5) {\sse$ A $}
  \put(21.7,-8.8){\sse$ A $}
  \put(37.5,46.5){\sse$ A $}
     } \put(340,6){
  \put(10,5)     {\Includeourbeautifulpicture 62d}
  \put(-25,14)   {$ \eps~= $}
  \put(8.8,-4.6) {\sse$ A $}
     }
  }
$A$ is an algebra and a coalgebra, i.e.\ the above morphisms obey
  \beaa
  m\circ(m\oti\id_A) = m\circ(\id_A\oti m) & \mbox{(associativity)}\\[.4em]
  m\circ(\eta\oti\id_A) = \id_A = m\circ(\id_A\oti\eta) ~~& \mbox{(unitality)}\\[.4em]
  (\Delta\oti\id_A)\circ\Delta = (\id_A\oti\Delta)\circ\Delta &
                                                    \mbox{(coassociativity)}\\[.4em]
  (\varepsilon\oti\id_A)\circ\Delta = \id_A = (\id_A\oti\varepsilon)\circ\Delta
                                                    & \mbox{(counitality)}
  \end{array}
  \labl{eq:asun}
That $A$ is furthermore symmetric special Frobenius means that
  \beaa
  \Delta\circ m = (m\oti\id_A)\circ(\id_A\oti\Delta) 
              = (\id_A\oti m)\circ(\Delta\oti\id_A) & \mbox{(Frobenius)}\\[.4em]
  ((\varepsilon\Circ m)\oti\id_{A^\vee})\circ(\id_A\oti b_A)
              = (\id_{A^\vee}\oti(\varepsilon\Circ m))\circ(\tilde b_A\oti\id_A)
                                                    & \mbox{(symmetry)}\\[.4em]
  \varepsilon\circ\eta = \dim(A)\,\id_\one \quad{\rm and}\quad
  m\circ\Delta = \id_A & \hspace*{-3em}\mbox{((normalised) specialness)}
  \end{array}
  \labl{eq:ssF}
with $\dim(A)\,{\ne}\,0$.
The relations in (\ref{eq:asun}) and (\ref{eq:ssF}) are shown graphically in 
equations (3.2), (3.27), (3.29), (3.31) and (3.33) of \cite{tft1}, respectively.

\medskip

Given a symmetric special\,\footnote{~%
  Specialness requires only that the last conditions in \erf{eq:ssF} hold
  up to nonzero complex numbers. By rescaling morphisms  one can choose
  normalisations such that these constants are $1$ and $\dim(A)$, respectively.
  In the sequel we assume that a special algebra is normalised in this way. Thus
  from here on `special' stands for `normalised special'.}
Frobenius algebra $A$ in a ribbon category $\Cc$, consider the element 
  \eqpic{pic_fjfrs2_65} {30} {54} {
     \put(0,-2){
  \put(0,0)        {\Includeourbeautifulpicture 65{}}
  \put(-48.6,55.5) {$ P^l_{\!\!A}~:= $}
  \put(8.2,30)     {\sse$ A $}
  \put(25.5,-8.8)  {\sse$ A $}
  \put(26.1,125)   {\sse$ A $}
  \put(73.2,67)    {\sse$ A $}
     }
  }
of $\Hom(A,A)$. This is an idempotent (see e.g.\ lemma 5.2 in \cite{tft1}).
It is used in the following 
\\[-2.3em]

\dtl{Definition}{def:left-centre}
Let $A$ be a symmetric special Frobenius algebra $A$ in an idempotent complete
ribbon category $\Cc$. The {\em left center\/} $C_l(A)$ of $A$ is a retract
$(C_l(A),e_C,r_C)$ of $A$ such that $e_C \Circ r_C \eq P^l_{\!\!A}$.

\medskip

The left center is unique up to isomorphism of retracts and satisfies
$m\Circ c_{A,A}\Circ(e_C\oti\id_A) \eq m
   $\linebreak[0]$
\Circ(e_C\oti\id_A)$, whence the name.
Analogously one defines a right center in terms of a right central idempotent,
but we will not need this concept here. More details and references on the
left and right centers can be found in \cite[sect.\,2.4]{ffrs}.

\medskip

To describe the space of closed states below, we need a certain algebra in 
$\CxCb$. Choose a basis $\{\lambda^\alpha_{(ij)k}\}_\alpha$ in each of the 
spaces $\Hom(U_i\otimes U_j, U_k)$. Denote by $\{\lambda^\alpha_{k(ij)}\}_\alpha$
the basis of $\Hom(U_k,U_i\otimes U_j)$ that is dual to the former in the 
sense that $\lambda^\alpha_{(ij)k} \Circ \lambda^\beta_{l(ij)} 
\eq \delta_{k,l}\, \delta_{\alpha,\beta}\, \id_{U_k}$.

\dtl{Definition}{def:triv-alg} 
For $\Cc$ a modular tensor category, The {\em canonical trivialising algebra\/} 
$T_\Cc\,{\equiv}\, (T_\Cc,m_T,\eta_T)$ in the product category $\CxCb$ is the 
algebra with underlying object
  \be
  T_\Cc := \bigoplus_{i\in \mathcal{I}}\, U_i\times\ol{U_i}
  \ee
and with unit morphism $\eta_T$ defined to be the obvious monic
$e_{\one\times\ol\one\prec T_\Cc}$, and multiplication $m_T$ defined through
its restrictions $m_{ij}^{\phantom{i.}k}$ to $\Hom_{\Cc\boxtimes\ol\Cc}
\big( (U_i\Times\ol{U_i})\oti(U_j\Times\ol{U_j}) , U_k\Times\ol{U_k} \big)$ by
  \be
  m_{ij}^{\phantom{ij}k} :=
  \sum_\alpha\lambda^\alpha_{(ij)k}\oti \overline{\lambda^{\alpha}_{k(ij)}} \,.
  \ee

As shown in section 6.3 of \cite{ffrs}, $T_\Cc$ extends to a haploid
commutative symmetric special Frobenius algebra in $\CxCb$. The qualification
`trivialising' derives from the fact that the category of local $T_\Cc$-modules 
in $\CxCb$ is equivalent to $\Vect$ (see proposition 6.23 of \cite{ffrs}), but 
this property will not play a role here. Instead, $T_\Cc$ is instrumental in the
\\[-2.3em]

\dtl{Definition}{def:full-centre} 
For $A$ a symmetric special Frobenius algebra in a modular tensor category
$\Cc$, the {\em full center\/} $Z(A)$ of $A$ is the object
  \be
  Z(A) := C_l( (A{\times}\one) \oti T_\Cc ) ~ \in \Obj(\CxCb) \,.
  \ee

For this definition to make sense, $(A \Times \one) \oti T_\Cc$ must itself be 
a symmetric special Frobenius algebra. This is indeed the case, see for example 
section 3.5 of \cite{tft1}. Moreover, as shown in the appendix of \cite{rffs}, 
we have
\\[-2.3em]

\dtl{Lemma}{lem:Z(A)-sFA} 
The full center $Z(A)$ is a commutative symmetric Frobenius algebra.

\dt{Remark}
(i)~\,In a braided monoidal category, there are in fact two inequivalent ways 
to endow the tensor product $B \oti C$ of two algebras $B$ and $C$ with an 
associative product; one can either take $(m_B \oti m_C) \Circ (\id_B \oti 
c_{C,B}^{~-1} \oti \id_C)$ or one can use the braiding itself instead of 
its inverse. Our convention is to use the inverse braiding.
\\[.3em]
(ii)~To be more precise, in definition \ref{def:full-centre} we should use 
the term `left full center.' There is analogously a right full center, defined 
in terms of the right center $C_r(\cdot)$, if one at the same time uses the 
other convention for the tensor product $B \oti C$ of algebras as mentioned in 
(i). Or one could use the algebra $A \iN \Obj(\Cc)$ to obtain an algebra 
$\ol A \iN \Obj(\ol\Cc)$, and consider $C_{l/r}((\one\Times\ol A)\oti T_\Cc)$. 
But these four algebras are related by isomorphism or by exchanging the roles of
$\Cc$ and $\ol\Cc$, and each one determines the other three up to isomorphism.
We will work with $Z(A)$ as given in definition \ref{def:full-centre}.
\\[.3em]
(iii)~In a symmetric tensor category the notions of left and right center
coincide, and in the category of vector spaces they also 
coincide with the notion of full center.

\medskip

Let now again $A$ be a symmetric special Frobenius algebra in a modular tensor 
category $\Cc$. Let the morphisms $e_Z \iN \Hom_{\Cc\boxtimes\ol\Cc}(Z(A),
(A\oti\Omeka)\Times\ol\Omeka)$ and $r_{\!Z} \iN \Hom_{\Cc\boxtimes\ol\Cc}( 
(A\oti\Omeka)\Times\ol\Omeka,Z(A))$ be given by
\void{
  \eqpic{eq:eZ-rZ-def} {340} {74} {
  \put(70,0)    {\Includeournicemediumpicture 25a }
  \put(260,0)   {\Includeournicemediumpicture 25b }
  \put(0,76)    {$ e_Z~=~\dsty\sum_{i\in\Ic} $}
  \put(190,76)  {$ r_Z~=~\dsty\sum_{i\in\Ic} $}
  \put(82,-7)   {\sse $Z(A)$}
  \put(55,160)  {\sse $(A\!\oti\!\Omeka)\!\times\!\ol\one$}
  \put(103,160) {\sse $\one\!\times\!\ol\Omeka$}
  \put(85,28)   {\sse $e_C$}
  \put(95,54)   {\sse $\tilde r_i$}
  \put(95,80)   {\sse \rm{id}}
  \put(82,107)  {\sse $e_i$}
  \put(109,107) {\sse $\ol r_i$}
  \put(100,40)  {\sse $T_\Cc$}
  \put(60,54)   {\sse $A\!\times\!\ol\one$}
  \put(100,67)  {\sse $U_i\!\times\!\ol U_i$}
  \put(87.5,98) {\tiny $U_i\!\!\times\!\ol\one$}
  \put(110,90)  {\tiny $\one\!\times\!\ol U_i$}
  \put(82,129)  {\sse $\Omeka\!\times\!\ol\one$}
  \put(245,-7)  {\sse $(A\!\oti\!\Omeka)\!\times\!\ol\one$}
  \put(293,-7)  {\sse $\one\!\times\!\ol\Omeka$}
  \put(272,160) {\sse $Z(A)$}
  \put(272,47)  {\sse $r_i$}
  \put(299,47)  {\sse $\ol e_i$}
  \put(285,72)  {\sse \rm{id}}
  \put(285,99)  {\sse $\tilde e_i$}
  \put(275,126) {\sse $r_C$}
}  }
  \eqpic{eq:eZ-rZ-def} {340} {74} {
  \put(70,0)    {\Includeournicemediumpicture 25a }
  \put(260,0)   {\Includeournicemediumpicture 25b }
  \put(0,76)    {$ e_Z~=~\dsty\sum_{i\in\Ic} $}
  \put(190,76)  {$ r_Z~=~\dsty\sum_{i\in\Ic} $}
  \put(82,-8)   {\sse $Z(A)$}
  \put(55,161)  {\sse $(A\!\oti\!\Omeka)\!\times\!\ol\one$}
  \put(103,161) {\sse $\one\!\times\!\ol\Omeka$}
  \put(85,28)   {\sse $e_C$}
  \put(96,53.7) {\sse $\tilde r_i$}
  \put(95,80)   {\sse \rm{id}}
  \put(82,107)  {\sse $e_i$}
  \put(109,106.6){\sse $\ol r_i$}
  \put(99,40)   {\sse $T_\Cc$}
  \put(60,54)   {\sse $A\!\times\!\ol\one$}
  \put(100,67)  {\sse $U_i\!\times\!\ol U_i$}
  \put(87.9,98) {\tiny $U_i\!\!\times\!\ol\one$}
  \put(110,90)  {\tiny $\one\!\times\!\ol U_i$}
  \put(82.5,132){\sse $\Omeka\!\times\!\ol\one$}
  \put(245,-8)  {\sse $(A\!\oti\!\Omeka)\!\times\!\ol\one$}
  \put(293,-8)  {\sse $\one\!\times\!\ol\Omeka$}
  \put(272,161) {\sse $Z(A)$}
  \put(272,47)  {\sse $r_i$}
  \put(299,47)  {\sse $\ol e_i$}
  \put(285,72.8){\sse \rm{id}}
  \put(285,99)  {\sse $\tilde e_i$}
  \put(275,126.5){\sse $r_C$}
  \put(272,20)  {\sse $\Omeka\!\times\!\ol\one$}
  \put(279,57)  {\tiny $U_i\!\!\times\!\ol\one$}
  \put(299,63)  {\tiny $\one\!\times\!\ol U_i$}
  \put(290,85)  {\sse $U_i\!\times\!\ol U_i$}
  \put(289,112) {\sse $T_\Cc$}
  \put(251.5,102){\sse $A\!\times\!\ol\one$}
  }
where $(U_i,e_i,r_i)$ realises $U_i$ as a retract of $\Omeka$ and
$(U_i\Times\ol{U_i},\tilde e_i,\tilde r_i)$ realises $U_i\Times\ol{U_i}$ as a 
retract of $T_\Cc$.

\dtl{Lemma}{lem:Z(A)-retract} 
$(Z(A),e_Z,r_Z)$ is a retract of $\BlxBr$ with $B_l \eq A\oti\Omeka$ and 
$B_r \eq \Omeka$.

\medskip\noindent
Proof:\\
We have to show that $r_Z \Circ e_Z \eq \id_{Z(A)}$. This can be done by writing
out the definitions of $r_Z$ and $e_Z$ and using the identity $r\Circ e\eq \id$
for the various embedding and restriction morphisms that appear in 
\erf{eq:eZ-rZ-def}, as well as $\sum_i e_i\Circ r_i \eq \id_\Omeka$.
\qed

\dt{Remark} In \cite{rffs} the objects $B_l$ and $B_r$ were both chosen to be 
$A\otimes\Omeka$. Since according to section \ref{sec:sew-equiv} it is
irrelevant how $\Hcl$ is realised as a retract, this does not affect any of
our results.

\bigskip

By choosing 
  \be
  \Hop :=A  \qquad{\rm and}\qquad \Hcl := Z(A)
  \ee
for a symmetric special Frobenius algebra $A$, we can construct a collection of 
correlators in the following way. Recall the definition of the connecting 
manifold $\Mr_\Xr$ in \erf{eq:conmf-def}. Let $\Xr$ be a world sheet and $x$ be 
a point in $\dot\Xr$. Take $p \iN \Xtil$ to be a point in the pre-image of 
$\pi_\Xr {:}\ \Xtil \To \dot\Xr$, and define a map $I {:}\ \dot\Xr \To \Mr_\Xr$
by setting $I(x) \,{:=}\, [x,0]$. The equivalence relation in \erf{eq:conmf-def}
ensures that $I$ is well defined and injective.

To construct the ribbon graph in $\Mr_\Xr$ we first need to choose a directed 
dual triangulation of $\dot\Xr$ not intersecting the images of 
$b_\Xr^{\text{in}}\,{\cup}\, b_\Xr^{\text{out}}$. Here by the qualification
{\em dual\/} we mean that all vertices are trivalent, while faces can have an 
arbitrary number of edges. The (dual) triangulation is constrained such as to 
cover the physical components of $\partial\dot\Xr$ with direction given by the 
orientation of $\partial\dot\Xr$ (taking this direction rather than the opposite
one is merely a convention), and such that at each vertex there are both inwards-
and outwards-directed edges. The ribbon graph is then constructed as follows. 
\\[-1.6em]

\def\leftmargini{1.4em}\begin{itemize}
\addtolength\itemsep{-6pt}
\item[1)] 
Each edge is covered by a ribbon labeled by the algebra object $A$, such that 
the core orientation of the ribbon is opposite to the direction of the 
corresponding edge, and
the 2-ori\-entation is opposite to that of $I(\dot\Xr)$. 
\item[2)]
A vertex is covered by a coupon labeled by $m$ (respectively, $\Delta$) if 
there are two in-going edges (respectively, out-going edges) meeting at the 
vertex. 
\item[3)]
In a neighbourhood of an open state boundary (an interval resulting from 
$a\iN b^{\text{in/out}}$ such that $\iota_*(a)\eq a$), a ribbon labeled by 
$A$ is inserted running from $\partial\Mr_\Xr$ towards $\partial\dot\Xr$. If 
the open state boundary is in-going (respectively, out-going), the core 
orientation is chosen inwards from (respectively, out towards) $\partial
\Mr_\Xr$. The ribbon is joined to the dual triangulation at $\partial\dot\Xr$     
by a coupon labeled $m$ ($\Delta$) for an in-(out-)going open 
state boundary. The two possibilities are displayed in the following picture:
  \eqpic{pic_fjfrs2_70} {350} {48} {
  \put(0,-6)     {\IncludeourniceMediumpicture 70a}
  \put(200,-6)   {\IncludeourniceMediumpicture 70b}
  \put(0,-6){
     \setlength{\unitlength}{.48pt}\put(-14,-13){
     \put(0,247)       {in-going:}
     \put(101,142)   {\sse$ A $}
     \put(181,169)   {\sse$ A $}
     \put(115, 99)   {\sse$ m $}
     \put(108, 65)   {\sse$ A $}
     }\setlength{\unitlength}{1pt}}
  \put(200,-6){
     \setlength{\unitlength}{.48pt}\put(-14,-13){
     \put(0,247)       {out-going:}
     \put(101,142)   {\sse$ A $}
     \put(161,181)   {\sse$ A $}
     \put(149,148)   {\sse$ \Delta $}
     \put(116, 80)   {\sse$ A $}
     }\setlength{\unitlength}{1pt}}
  }
\item[4)]
For closed state boundaries (circles corresponding to $a\iN b^{\text{in/out}}$ 
such that $\iota_*(a)\,{\ne}\,a$), the prescription is somewhat more involved. 
Consider the \disc s $D_a$ and $D_b$ glued to closed state boundary components
$a$ and $b\eq\iota_*(a)$. By definition of the three-manifold $\Mr_\Xr$, the two 
cylinders $\{ (p,t) \,|\, p \iN D_a,\, t \iN [-1,1] \}$ and $\{ (p,t) \,|\, 
p \iN D_b,\, t \iN [-1,1] \}$ are identified. In this cylinder there has to be 
inserted one of the ribbon graphs shown below, depending on whether the closed 
state boundary is in- or out-going.
  \eqpicc{pic_fjfrs2_68} {410} {80} {
  \put(-15,-16)  {\IncludeourniceMediumpicture 68a}
  \put(225,-16)  {\IncludeourniceMediumpicture 68b}
  \put(-15,173)  {in-going:}
  \put(225,173)  {out-going:}
  \put(0,75)     {\sse $A$}
  \put(120,95)   {\sse $A$}
  \put(162,75)   {\sse $A$}
  \put(30,42)    {\sse $A$}
  \put(116,52)   {\sse $A$}
  \put(77,70)    {\sse $A$}
  \put(48,85)    {\sse $\Omeka$}
  \put(71,117)   {\sse $A\oti\Omeka$}
  \put(63,101)   {\sse \rm{id}}
  \put(22,75)    {\sse \begin{turn}{-120}$\Delta$\end{turn}}
  \put(86,53)    {\sse \begin{turn}{-120}$m$\end{turn}}
  \put(139,75)   {\sse \begin{turn}{-120}$m$\end{turn}}
  \put(242,75)   {\sse $A$}
  \put(404,75)   {\sse $A$}
  \put(295,72)   {\sse \begin{turn}{-20}$A$\end{turn}}
  \put(317,101)  {\sse \rm{id}}
  \put(262,75)   {\sse \begin{turn}{-120}$\Delta$\end{turn}}
  \put(271,53)   {\sse \begin{turn}{-120}$\Delta$\end{turn}}
  \put(379,75)   {\sse \begin{turn}{-120}$m$\end{turn}}
  \put(331,35)   {\sse $\Omeka$}
  \put(320.6,68) {\sse $w_\Omeka$}
  \put(325,117)  {\sse $A\oti\Omeka$}
  }
\end{itemize}

For a given triangulation $\Tri$, denote the resulting cobordism of extended 
surfaces by $\Mr_\Xr(\Tri)$. Define a linear map 
$\corA_{\Xr,\Tri}{:}\ \Cb\To \bl(\Cc,A,Z(A),A\oti\Omeka,\Omeka,e,r)$ by
  \be
  \corA_{\Xr,\Tri} := \tftC(\Mr_\Xr(\Tri)) \,.
  \ee
It was established in \cite{tft5}\,\footnote{~%
  Some of the conventions in \cite{tft5} differ from those used here.
  In \cite{tft5} every edge at a vertex is directed outwards, and subsequently 
  the prescription for constructing the ribbon graph differs from the one given
  here.  Using that $A$ is symmetric, special and Frobenius, it is, however, 
  easily realised that the linear maps obtained after applying the 3-d TFT functor
  to the respective ribbon graphs are equal.}
that $\corA_{\Xr,\Tri}\eq\corA_{\Xr,\Tri'}$ for any two dual triangulations 
$\Tri$ and $\Tri'$, and it therefore makes sense to abbreviate $\corA_{\Xr,\Tri}$
by $\corA_{\Xr}$.
As further shown in \cite{rffs}, the tuple $(\Cc,A,Z(A),A\oti\Omeka,A,e_Z,r_Z,
\corA)$  is a solution to the sewing constraints, i.e.\ the collection of 
correlators as defined above gives a monoidal natural transformation. We 
therefore arrive at the following\,\footnote{~%
  As mentioned in the previous footnote, there are slight differences between
  the prescriptions in \cite{tft5} and the present paper. But it is
  straightforward to adapt the proofs of \cite{tft5}. We refrain from giving the
  details here; an outline can be found in section 3 and the appendix of
  \cite{rffs}.}
\\[-2.3em]

\dtl{Theorem}{thm:Frob-SCA}
For any symmetric special Frobenius algebra $A$ in a modular tensor category $\Cc$,
the tuple
  \be
  \sfc(\Cc,A) := (\Cc,A,Z(A),A\oti\Omeka,\Omeka,e_Z,r_Z,\corA)
  \labl{eq:sfc-def}
is a solution to the sewing constraints.

\medskip

Given the solution $\sfc(\Cc,A)$ to the sewing constraints, we can express the 
fundamental correlators with the help of morphisms involving the algebra object 
$A$.  Carrying out the construction described above results in the expressions
  \be
  \corA_{\Xr_a} = \tftC( \Fr(\Xr_a,f_a) )
  \qquad{\rm for}\quad a \in \{\eta,\eps,m,\Delta,Bb\} \,,
  \ee
where the cobordisms $\Fr(\Xr_a,f_a)$ are those given in figure 
\ref{fig:fund-cobord} and the morphisms $f_a$ are determined by $A$ as
  \be
  f_\eta = \eta \,,\qquad f_\eps = \eps \,,\qquad f_m = m \,,\qquad
  f_\Delta = \Delta \qquad{\rm and}\qquad f_{Bb} = \Phi_{\!A} \,,
  \labl{eq:disc-corr-by-A}
with the morphism $\Phi_{\!A}\iN \Hom(A\oti A\oti\Omeka, A\oti\Omeka)$ given by
  \eqpic{eq:PhiA} {140} {81} {
  \put(50,0)    {\Includeournicemediumpicture 38{}}
  \put(0,83)    {$ \Phi_{\!A}~= $}
  \put(66.5,-8) {\sse $A$}
  \put(94,-8)   {\sse $A$}
  \put(110,-8)  {\sse $\Omeka$}
  \put(67,177)  {\sse $A$}
  \put(110,177) {\sse $\Omeka$}
  \put(52,115)  {\sse $A$}
  \put(127,70)  {\sse $A$}
  }
The expressions for the correlators on $\Xr_{B(1)}$, $\Xr_{B(3)}$ and $\Xr_{oo}$
in terms of $A$ are not required for the calculations below, but we include them
here for completeness. It is convenient to use the cobordisms
  \eqpiczz{pic-fjfrs2_52} {420} {122} {
  \put(73,-6)    {\Includeournicelargepicture 52a}
  \put(-12,52)   {$ G_{B(1)}(g_{B(1)})~= $}
  \put(108.7,56) {\tiny$ g_{\!B(1)}^{} $}
  \put(121,35)   {\sse$ K $}
  \put(121,71)   {\sse$ A{\otimes}K $}
  \put(300,-6)   {\Includeournicelargepicture 52b}
  \put(213,52)   {$ G_{B(3)}(g_{B(3)})~= $}
  \put(355,56)   {\sse$ g_{B(3)}^{} $}
  \put(387,35)   {\sse$ K $}
  \put(387,69)   {\sse$ A{\otimes}K $}
  } {
  \put(290,-11)  {\Includeournicelargepicture 52c}
  \put(217,47)   {$ G_{oo}(g_{oo})~= $}
  \put(337,50)   {\sse$ g_{oo}^{} $}
  \put(357,28)   {\sse$ \Omeka $}
  \put(357,68)   {\sse$ A{\otimes}\Omeka $}
  }
In terms of these cobordisms we have
$\corA_{\Xr_a} \eq \tftC( G_a(g_a) )$ for $a \iN \{B(1), B(3), oo\}$, with
  \eqpic{pic-fjfrs2_69} {130} {77} {
  \put(60,0)     {\Includeourbeautifulpicture 69b}
  \put(0,79)     {$ g_{B(1)}~= $}
  \put(86,-8)    {\sse $A$}
  \put(100,-8)   {\sse $\Omeka$}
  \put(100,169.5){\sse $\Omeka$}

  }
and
  \eqpicc{pic-fjfrs2_69-2} {420} {157} {
  \put(32,-21)   {\Includeourbeautifulpicture 69d}
  \put(-20,151)  {$ g_{B(3)}~= $}
  \put(77,-29)   {\sse $A$}
  \put(90,-29)   {\sse $\Omeka$}
  \put(90,335)   {\sse $\Omeka$}
  \put(144,-29)  {\sse $A$}
  \put(157,-29)  {\sse $\Omeka$}
  \put(157,335)  {\sse $\Omeka$}
  \put(211,-29)  {\sse $A$}
  \put(224,-29)  {\sse $\Omeka$}
  \put(224,335)  {\sse $\Omeka$}
  \put(316,30)   {\Includeourbeautifulpicture 69e}
  \put(271,151)  {$ g_{oo}~= $}
  \put(347,22)   {\sse $\Omeka$}
  \put(412,22)   {\sse $\Omeka$}
  \put(323,309.6){\sse $A$}
  \put(349,309.6){\sse $\Omeka$}
  \put(384,309.6){\sse $A$}
  \put(412,309.6){\sse $\Omeka$}
  \put(345,151)  {\sse $w_\Omeka$}
  \put(409,192)  {\sse $w_\Omeka$}
  }  

\dt{Remark}
(i)~~The class of two-di\-men\-sional conformal field theories contains in 
particular the two-di\-men\-sional topological field theories. For topological
field theories the modular tensor category $\Cc$ of the present setup is 
equivalent to $\Vect$. The data collected in a solution to the sewing 
constraints can in this case be compared to those encoded in a `knowledgeable 
Frobenius algebra', that is \cite{lapf}, a quadruple $(A,C,\iota,\iota^*)$
consisting of a symmetric Frobenius algebra $A$, a commutative Frobenius algebra
$C$, an algebra homomorphism $\iota{:}\ C \To A$ from $C$ to the center of $A$
and a linear map $\iota^*{:}\ A \To C$ that is uniquely determined by $A$, $C$
and $\iota$. It has been shown \cite{lapf} that specifying an open/closed 2-d TFT
(in the sense of \cite{lapf}) is equivalent to giving a knowledgeable Frobenius
algebra. A 2-d TFT (in the sense of \cite{lapf}) with target category $\Vect$
gives rise to a solution to the sewing constraints for $\Vect$. 
However, in general not every solution can be obtained this way, as it is not
required that the correlators for $\Xr_p$ and $\Xr_{Bp}$ correspond to invertible 
morphisms of $\Cc$, whereas for a 2-d TFT they get automatically mapped to the
identity because $\Xr_p$ and $\Xr_{Bp}$ are the identity morphisms in the
relevant cobordism category.
\\[.3em]
(ii)~\,Going from the special case $\Cc \,{\simeq}\, \Vect$ to the general 
situation, we see that $A$ and $Z(A)$ in the solution $\sfc(\Cc,A)$ remain a 
symmetric and a commutative Frobenius algebra, respectively. However, $A$ and 
$Z(A)$ are now objects of different categories, namely of $\Cc$ and of $\CxCb$,
respectively (in the topological case $\Cc\,{\simeq}\,\Vect$ the difference is
not noticeable because 
$\Vect\hspace*{+.05em}{\boxtimes}\hspace*{+.1em}\overline{\Vect} \,{\simeq}\, 
\Vect\hspace*{+.05em}{\boxtimes}\hspace*{+.1em}\Vect \,{\simeq}\, \Vect$).
\\[.3em]
(iii)~Due to the presence of the scale parameter $\gamma \iN \Cb^\times$ in the
definition \erf{eq:sol-equiv}, which is motivated by the physical considerations
made around \erf{f.gamma}, the notion of equivalence for solutions to the 
sewing constraints is broader than isomorphy of knowledgeable Frobenius algebras 
(in the case $\Cc\eq\Vect$, when both structures are defined).
\\[.3em]
(iv)~As pointed out first in \cite{tft1} (sections 3.2 and 5.1), the symmetric 
special Frobenius algebra $A$ used to decorate the triangulation of the world 
sheet is in fact the same as the algebra of boundary fields for the boundary 
condition labeled $A$. In the present formulation this manifests itself in 
the fact that the correlators on the \disc s $\Xr_\eta$, $\Xr_\eps$, $\Xr_m$ 
and $\Xr_\Delta$ are directly given by the (co)unit
and (co)multiplication of $A$, see formula \erf{eq:disc-corr-by-A}.
\\
This effect has also been observed in the special case of two-dimensional 
topological field theory \cite{lapf2}. On the other hand, the treatment in 
\cite{lapf2} is more general than what is obtained by restricting our formalism to 
$\Cc\eq\Vect$. Namely, it is not required that one works over an algebraically 
closed field, the category $\Vect$ can be replaced by a more general symmetric 
monoidal category, and the Frobenius algebra used to decorate the triangulation
is only demanded to be strongly separable, a slightly weaker condition than 
symmetric special.


\subsection{From a solution to the sewing
            constraints to a Frobenius algebra}\label{sec:sew-to-frob}

It will be useful to have at our disposal a way to `cut' a world
sheet into simpler pieces without having to specify explicitly the
parametrisation of the newly arising state boundaries of the individual
pieces. This is achieved by the next two definitions.
\\[-2.2em]

\dtl{Definition}{def:cutting}
(i) A {\em cutting of a world sheet\/} $\Xr$ is a subset $\gamma$ of $\Xtil$ 
such that $\gamma \,{\cap}\, \partial\Xtil \eq \emptyset$,
$\iota_\Xr(\gamma) \eq \gamma$, and each connected component of $\gamma$ is
homeomorphic to the half-open annulus $1 \,{\le}\, |z| \,{<}\, 2 \subset \Cb$.
\\[2pt]
(ii) Two cuttings $\beta$ and $\gamma$ of a world sheet are {\em equivalent\/},
denoted by $\beta \sim \gamma$, iff they contain the same boundary circle, 
i.e.\ iff $\partial\beta{\cap}\beta \eq \partial\gamma{\cap}\gamma$. 

\medskip

Note that every connected component of the projection of a cutting to the
quotient $\dot\Xr$ for a
world sheet $\Xr$ either has the topology of an annulus or of a semi-annulus.

Given world sheets $\Xr$ and $\Yr$ and a morphism $\xm{:}\ \Xr \To \Yr$, one 
obtains a cutting $\Gamma(\xm)$ of $\Yr$ as follows. Write $\xm \eq (\sew,f)$
and choose a small open neighbourhood $U$ of the union of all boundary 
components $b$ of $\partial\Xtil$ for which $(a,b) \iN \sew$. By replacing $U$ 
by $U \,{\cup}\, \iota_\Xr(U)$ if necessary, one can ensure that $\iota_\Xr(U) 
\eq U$. We denote by $\Gamma(\xm)$ the corresponding subset of $\Yr$, i.e.\ 
set $\Gamma(\xm) \,{:=}\, f\Circ\pi_{\sew,\Xr}(U) \subset \Yr$. 
Different choices for $U$ lead to equivalent cuttings $\Gamma(\xm)$.
Using the operation $\Gamma(\,\cdot\,)$ we can formulate
\\[-2.3em]

\dtl{Definition}{def:realise-cut}
A {\em realisation of a cutting\/} $\gamma$ of a world sheet $\Xr$ is a world 
sheet $\Xr|_\gamma$ together with a morphism $c_\gamma{:}\ \Xr|_\gamma\To\Xr$ 
such that $\Gamma(c_\gamma) \,{\sim}\, \gamma$.

\medskip

Similarly to the isomorphism $\Psi_m$ in \erf{eq:Xm-iso}, for any world sheet 
$\Xr_a$ of the type a \disc\ with $p$ in- and $q$ out-going open state 
boundaries we are given an isomorphism 
  \be
  \Psi_a{:}\quad \Hom(\Hop^{\otimes p}, \Hop^{\otimes q})\to
  \Htft(\widehat \Xr_a) \,.
  \labl{eq:Psi_a-def}
Given a solution $\sfc$ to the sewing constraints, we define $m _\sfc$ to be 
the unique element of $\Hom(\Hop\oti\Hop,\Hop)$ such that 
$\cor_{\Xr_m} \eq \tftC(\Fr(\Xr_m;m_\sfc))$ or, equivalently,
  \be
  m_\sfc = \Psi_m^{-1}\big( \cor_{\Xr_m} \big) \,.
  \ee 
Analogously we can use the isomorphisms $\Psi_x$, $x\iN\{\eta,\Delta,\eps,p\}$,
coming from the corresponding cobordisms in figure \ref{fig:fund-cobord} to 
define morphisms $\eta_\sfc \iN \Hom(\one,\Hop)$, $\Delta_\sfc \iN
\Hom(\Hop,\Hop\oti\Hop)$, $\eps_\sfc \iN \Hom(\Hop,\one)$ and 
$Q_\sfc \iN \Hom(\Hop,\Hop)$ via
  \bea
  \eta_\sfc := \Psi_\eta^{-1}\big( \cor_{\Xr_\eta} \big) \,,\qquad
  \Delta_\sfc := \Psi_\Delta^{-1}\big( \cor_{\Xr_\Delta} \big) \\{}\\[-.6em]
  \eps_\sfc := \Psi_\eps^{-1}\big( \cor_{\Xr_\eps} \big)\,,\qquad
  Q_\sfc := \Psi_p^{-1}\big( \cor_{\Xr_p} \big)\,.
  \end{array}\labl{eq:algdef}

\dtl{Lemma}{lem:ps-identity} 
Let $\sfc$ be a solution to the sewing constraints such that $Q_\sfc \iN
\Hom(\Hop,\Hop)$ is an isomorphism. Then $Q_\sfc \eq \id_{\Hop}$.

\medskip\noindent
Proof:\\
Consider the world sheet $\Xr_p$ with a cutting $\alpha$ such that 
$\Xr_p|_\alpha \Cong \Xr_p\sqcup\Xr_p$. Choose a realisation of $\alpha$ of the 
form $c_\alpha{:}\ \Xr_p\sqcup\Xr_p \To \Xr_p$. Naturality of $\cor$ implies
  \be
  \cor_{\Xr_p} = \bl(c_\alpha)\circ(\cor_{\Xr_p}\oti\cor_{\Xr_p})
  = \bl(c_\alpha)\circ\tftC(\Fr(\Xr_p;Q_\sfc)\sqcup\Fr(\Xr_p;Q_\sfc)) \,.
  \labl{CorXp}
Expressing $\bl(c_\alpha)$ through a cobordism, and implementing the composition
by gluing of cobordisms, implies that the right hand side of \erf{CorXp} equals 
$\tftC(\Fr(\Xr_p;Q_\sfc\circ Q_\sfc))$. Applying $\Psi_p^{-1}$ results in 
$Q_\sfc\Circ Q_\sfc \eq Q_\sfc$, i.e.\ $Q_\sfc$ is an idempotent. But by 
assumption $Q_\sfc$ is also invertible, hence $Q_\sfc\eq\id_{\Hop}$. 
\qed

\dtl{Proposition}{prop:openalg} 
Let $\sfc$ be a solution to the sewing constraints such that $Q_\sfc$ is 
invertible. Then 
  \be
  A_\sfc := (\Hop,m_\sfc,\eta_\sfc,\Delta_\sfc,\eps_\sfc)
  \labl{eq:Asfc-def}
is a symmetric Frobenius algebra in $\Cc$. 

\medskip\noindent
Proof:\\
The proof follows the standard route to extract properties of the generating
world sheets from different ways of decomposing more complex ones.
\\[.3em]
\nxt {Unit property:}\\[2pt]
The relation to be shown is the second relation of (\ref{eq:asun}). 
Consider the cutting $\alpha$ of $\Xr_p$ indicated in 
  \eqpic{pic_fjfrs2_61} {100} {30} {
  \put(0,1)      {\Includeournicemediumpicture 61{}}
  \put(-11,35)   {\small$ in$}
  \put(132,35)   {\small$ out$}
  \put(60,28)    {$\alpha$}
  }
Choose a realisation $c_\alpha{:}\ \Xr_\eta\sqcup\Xr_m \To \Xr_p$ of 
this cutting $\alpha$. Naturality of $\cor$ implies
  \be
  \cor_{\Xr_p} = \bl(c_\alpha)\circ
  \tftC(\Fr(\Xr_m;m_\sfc)\,{\sqcup}\,\Fr(\Xr_\eta;\eta_\sfc)) \,.
  \ee
Implementing the composition with $\bl(c_\alpha)$ by gluing cobordisms yields
  \eqpic{pic-fjfrs2_63} {170} {48} {
  \put(70,0)     {\Includeourbeautifulpicture 63{}}
  \put(0,51)     {$ \cor_{\Xr_p}~= $}
  \put(70,0){
     \setlength{\unitlength}{.38pt}\put(-98,-407){
     \put(228,565)   {\sse \begin{turn}{-12}$ m_{\sfc} $\end{turn}}
     \put(252,506)   {\sse \begin{turn}{25}$ \eta_{\sfc}^{} $\end{turn}}
     \put(142,531)   {\sse$ \Hop $}
     \put(218,593)   {\sse$ \Hop $}
     }\setlength{\unitlength}{1pt}}
  }
Applying $\Psi_p^{-1}$ to both sides of this equality and using lemma 
\ref{lem:ps-identity} yields
$\id_{\Hop} \eq m_\sfc\Circ(\id_{\Hop}\oti\eta_\sfc)$. That the equality 
$\id_{\Hop} \eq m_\sfc\Circ(\eta_\sfc\oti \id_{\Hop})$ holds as well can be seen
analogously. This establishes the unit property.
\\[.3em]
\nxt {Associativity:}\\[2pt]
Next we show the first relation in (\ref{eq:asun}). Consider the world sheet
$\Xr_q$ for which $\dot\Xr_q$ is a \disc\ with three in-going and one out-going 
open state boundaries, and with cuttings $\alpha$ and $\beta$ as indicated in 
  \eqpic{pic_fjfrs2_64} {166} {28} {
  \put(45,0)     {\Includeourbeautifulpicture 64{}}
  \put(0,37)     {$ \Xr_q~= $}
  \put(96,24)    {$ \alpha $}
  \put(133,28)   {$ \beta $}
  }
Let $c_\alpha, c_\beta{:}\ \Xr_m\,{\sqcup}\,\Xr_m \To \Xr_m$ be realisations of 
$\alpha$ and $\beta$, respectively. Naturality implies
  \be
  \cor_{\Xr_q} = \bl(c_{\delta})\circ\tftC(\Fr(\Xr_m;m_\sfc)
  \,{\sqcup}\,\Fr(\Xr_m;m_\sfc))
  \ee
for $\delta \eq \alpha,\beta$. Evaluating the right hand side by gluing 
cobordisms, followed by applying $\Psi_q^{-1}$, yields the equality
  \be
  m_\sfc\circ(m_\sfc\oti\id_{\Hop})=m_\sfc\circ(\id_{\Hop}\oti m_\sfc)
  \ee
which is the condition of associativity.
\\[.3em]
\nxt {Counit property and coassociativity:}\\[2pt]
The proof of the counit property (the third relation in (\ref{eq:asun})) is
analogous to the proof of the unit property, considering instead cuttings 
$\alpha$ such that ${\Xr_p}|_{\alpha}\Cong\Xr_\varepsilon\,{\sqcup}\,\Xr_\Delta$.
\\[2pt]
The proof of coassociativity (the last relation in (\ref{eq:asun})) follows 
closely the proof of associativity, starting instead with the world sheet of a 
\disc\ with one in-going and three out-going open state boundaries, and cutting 
it in two components, each isomorphic to $\Xr_\Delta$.  
\\[.3em]
\nxt {Frobenius property:}\\[2pt]
The Frobenius condition is the first relation in (\ref{eq:ssF}). Denote the 
world sheet of a \disc\ with two in- and two out-going open state boundaries 
by $\Xr_F$. Consider two cuttings $\alpha,\,\beta$
of $\Xr_F$, as indicated in 
  \eqpic{pic_fjfrs2_66} {180} {48} {
  \put(52,0)     {\Includeournicemediumpicture 66{}}
  \put(0,56)     {$ \Xr_F~= $}
  \put(117.5,73) {$ \alpha $}
  \put(80,48)    {$ \beta $}
  }
Note that these cuttings show that ${\Xr_F}|_\delta^{}$ is isomorphic to 
$\Xr_m\,{\sqcup}\,\Xr_\Delta$ for $\delta\eq\alpha,\beta$. Consider realisations
$c_\delta{:}\ \Xr_m\,{\sqcup}\,\Xr_\Delta \To \Xr_F$ of the two cuttings
$\delta\eq\alpha,\beta$. Again, by definition of the correlator we have the 
relation $\cor_{\Xr_F} \eq \bl(c_\delta)\Circ\tftC(\Fr({\Xr_m};m_\sfc)
{\sqcup}\Fr(\Xr_\Delta;\Delta_\sfc))$ for $\delta\eq\alpha,\beta$.
Representing the compositions on the right hand side as gluing of
cobordisms yields the extended cobordisms
  \eqpic{pic_fjfrs2_71} {420} {44} {
  \put(46,-4)    {\Includeourbeautifulpicture 71a}
  \put(46,-4){
     \setlength{\unitlength}{.38pt}\put(-98,-407){
     \put(155,534)   {\sse\begin{turn}{-10}$\Delta_\sfc$\end{turn}}
     \put(277,570)   {\sse\begin{turn}{-10}$m_\sfc$\end{turn}}
     \put(121,482)   {\sse$ \Hop $}
     \put(158,582)   {\sse$ \Hop $}
     \put(215,567)   {\sse$ \Hop $}
     \put(254,499)   {\sse$ \Hop $}
     \put(317,608)   {\sse$ \Hop $}
     }\setlength{\unitlength}{1pt}}
  \put(281,-4)   {\Includeourbeautifulpicture 71b}
  \put(281,-4){
     \setlength{\unitlength}{.38pt}\put(-98,-407){
     \put(238,575)   {\sse\begin{turn}{-10}$\Delta_\sfc$\end{turn}}
     \put(196,533)   {\sse\begin{turn}{-10}$m_\sfc$\end{turn}}
     \put(135,531)   {\sse$ \Hop $}
     \put(235.5,494) {\sse$ \Hop $}
     \put(243,545)   {\sse$ \Hop $}
     \put(222,625)   {\sse$ \Hop $}
     \put(314,609)   {\sse$ \Hop $}
     }\setlength{\unitlength}{1pt}}
  \put(0,44)     {$ \Mr_\sew~=~ $}
  \put(193,44)   {and~~~~ $ \Mr_{\sew'}~=~ $}
  }
Applying $\Psi_F^{-1}$ to each of these cobordisms yields one half of the 
Frobenius property, namely
  \be
  (\id_{\Hop}\oti m_\sfc)\circ(\Delta_\sfc\oti\id_{\Hop})
  = \Delta_\sfc\circ m_\sfc \,.
  \ee
The other half of the Frobenius property in \erf{eq:ssF} can be seen 
analogously, by changing the direction of the cutting $\alpha$ in 
\erf{pic_fjfrs2_66}.
\\[.3em]
\nxt {Symmetry:}\\[2pt]
The symmetry condition is the second relation in (\ref{eq:ssF}). Denote by 
$\Xr_{p'}$ the world sheet for which $\dot\Xr_{p'}$ consists of a \disc\ with 
two in-going open state boundaries. We make use of the isomorphism
$\Psi_{p'}{:}\ \Hom(\Hop\oti\Hop,\one)\To \Htft(\widehat \Xr_{p'})$. 
Choose two cuttings $\alpha$ and $\beta$ according to
  \eqpic{pic_fjfrs2_67} {175} {34} {
  \put(36,0)     {\Includeournicemediumpicture 67{}}
  \put(25,35)    {\small$ in $}
  \put(168,35)   {\small$ in $}
  \put(62,45)    {$ \alpha $}
  \put(62,22)    {$ \beta $}
  }
implying that ${\Xr_{p'}}|_{\alpha,\beta}$ are both isomorphic to 
$\Xr_m\,{\sqcup}\,\Xr_\varepsilon$. The same procedure as
in the previous demonstrations results in the equality
  \be
  \varepsilon_\sfc\circ m_\sfc = d_{\Hop}\circ(\id_{\Hop^\vee}\oti
  (\varepsilon_\sfc\circ m_\sfc)\oti\id_{\Hop})\circ
  (\tilde b_{\Hop}\oti\id_{\Hop}\oti\id_{\Hop}) \,.
  \ee
By composing these morphisms with $\id_{\Hop}\oti b_{\Hop}$ and using the 
duality axiom, the result is precisely the symmetry condition in (\ref{eq:ssF}).
\qed

\medskip

\dtl{Definition}{def:openalg} 
The Frobenius algebra $A_\sfc$ described in proposition \ref{prop:openalg} is 
called the {\em algebra of open states of\/} $\sfc$.

\bigskip

To fix our notation, let us briefly recall the notion of bimodules and bimodule
intertwiners.
\\[-2.2em]

\dtl{Definition}{def:bimod} 
For $A$ an algebra in a tensor category $\Cc$, an $A$-{\em bimodule\/} $B\eq 
(\dot B,\rho_l,\rho_r)$ is a triple consisting of an object $\dot B$ and of two 
morphisms $\rho_l\iN\Hom(A\oti\dot B,\dot B)$ and $\rho_r\iN\Hom(\dot B\oti A,
\dot B)$ such that
  \be
  \begin{array}{ll}
  \rho_l\circ(\id_A\oti\rho_l ) = \rho_l\circ(m\oti\id_{\dot B}) \,, \qquad&
  \rho_l\circ(\eta\oti\id_{\dot B}) = \id_{\dot B} \,,
  \\{}\\[-.8em]
  \rho_r\circ(\rho_r\oti\id_A) = \rho_r\circ(\id_{\dot B}\oti m) \,, &
  \rho_r\circ(\id_{\dot B}\oti \eta) = \id_{\dot B} \,,
  \\{}\\[-.8em]
  \rho_l\circ(\id_A\oti\rho_r) = \rho_r\circ(\rho_l\oti\id_A) \,.
  \end{array}\ee

In other words, an $A$-bimodule is simultaneously a left $A$-module and a right 
$A$-module, with commuting left and right actions of $A$. The category \CAA\ of 
$A$-bimodules has bimodules as objects and intertwiners as
morphisms, i.e.\ the morphism spaces are 
  \be
  \HomAA(B,C) := \{f\iN\Hom(\dot B,\dot C) \,|\, f\Circ\rho^B_l\eq\rho^C_l\Circ
  (\id_A\oti f),~f\Circ\rho^B_r\eq\rho^C_r\Circ(f\oti\id_A)\} \,.
  \ee
An algebra $A$ is called {\em absolutely simple\/} iff $\HomAA(A,A)$ is 
one-dimensional.

\dtl{Proposition}{prop:open-alg-simple} 
Let $\sfc \eq (\Cc,\Hop,\Hcl,B_l,B_r,e,r,\cor)$ be a solution to the
sewing constraints such that $Q_\sfc$ is invertible.
If $\dimc(\Hom_{\Cc \boxtimes \ol\Cc}( \one {\times} \ol\one, \Hcl)) \eq 1$,
then the algebra $A_\sfc$ of open states of $\sfc$ is absolutely simple.

\medskip\noindent
Proof:\\
For the sake of brevity, in this proof we write $A$ for $A_\sfc$. From $A$ one 
obtains a $\Cb$-algebra $A_\text{top}\eq\Hom_\Cc(\one,A)$ by choosing 
$\eta_\text{top}{:}\ 1\,{\mapsto}\, \eta$ as unit and 
$m_\text{top}{:}\ \alpha\oti\beta\,{\mapsto}\,m\Circ(\alpha\oti\beta)$ as 
multiplication, see \cite[sect.\,3.4]{tft1}. The subalgebra
$\text{cent}_{\!A}(A_\text{top}) \,{:=}\, \{\alpha\iN A_\text{top}\,|\,
m\Circ(\alpha\Oti\id_A) \eq m\Circ(\id_A\Oti\alpha)\}$ of $A_\text{top}$ is
called the relative center \cite[definition 3.15]{tft1}; we abbreviate it by
$\text{cent}_{\!A}(A_\text{top})\,{=:}\,C$. It is not difficult to see that the
mapping $\alpha \,{\mapsto}\, m \Circ (\id_A \oti \alpha)$ is an isomorphism
(with inverse $\varphi \,{\mapsto}\, \varphi \Circ \eta$) from $C$ to 
$\HomAA(A,A)$ as vector spaces, so that $A$ is absolutely simple if and only if 
$C$ is one-dimensional.
\\[2pt]
Assume now that $\dimc(C)\,{>}\,1$. Then one has $m_\text{top}(x,y)\eq0$ for 
suitable nonzero elements $x,y\iN C$. This is seen by noting that $C$, being a 
finite-dimensional commutative associative unital algebra over $\Cb$, can be 
written as a sum of its semisimple part and its Jacobson ideal, see
e.g.\ \cite[chapters I.4 and II.5]{AUrs}. If the Jacobson ideal of $C$ is 
non-trivial, it contains at least one nilpotent element $n \iN C$, so that
$m_\text{top}(n',n')\eq0$ for a suitable nonvanishing power $n'$ of $n$.
If, on the other hand, $C$ is semisimple, then it has a basis 
$\{p_i \,|\, i\eq1,...\,,\dimc(C)\}$ consisting of orthogonal idempotents,
and we can choose $x\eq p_1$ and $y\eq p_2$. It is not hard to see that 
$A_\text{top}$ is itself a symmetric Frobenius algebra in $\Vect$ (see lemma 
3.14 of \cite{tft1}), and $\eps \Circ m$ provides a nondegenerate bilinear form 
on $A_\text{top}$; thus there exists a morphism $\psi_1 \iN \Hom(\one,A)$ such 
that $\eps \Circ m \circ (x\oti\psi_1) \,{\ne}\, 0$, or in other words,
$\eps \Circ q_1 \,{\ne}\, 0$ for $q_1 \,{:=}\, m \Circ (x\oti\psi_1)$. Similarly
there is a $q_2 \eq m \Circ (y\oti\psi_2)$ such that $\eps\Circ q_2\,{\ne}\,0$.
\\[2pt]
Consider a world sheet $\Xr$ which is an annulus with one in-going open state 
boundary on either side, 
  \eqpic{eq:open-simple-aux1} {140} {50} {
  \put(45,0)     {\Includeourbeautifulpicture 12{}}
  \put(0,59)     {$ \Xr~= $}
  \put(46,17)    {$ \beta $}
  \put(41,86)    {$ \alpha $}
  \put(42,60)    {\small$ in $}
  \put(130.4,60) {\small$ in $}
  \put(144,100)  {\small identify}
  } 
Also indicated in this picture are two cuttings $\alpha,\beta$
which will be used in the sequel. Construct a cobordism
  \eqpic{pic-fjfrs2_13} {170} {42} {
  \put(69,0)     {\Includeourbeautifulpicture 13{}}
  \put(0,45)     {$ \Mr_{q_1,q_2}~:= $}
  \put(79,47)    {\sse$ q_1^{} $}
  \put(167,47)   {\sse$ q_2^{} $}
  \put(90,53)    {\sse$ A $}
  \put(156,53)   {\sse$ A $}
  } 
by inserting the indicated ribbon graph in the cylinder over $\Xhat$ and 
removing the arc on the out-going boundary. One then finds
  \be
  \tftC(\Mr_{q_1,q_2})\circ \cor_\Xr = 0 \,.
  \labl{eq:open-simple-aux2}
To see this, choose a realisation $c_\alpha{:}\ \Xr|_\alpha \To \Xr$ of the 
cutting $\alpha$. The world sheet $\Xr|_\alpha$ is a \disc\ with four open 
state boundaries, and the correlator can be represented by \erf{eq:Psi_a-def} 
using the multiplication $m$ of $A$ as described in section \ref{sec:frob-to-sew}.
The composition with $\Mr_{q_1,q_2}$ then results in the morphism in the first
line of the chain of equalities
  \bea
  m \circ (m \oti \id_A) \circ (q_1 \oti \id_A \oti q_2)
  \enL\hspace*{4em}
  = m \circ (m\oti\id_A) \circ ((m\Circ (x\oti \psi_1))
  \oti \id_A \oti (m\Circ (y\oti \psi_2)))
  \enL\hspace*{4em}
  = m \circ (m \oti \id_A) \circ (\psi_1 \oti \id_A \oti \psi_2) \circ
  m \circ ( (m \Circ (x\oti y)) \oti \id_A ) = 0 \,.
  \end{array}\ee
Here in the first step the definitions of $q_1$ and $q_2$ are 
inserted. The second step uses associativity of $A$ and the fact that $x,y\iN C$
so that they commute with all of $A$. The last step follows since by 
construction $m \Circ (x\oti y) \eq m_\text{top}(x,y)\eq0$.
\\[2pt]
On the other hand, owing to $\dimc(\Hom_{\Cc \boxtimes \ol\Cc}( \one{\times}\ol
\one, \Hcl)) \eq 1$ we can project to the closed state vacuum on the circle 
indicated by the cutting $\beta$. Let $c_\beta{:}\ \Xr|_\beta \To \Xr$ be a 
realisation of the cutting $\beta$. Choose $\Xr|_\beta$ such that $c_\beta$ is 
of the form $(\sew_\beta,\id)$. Then according to proposition 
\ref{prop:proj-vac} we obtain
  \be
  P^{\text{vac}}_{\sew_\beta,\Xr|_\beta}\circ\cor_\Xr
  = \Lambda_\sfc^{\,-1} E^{\text{vac}}_{\sew_\beta,\Xr|_\beta}\circ
  \cor_{\fill_{\sew_\beta}(\Xr|_\beta)} \,.
  \labl{PbetaC}
Let $\Xr_A$ be the annulus-shaped world sheet that is obtained by omitting the 
two open state boundaries from the world sheet \erf{eq:open-simple-aux1}, so 
that $\Mr_{q_1,q_2}$ is a cobordism from $\Xr$ to $\Xr_A$. It is easy to see 
that $\tftC(\Mr_{q_1,q_2}) \Circ P^{\text{vac}}_{\sew_\beta, \Xr|_\beta} \eq 
P^{\text{vac}}_{\sew_\beta,\Xr_A|_\beta} \Circ\tftC(\Mr_{q_1,q_2})$.
Combining this equality with \erf{eq:open-simple-aux2} and denoting the left and
right hand sides of \erf{PbetaC} by $L$ and $R$, respectively, we obtain
$0 \eq \tftC(\Mr_{q_1,q_2}) \Circ L \eq \tftC(\Mr_{q_1,q_2}) \Circ R$. But the 
world sheet $\fill_{\sew_\beta}(\Xr)$ consists of two \disc s $\Xr_\eps$ with one 
in-going open state boundary each. Their correlators are $\cor_{\Xr_\eps} \eq 
\tftC(\Fr(\Xr_\eps;\eps))$. The co\-bor\-dism for $\tftC(\Mr_{q_1,q_2})\Circ R$ is 
thus
  \eqpic{eq:open-simple-aux3} {235} {44} {
  \put(69,0)     {\Includeourbeautifulpicture 24{}}
  \put(0,49)     {$ \Mr_{q_1,q_2}~:= $}
  \put(83,69)    {\sse$ q_1^{} $}
  \put(125,68)   {\sse$ \eps $}
  \put(125,35)   {\sse$ \eps $}
  \put(163,36)   {\sse$ q_2^{} $}
  \put(103,60)   {\sse$ A $}
  \put(145,41)   {\sse$ A $}
  \put(198,40)   {\small$ \Fr(\Xr_\eps;\eps)$}
  \put(203,80)   {\small$ \Mr_{q_1,q_2}$}
  \put(203,59)   {\small cobordism for $E^{\text{vac}}_{\sew,\Xr|\beta}$}
  }
The two morphisms $\eps \Circ q_1$ and $\eps \Circ q_2$ are nonzero by 
construction, so that \erf{eq:open-simple-aux3} is a nonzero constant times the
invariant of a solid torus with empty ribbon graph, which is nonzero as well,
implying that $\tftC(\Mr_{q_1,q_2}) \Circ R \,{\ne}\, 0$.
\\
Thus assuming that $\dimc(C)$ is larger than 1 leads to a contradiction, and 
hence indeed $\dimc(C) \eq 1$, i.e.\ $A$ is absolutely simple.
\qed

\medskip

In a category that is \koerper-linear, with \koerper\ a field, and sovereign 
(i.e., is monoidal and has left and right dualities which coincide both on 
objects and on morphisms) and for which $\Hom(\one,\one)\eq \koerper\,\id_\one$,
there are two notions of dimension for an object $U$, the left and the right 
dimension $\dim_{l,r}(U) \iN \koerper$. In a ribbon category, these two 
dimensions coincide (see e.g.\ \cite[sect.\,2.1]{ffrs} for more details).
Part (ii) of the following statement will be used when proving the properties of 
$A_\sfc$ below.
\\[-2.2em]

\dtl{Lemma}{lem:simple-special} 
Let $A$ be a symmetric Frobenius algebra in a
sovereign \koerper-linear category with $\Hom(\one,\one) \eq \koerper\,\id_\one$.
\\[2pt]
(i)~\,$\dim_l(A) \eq \dim_r(A)$.
\\[2pt]
(ii) Write  $\dim(A)$ for $\dim_l(A) \eq \dim_r(A)$.
If $A$ is absolutely simple and $\dim(A) \,{\ne}\, 0$, then $A$ is also special.

\medskip\noindent
Proof:\\
(i)~\,Consider the equalities
  \eqpicc{eq:simple-special-aux} {430} {55} {
  \put(95,13)    {\Includeourbeautifulpicture 27a }
  \put(195,13)   {\Includeourbeautifulpicture 27b }
  \put(295,13)   {\Includeourbeautifulpicture 27c }
  \put(398,13)   {\Includeourbeautifulpicture 27d }
  \put(0,65)     {$ \eps\circ m\circ\Delta\circ\eta~= $}
  \put(170,65)   {$ \overset{(1)}= $}
  \put(270,65)   {$ \overset{(2)}= $}
  \put(375,65)   {$ \overset{(3)}= $}
  \put(270,-14)  {$ \overset{(4)}{=}~\dim_l(A)\, \id_{\one} \,, $}
  }
where (1) is symmetry of $A$, (2) is the Frobenius property, (3) uses the unit 
and counit properties, and (4) is the definition of the left dimension. Thus
one has $\dim_l(A)\, \id_\one \eq \eps \Circ m \Circ \Delta \Circ \eta$. 
A version of the calculation \erf{eq:simple-special-aux} in which all pictures 
are left-right-reflected yields analogously
$\dim_r(A)\, \id_\one \eq \eps \Circ m \Circ \Delta \Circ \eta$.
\\[2pt]
(ii)~Since for a Frobenius algebra $A$ one has $m \Circ \Delta \iN \HomAA(A,A)$,
and the latter space is one-dimensional by assumption,
we have $m \Circ \Delta \eq \xi \id_{A}$ for some $\xi \iN \koerper$.
Composing both sides of this equality with $\eps \,{\circ}\cdots{\circ}\,\eta$
gives $\eps \Circ m \Circ \Delta \Circ \eta \eq \xi \, \eps\Circ\eta$.
Thus by (i) we have $\dim(A)\, \id_\one \eq \xi\, \eps\Circ\eta$.
Since $\dim(A) \,{\ne}\, 0$, also $\xi$ and $\eps \Circ \eta$ are nonzero.
Thus $A$ is special. 
\qed

With the help of this result we obtain the following corollary to proposition 
\ref{prop:open-alg-simple}.
\\[-2.2em]

\dtl{Corollary}{cor:A-sssFA} 
Let $\sfc$ be a solution to the sewing constraints with 
$\dimc(\Hom_{\Cc \boxtimes \ol\Cc}(\one{\times}\ol\one, \Hcl)) \eq 1$
such that $Q_\sfc$ is invertible. If $\dim(A_\sfc) \,{\ne}\,0$, then $A_\sfc$ 
is an absolutely simple symmetric special Frobenius algebra.

\medskip\noindent
Proof:\\
By proposition \ref{prop:openalg}, $A_\sfc$ is a symmetric Frobenius algebra, by
proposition \ref{prop:open-alg-simple} it is absolutely simple, and therefore 
by lemma \ref{lem:simple-special} it is also special.  
\qed

\medskip

To apply the construction of section \ref{sec:frob-to-sew} for obtaining a 
solution to the sewing constraints in terms of a symmetric special Frobenius 
algebra, we need to impose the normalisation condition $m \Circ \Delta \eq
\id_A$ on product and coproduct. To account for this condition we introduce 
the following notion.
\\[-2.2em]

\dtl{Definition}{def:norm-open-state} 
If $\sfc$ is a solution to the sewing constraints such that the algebra $A_\sfc$
of open states is special with $m_\sfc \Circ \Delta_\sfc \eq \xi\, \id_{A_\sfc}$,
then an algebra $A$ satisfying
  \be
  A \equiv (A,m,\eta,\Delta,\eps) = (A_\sfc,\gamma m_\sfc,\gamma^{-1} \eta_\sfc,
  \gamma \Delta_\sfc,\gamma^{-1} \eps_\sfc)
  \qquad {\rm with} \quad \gamma^2 \eq \xi^{-1}
  \labl{eq:norm-open-state}
is called a {\em normalised algebra of open states\/}.
\\[3pt] 
Note that there are two normalised algebras $A_\pm$ of open states, which
differ in the choice of sign for $\gamma$. However, the corresponding solutions
$\sfc(\Cc,A_+)$ and  $\sfc(\Cc,A_-)$ to the sewing constraints are equivalent, and
accordingly we will below also speak of `the' normalised algebra of open states.

\medskip

For a normalised algebra of open states one computes that indeed 
$m \Circ \Delta \eq (\gamma m_\sfc)\Circ(\gamma\Delta_\sfc) \eq 
  $\linebreak[0]$
\gamma^2 \xi\, \id_{A} \eq \id_A$.


\subsection{The uniqueness theorem}\label{sec:unique-thm}

We have now gathered all the ingredients needed to formulate the following 
uniqueness result: under natural conditions the algebra of open states of a 
solution to the sewing constraints determines the solution up to equivalence.
In more detail:
\\[-2.3em]

\dtl{Theorem}{thm:unique} 
Let $\Cc$ be a modular tensor category
and let $\sfc\eq (\Cc,\Hop,\Hcl,B_l,B_r,e,r,\cor)$
be a solution to the sewing constraints with the following properties:
  \\[-1.5em]~
\def\leftmargini{2.2em}\begin{itemize}\addtolength\itemsep{-5pt}
\item[(i)]
({\em Uniqueness of closed state vacuum\/})\\[2pt]
$\dim_\Cb \Hom_{\CxCb}(\one {\times} \ol\one, \Hcl) \eq 1$.
\item[(ii)]
({\em Non-degeneracy of \disc\ two-point function\/})\\[2pt]
For $\Xr_p$ the unit \disc\ with one in-going and one out-going open state 
boundary, the correlator $\cor_{\Xr_p}$ corresponds, via the 
distinguished isomorphism $\bl(\Xr_p) \Cong \Hom_\Cc(\Hop,\Hop)$,
to an invertible element in $\Hom_\Cc(\Hop,\Hop)$.
\item[(iii)]
({\em Non-degeneracy of sphere two-point function\/})\\[2pt]
For $\Xr_{Bp}$ the unit sphere with one in-going and one out-going closed 
state boundary, the correlator $\cor_{\Xr_{Bp}}$ corresponds, via the 
distinguished isomorphism $\bl(\Xr_{Bp}) \Cong \Hom_{\CxCb}(\Hcl,\Hcl)$,
to an invertible element in $\Hom_{\CxCb}(\Hcl,\Hcl)$.
\item[(iv)]
({\em Quantum dimensions\/})\\[2pt]
$\Hop$ obeys $\dim(\Hop)\,{\ne}\,0$. Further, let $A$ be the normalised algebra 
of open states for $\sfc$; then for each subobject $U_i\Times\ol{U_j}$ of the 
full center $Z(A)$ (see definition \ref{def:full-centre})
we have $\dim(U_i)\hspace*{.5pt}\dim(U_j) \,{>}\, 0$.
\end{itemize}~\\[-1.5em]
Then $\sfc$ is equivalent to $\sfc(\Cc,A)$, with $A$ the 
normalised algebra of open states of $\sfc$.

\dtl{Remark}{rem:unique}
(i)~\,Condition (iv) in the theorem is a technical requirement needed in the 
proof.  It might be possible that by a different method of proof the above 
theorem could be established without imposing (iv).\\
Also note that (iv) is always fulfilled if all quantum dimensions in $\Cc$ are 
positive. But it is in fact a much weaker condition. Consider for example the 
case of non-unitary Virasoro minimal models, and let $\Cc$ be the relevant
representation category. In this category some simple objects have negative
quantum dimension (given e.g.\ in \cite[sect.\,10.6]{DIms}). Nonetheless theorem 
\ref{thm:unique} can be applied e.g.\ in the case $\Hop\eq\one$, where it 
implies that $\sfc(\Cc,\one)$ is the unique solution to the sewing constraints 
with $\Hop\eq\one$ and obeying (i)\,--\,(iii). Condition (iv) is satisfied
because $Z(A)$ only contains objects of the form $U_i \Times \ol U_i$, and 
$\dim(U_i)^2 \,{>}\, 0$ holds trivially (as quantum dimensions of simple 
objects in a modular tensor category over $\Cb$ are 
nonzero and real, see section 2.1 and Corollary 2.10 in \cite{etno}).
\\[.3em]
(ii)~\,Suppose the modular tensor category $\Cc$ has the property that 
$\theta_i \eq \theta_j$ implies $\dim(U_i)\hspace*{.5pt}\dim(U_j)
      $\linebreak[0]${>}\,0$. 
Then the second part of condition (iv) is automatically satisfied. Indeed,
it follows from theorem 5.1 of \cite{tft1} (see the explanation before lemma
\ref{lem:phicl-epi} below for more details) that $U_i\Times\ol{U_j}$ can be 
a subobject of $Z(A)$ only if $\theta_i\eq\theta_j$.
\\[.3em]
(iii)~%
If any one of the conditions (i), (ii) or (iii) in the theorem is removed, the 
conclusion does not hold any longer. When omitting (i), one can construct a 
solution such that correlators on world sheets that contain a closed state 
boundary and a physical boundary vanish identically on some subobject of $\Hcl$.
$\Hcl$ then has a part that entirely decouples from the boundary and that 
therefore cannot be reconstructed from the algebra of open states. An example 
of this type for $\Cc\eq\Vect$ can be found in \cite[example\,2.19]{lapf2}. 
\\
When dropping either (ii) or (iii), one can choose $\Hop$ or $\Hcl$ `too big' at 
the cost of $\cor(\Xr_p)$ or $\cor(\Xr_{Bp})$ not corresponding to invertible 
morphisms (i.e.\ they are idempotents with non-trivial kernel). For example, 
given a simple symmetric Frobenius algebra $A$ in a category $\Cc$ with 
properties as in theorem \ref{thm:unique}, the solution $\sfc(\Cc,A)$ obeys 
(i)\,--\,(iii). From $\sfc(\Cc,A)$ we can construct, for any
nonzero $U \iN \Obj(\Cc)$, another solution $\sfc' \eq (\Cc,A\,{\oplus}\,U,Z(A),
A\oti\Omeka,\Omeka,e_Z,r_Z,\corP)$ with $\corP$ defined as follows. Let 
$(A,e_A,r_A)$ be the realisation of $A$ as a retract of $A\,{\oplus}\,U$. Then 
we set $\corP_\Xr \,{:=}\, F_\Xr(r_A,e_A,\id,\id) \Circ \corA_\Xr$ with 
$F_\Xr(\cdot)$ as defined in \erf{eq:F_X-def}. One verifies that $\sfc'$ is again
a solution, that it obeys (i) and, since it coincides with $\sfc(\Cc,A)$ in the 
absence of open state boundaries, satisfies (iii) as well. However, $\sfc'$ 
violates (ii), because $\Psi_p^{-1}(\corP_{\Xr_p}) \eq e_A \Circ r_A$, which is 
not invertible. An example that violates (iii) but not (ii) can be constructed 
along similar lines with a little more work.
\\[.3em]
(iv)~The analysis of \cite{lewe3}, and in the case of 2-d TFT the results of
\cite{lapf,kaPe}, show that in order to ensure that a given assignment of 
correlators to the fundamental world sheets results in a solution to the sewing 
constraints, only a finite number of relations needs to be verified. This set 
of relations arises by cutting certain genus-zero and genus-one world sheets in 
different ways into fundamental world sheets. In particular, one needs relations 
from genus-one world sheets (but no relations from genus two or higher). If one 
is interested only in solutions that satisfy the physically meaningful
conditions (i)\,--\,(iii) 
above, then theorem \ref{thm:unique} implies that it is enough to fix a simple 
symmetric Frobenius algebra $A$ as the algebra of open states. Note that the 
data and the relations for $A$ involve only \disc\ correlators at genus zero 
with up to four open state boundaries. The correlators on world sheets of higher
genus and/or with closed state boundaries are then determined by $A$ up to 
equivalence, and they are guaranteed to also solve
the genus one relations, and the relations involving closed states boundaries.
\\[.3em]
(v)~\,In the case of two-dimensional topological field theory, i.e.\ for 
$\Cc\,{\simeq}\,\Vect$, the statement of the theorem becomes trivial. Indeed,
let $(A,C,\iota,\iota^*)$ be a knowledgeable Frobenius algebra over $\Cb$ 
satisfying condition (i), which simply means that $C \Cong \Cb$ (conditions 
(ii) and (iii) are implicit in the definition of the 2-d TFT associated to a 
knowledgeable Frobenius algebra). Then by proposition \ref{prop:open-alg-simple}
$A$ is absolutely simple and hence has trivial center, $Z(A)\eq\Cb\,\eta$. It 
follows that there is a unique choice for $\iota$, and thereby also for 
$\iota^*$. Thus for absolutely
simple $A$ any two knowledgeable Frobenius algebras $(A,C,\iota,\iota^*)$ and
$(A,C',\iota',{\iota'}^*)$ with one-dimensional $C$ and $C'$ are isomorphic.
\\[.3em]
(vi)~A result analogous to theorem \ref{thm:unique} has been obtained for CFTs
on $1{+}1$-dimensional Minkowski space in the framework of local quantum field
theory \cite{lore2}. According to \cite[Propos.\,2.9]{lore2} a local net of
observables on the Minkowski half-space $M_+ \eq \{\, (x,t) \,|\, x\,{\ge}\,0 \,\}$
gives rise to a (non-local) net of observables on the boundary $\{ x{=}0 \}$.
Conversely, given such a net of observables on the boundary, there
is a maximal compatible local net on $M_+$. In fact, there can be more
than one compatible net on $M_+$, but they are all contained in the
maximal one. This non-uniqueness stems from the fact that in this setting there
is no reason to impose modular invariance.
\\
In our context, i.e.\ treating the combinatorial aspects of constructing
a euclidean CFT from a given chiral one, we start from assumptions which are 
much weaker than those used in local quantum field theory.
In particular, the category $\Cc$ is not a $C^*$-category, and it
is not concretely realised in terms of a net of subfactors. Consequently
the methods from operator algebra which are instrumental in \cite{lore2} are 
not applicable. On the other hand, in the $1{+}1$-dimensional setting the 
analogues of the assumptions of theorem \ref{thm:unique} are consequences of 
the common axioms of local quantum field theory.


\sect{Proof of the uniqueness theorem}\label{sec:unique-proof}

Throughout this section we fix a solution 
$\sfc \eq (\Cc,\Hop,\Hcl,B_l,B_r,e_\sfc,r_\sfc,\cor)$ to the sewing constraints
and assume that $\sfc$ obeys the conditions of the uniqueness theorem 
\ref{thm:unique}. Then, due to conditions (i) and (ii) in theorem 
\ref{thm:unique}, proposition \ref{prop:open-alg-simple} and corollary 
\ref{cor:A-sssFA} apply, so that the algebra of open states for $\sfc$ is 
special. It therefore makes sense to consider the normalised algebra of open 
states, as in definition \ref{def:norm-open-state}. In the rest of this section 
we denote the normalised algebra of open states for $\sfc$ by $A$. In 
particular, as objects in $\Cc$ we have $A \eq \Hop$.

\medskip       

Let us first give a brief outline of the proof. We want to establish equivalence
of the given solution $\sfc \,{\equiv}\, (\Cc,\Hop,\Hcl,B_l,B_r,e_\sfc,r_\sfc,
\cor)$ and $\sfc(\Cc,A)$. According to definition \ref{def:sol-equiv} this 
amounts to the construction of isomorphisms $\varphi_\text{op}^A$ between $\Hop$
and $A$ as objects of \C\ and $\varphi_\text{cl}^A$ between $\Hcl$ and $Z(A)$ 
as objects of $\CxCb$, and to showing the equality 
$\cor \eq G^\gamma\Circ\aleph(\varphi_\text{op}^A,\varphi_\text{cl}^A)
\Circ \corA$, with $G^\gamma$ the natural transformation introduced in lemma 
\ref{lem:Euler-nat} and a normalisation factor $\gamma$ as given in definition 
\ref{def:norm-open-state}.
We construct a candidate morphism $\varphi_\text{cl}^A\iN\Hom(\Hcl,Z(A))$ 
in section \ref{sec:phicl-def}, and then show in sections \ref{sec:phicl-mono} 
and \ref{sec:phicl-epi}, respectively, that it is both a monomorphism and an 
epimorphism. Furthermore, it turns out that for $\varphi_\text{op}^A$ we may
simply take the identity $\id_{\Hop}$, so that it remains to show that
$\cor \eq G^\gamma\Circ\aleph(\id_{\Hop},\varphi_\text{cl}^A) \Circ \corA$.
In section \ref{sec:equiv-of-sol} we demonstrate that indeed we have
  \be
  \cor_{\Yr_\mu} = G^\gamma\circ\aleph(\id_{\Hop},\varphi_\text{cl}^A)\circ
  \corA_{\Yr_\mu}
  \ee
for all {\em fundamental\/} world sheets $\Yr_\mu$ as given in table 
\ref{table:fuwosh}. By lemma \ref{lem:compare} the equality then holds 
in fact for {\em all\/} world sheets, thus completing the proof.


\subsection{A morphism $\varphi_\text{cl}^A$ 
            from $\Hcl$ to $Z(A)$}\label{sec:phicl-def}

To define the morphism $\varphi_\text{ cl}^A$, we first need to recall the 
concept of $\alpha$-induced bimodules.

\dtl{Definition}{def:alphaind} 
Let $A$ be an algebra in a braided tensor category $\Cc$ and $U$ an object in 
$\Cc$. The two $A$-bimodules $A\,{\otimes^\pm}\,U\,{\equiv}\,(A\oti U, m\oti
\id_U,\rho^\pm_r)$ are obtained by defining the left $A$-action via the product
$m$ and the right $A$-action via $m$ and the braiding, according to
  \be
  \rho^+_r := (m\oti\id_U)\circ(\id_A\oti c_{U,A}^{}) \qquad\text{and}\qquad
  \rho^-_r := (m\oti\id_U)\circ(\id_A\oti c_{A,U}^{-1}) \,.
  \ee

Two tensor functors $\alpha^\pm_{\!A}{:}\ \Cc\To\CAA$, are obtained by
mapping objects $U\iN\Obj(C)$ to $A\,{\otimes^\pm}\,U$ and morphisms
$f\iN \Hom(U,V)$ to $\id_A\oti f\iN\HomAA(A{\otimes^\pm} U,A{\otimes^\pm} V)$.
These functors have been dubbed {\em $\alpha$-induction\/}, and accordingly
the bimodules $A\,{\otimes^\pm}\,U \eq \alpha^\pm_{\!A}(U)$ are called
$\alpha$-induced bimodules.

The following result will be used below to prove properties of 
$\varphi_\text{ cl}^A$.
\\[-2.2em]

\dtl{Lemma}{lem:leave-proj}
Let $A$ be a symmetric special Frobenius algebra in a modular tensor category 
$\Cc$ and $U,V$ objects of $\Cc$. Then for any morphism
$\Phi \iN \HomAA(A \,{\otimes^+}\, U, A \,{\otimes^-}\, V)$ one has
  \eqpic{eq:leave-proj} {140} {34} {
  \put(0,0)     {\Includeournicemediumpicture 57a}
  \put(100,0)   {\Includeournicemediumpicture 57b}
  \put(60,34)   {$ = $}
  \put(22.6,-7.6){\sse $U$}
  \put(23,80)	{\sse $V$}
  \put(7,80)	{\sse $A$}
  \put(2.3,17.3){\sse $A$}
  \put(15,25)	{\sse $\Phi$}
  \put(115,-7.6){\sse $U$}
  \put(116,80)	{\sse $V$}
  \put(100,80)	{\sse $A$}
  \put(94.3,17.3){\sse $A$}
  \put(108,25)	{\sse $\Phi$}
  }
Proof:\\
Similarly as in the proof of proposition 3.6 of \cite{ffrs} we can write
  \eqpic{58} {370} {100} {
    \put(0,115){
  \put(3,0)     {\Includeournicemediumpicture 58a}
  \put(96,0)    {\Includeournicemediumpicture 58b}
  \put(206,0)   {\Includeournicemediumpicture 58c}
  \put(313,0)   {\Includeournicemediumpicture 58d}
  \put(63,44)   {$ \overset{(1)}= $}
  \put(169,44)  {$ \overset{(2)}= $}
  \put(279,44)  {$ \overset{(3)}= $}
  \put(26,-7)	 {\sse $U$}
  \put(26,92)	 {\sse $V$}
  \put(11,92)	 {\sse $A$}
  \put(33,63)	 {\sse $A$}
  \put(18.5,37.7){\sse $\Phi$}
  \put(119,-7)	 {\sse $U$}
  \put(119,92)	 {\sse $V$}
  \put(104,92)	 {\sse $A$}
  \put(135,63)	 {\sse $A$}
  \put(111.5,37.7){\sse $\Phi$}
  \put(229,-7)	 {\sse $U$}
  \put(229,92)	 {\sse $V$}
  \put(214,92)	 {\sse $A$}
  \put(245,63)	 {\sse $A$}
  \put(222,56.5) {\sse $\Phi$}
  \put(336,-7)	 {\sse $U$}
  \put(336,92)	 {\sse $V$}
  \put(321,92)	 {\sse $A$}
  \put(329,56.5)	 {\sse $\Phi$}
  \put(326,36)	 {\sse $A$}
    }
  \put(104,0)   {\Includeournicemediumpicture 58e}
  \put(220,0)   {\Includeournicemediumpicture 58f}
  \put(320,0)   {\Includeournicemediumpicture 58g}
  \put(63,44)   {$ \overset{(4)}= $}
  \put(169,44)  {$ \overset{(5)}= $}
  \put(279,44)  {$ = $}
  \put(127,-7)	 {\sse $U$}
  \put(127,92)	 {\sse $V$}
  \put(111,92)	 {\sse $A$}
  \put(119,65.5)	{\sse $\Phi$}
  \put(117,36)	 {\sse $A$}
  \put(238,-7)	 {\sse $U$}
  \put(238,92)	 {\sse $V$}
  \put(223,92)	 {\sse $A$}
  \put(230,65.6)	{\sse $\Phi$}
  \put(217,50)	 {\sse $A$}
  \put(335,-7)	 {\sse $U$}
  \put(335,92)	 {\sse $V$}
  \put(319,92)	 {\sse $A$}
  \put(328,65.5)	{\sse $\Phi$}
  \put(314,40)	 {\sse $A$}
  }
The first step follows by using the unit property followed by the Frobenius 
property. In the second step the intertwining property of $\Phi$ is implemented,
and the third step uses again the Frobenius and unit properties. In the fourth 
step the symmetry property is applied, and the fifth step follows by first 
pulling the right $A$-line below the left one to the left and then canceling 
the resulting twist and inverse twist. The final equality holds by specialness.
\qed

We denote by $\Phi_\sfc \iN \Hom(A \oti B_l, A \oti B_r)$ the unique morphism 
such that
  \be
  \cor_{\Xr_{Bb}} = \tftC\big( F(\Xr_{Bb};\Phi_\sfc) \big) \,.
  \labl{eq:PhiS-def}
The following properties of $\Phi_\sfc$ prove to be important:
\\[-2.2em]

\dtl{Lemma}{lem:PhiS-HomAA}
(i)~\,$\Phi_\sfc$ is a bimodule morphism, 
$\,\Phi_\sfc \iN \HomAA(A \,{\otimes^+} B_l, A \,{\otimes^-} B_r)$.
\\[3pt]
(ii) Expanding $p_\sfc \eq e_\sfc \Circ r_\sfc$ as $p_\sfc\eq\sum_\alpha 
p_\alpha' \oti p_\alpha''$ we have $\,\sum_\alpha (\id_A \oti p_\alpha'') 
\circ \Phi_\sfc \circ (\id_A \otimes p_\alpha') = \Phi_\sfc$.

\medskip\noindent
Proof:\\
(i) Consider the world sheet 
  \eqpic{40} {210} {78} {
  \put(55,2)    {\IncludeourniceMediumpicture 40{}}
  \put(0,91)    {$ \Xr~:= $}
  \put(55,30)	{\small$ in $}
  \put(179,30)  {\small$ in $}
  \put(118,80)	{\small$ in $}
  \put(114,172) {\small$ out $}
  \put(120,0)	{$ \beta $}
  \put(149,2)	{$ \gamma $}
  \put(196,113) {$ \alpha $}
  }
i.e.\ a disk with two in-going and one out-going open state boundaries and one 
in-going closed state boundary. In the picture we also indicate three different
cuttings $\alpha$, $\beta$ and $\gamma$. With the help of the cobordism
  \eqpic{eq:PhiS-HomAA-aux1} {180} {40} {
  \put(57,0)     {\Includeourbeautifulpicture 41{}}
  \put(0,48)     {$ \Fr(f)~:= $}
  \put(108,52)   {\sse$ f $}
  \put(70,50)	 {\sse$ A $}
  \put(85,37)	 {\sse$ A $}
  \put(124,65)	 {\sse$ A $}
  \put(110,85)	 {\sse$ B_l $}
  \put(137.6,40) {\sse$ B_r $}
  }
we obtain an isomorphism $f \,{\mapsto}\, \tftC(\Fr(f))$ from 
$H \,{:=}\, \Hom(A \oti A \oti B_l, A \oti B_r)$ to $\Htft(\Xhat)$. Let $c \iN H$ 
be the unique morphism such that $\tftC(\Fr(c))\eq\cor_\Xr$.
\\[2pt]
For $\delta \iN \{\alpha, \beta, \gamma\}$ every realisation $X|_\delta$ of the 
cutting $\delta$ is isomorphic to $\Xr_m \,{\sqcup}\, \Xr_{Bb}$. Denote by 
$q_\delta{:}\, \Xr_m\,{\sqcup}\,\Xr_{Bb} \To \Xr$ a choice of realisation. Then
  \be
  \cor_\Xr = \bl(q_\delta) \circ \big( \cor_{\Xr_m} \oti \cor_{\Xr_{Bb}} \big)
  = \bl(q_\delta) \circ
  \tftC\big(\Fr(\Xr_m;m) \,{\sqcup}\, \Fr(\Xr_{Bb};\Phi_\sfc) \big)
  \ee 
Expressing also $\bl(q_\delta)$ on the right hand side of this equality as a 
cobordism and comparing with \erf{eq:PhiS-HomAA-aux1} yields three different
expressions for $c$:
  \be
  \begin{array}{ll} \displaystyle
  \text{from }q_\alpha :& c = \Phi_\sfc \circ (m \oti \id_{B_l}) \,,
  \enL 
  \text{from }q_\beta  :& c = (m \oti \id_{B_r}) \circ
  (\id_A \oti c_{A,B_r}^{-1}) \circ (\Phi_\sfc \oti \id_A)
  \circ (\id_A \oti c_{B_l,A}^{-1}) \,,
  \enL
  \text{from }q_\gamma  :& c = (m\oti\id_{B_r}) \circ (\id_A\oti\Phi_\sfc) \,.
  \end{array}
  \ee 
It is not difficult to see that equality of these three expressions is
equivalent to assertion (i).
\\[3pt]
(ii) By definition we have $\cor_{\Xr_{Bb}} \iN \bl(\Xr_{Bb})$, so that 
$P_{\Xr_{Bb}} \Circ \cor_{\Xr_{Bb}} \eq \cor_{\Xr_{Bb}}$. The statement then 
follows by substituting the explicit form of $P_{\Xr_{Bb}}$ in terms of the
cobordism \erf{eq:PX-def}. 
\qed

The following construction will be useful when working with morphisms in 
$\CxCb$. Let $V,V'$ be objects of $\Cc$ and $U_k$ a simple object of $\Cc$. 
Choose bases $e_\alpha \iN \Hom(U_k,V)$ and $r_\alpha \iN \Hom(V,U_k)$ such 
that $(U_k,e_\alpha,r_\alpha)$ is a retract of $V$ and $r_\alpha \circ e_\beta
\eq \delta_{\alpha,\beta}\,\id_{U_k}$. Similarly choose $e_\alpha' \iN \Hom
(U_k,V')$ and $r_\alpha' \iN \Hom(V',U_k)$ to be bases of retracts.
\\[-2em]

\dtl{Lemma}{lem:shift-hom} 
For every $f \iN \Hom(V,V')$ we have the identity 
  \be
  \sum_{\alpha} (r_\alpha'\Times \ol{e_\alpha'}) \circ (f \Times \id_{\ol{V'}})
  = \sum_{\beta} (r_\beta \Times \ol{e_\beta}) \circ (\id_V \Times \ol f) \,.
  \labl{eq:shift-hom}
for morphisms in $\CxCb$.

\medskip\noindent
Proof:\\
Since $U_k$ is simple, there are constants $\lambda(f)_{\delta\gamma} \iN \Cb$ 
such that
  \be
  r_\delta' \Circ f \Circ e_\gamma = \lambda(f)_{\delta\gamma} \, \id_{U_k} \,.
  \ee 
Composing the left hand side of \erf{eq:shift-hom} from the right with 
$e_\gamma \Times \ol{r_\delta'}$ one finds
  \be
  (r_\delta' \Circ f \Circ e_\gamma) \times \id_{\ol{U_k}}
  = \lambda(f)_{\delta\gamma} \, \id_{U_k} \Times \id_{\ol{U_k}} \,,
  \ee 
while the same manipulation of the right hand side results in
  \be
  \id_{U_k} \times (\ol{e_\gamma} \Circ \ol f \Circ \ol{r_\delta'})
  = \id_{U_k} \times (\ol{r_\delta' \Circ f \Circ e_\gamma})
  = \lambda(f)_{\delta\gamma} \, \id_{U_k} \Times \id_{\ol{U_k}} \,,
  \ee 
where the second equality uses the definition of composition in $\ol\Cc$. Thus 
the left and right sides of formula \erf{eq:shift-hom} are equal when composed 
with $e_\gamma \Times \ol{r_\delta'}$, for any choice of $\gamma$ and $\delta$. 
Since the latter morphisms form a basis of $\Hom_{\Cc\boxtimes\ol\Cc}(U_k\Times
\ol{U_k}, V \Times \ol{V'})$, we have indeed equality already in the form 
\erf{eq:shift-hom}. 
\qed

Let $e_{i\alpha}$ and $r_{i\alpha}$, for $\alpha\eq 1,...\,,\dimc(\Hom(U_i,B_r)
)$, be embedding and restriction morphisms for the various ways to realise 
$U_i$ as a retract of $B_r$. To define $\varphi_\text{cl}^A
\iN \Hom_{\Cc\boxtimes\ol\Cc}(\Hcl,Z(A))$, the essential ingredient is the 
morphism $\Phi_\sfc$ which allows one to replace $B_l$ by $B_r$; we set
  \eqpic{eq:phi_cl-def} {140} {113} {
    \put(0,4){
  \put(87,0)		{\Includeournicesmallpicture 26{}}
  \put(-9,116)          {$\varphi_\text{cl}^A~:=~\gamma^2\dsty\sum_{i,\alpha}$}
  \put(121,-9)		{\sse $\Hcl$}
  \put(123.5,21.7)	{\sse $e_\sfc^{}$}
  \put(129,35)		{\sse $\BlxBr$}
  \put(122.3,51.2)	{\sse $\id$}
  \put(73,65)		{\sse $A{\times}\overline{\one}$}
  \put(110,70)		{\sse $B_l{\times}\overline{\one}$}
  \put(141,83)		{\sse $\one{\times}\overline{B_r}$}
  \put(93.2,83.5)	{\small $\Phi_\sfc$}
  \put(108,97)		{\sse $B_r{\times}\overline{\one}$}
  \put(105,115.5)       {\sse $r_{i\alpha}$}
  \put(137,114.8)       {\sse $\bar{e}_{i\alpha}$}
  \put(113,125)		{\sse $U_i{\times}\overline{\one}$}
  \put(138.5,133)	{\sse $\one{\times}\overline{U_i}$}
  \put(122.3,145)	{\sse $\id$}
  \put(129,160)		{\sse $U_i{\times}\overline{U_i}$}
  \put(78,145)		{\sse $A{\times}\overline{\one}$}
  \put(123.2,177)	{\sse $\tilde{e}_i$}
  \put(127,190)		{\sse $T_\Cc$}
  \put(111.5,210.1)	{\sse $r_{\!C}^{}$}
  \put(106,234)		{\sse $Z(A)$}
    }
  } 
where $\gamma^2$ is the normalisation constant from definition 
\ref{def:norm-open-state} and $r_C$ is the restriction morphism in the 
realisation of $Z(A)$ as a retract of $(A\Times\ol\one)\oti T_\Cc$, see 
definitions \ref{def:left-centre} and \ref{def:full-centre}. To define the 
natural isomorphism \erf{eq:alphaX-def} we also need the corresponding 
morphism in $\Hom(\BlxBr,(A\oti\Omeka)\Times\ol\Omeka)$. Let us abbreviate
  \be
  \phi_\text{cl}^A := e_Z \circ \varphi_\text{cl}^A \circ r_\sfc \,,
  \labl{eq:phicl-phidef} 
with $e_Z$ as introduced in \erf{eq:eZ-rZ-def}. Recalling that 
$\Hom(\BlxBr, (A\oti\Omeka)\Times\ol\Omeka) \eq \Hom(B_l,A\oti\Omeka) 
  $\linebreak[0]$
\otic \Hom(\Omeka,B_r)$, we have 
\\[-2.3em]

\dtl{Lemma}{lem:phiS-tensor} 
We have $\phi_\text{cl}^A \eq \gamma^2 \sum_{i,\alpha}\phi_{i\alpha}' \oti
\phi_{i\alpha}''$ with $\phi_{i\alpha}' \eq (\id_A \oti (e_i \Circ r_{i\alpha}))
\Circ \Phi_\sfc \Circ (\eta \oti \id_{B_l})$ and $\phi_{i\alpha}'' \eq 
e_{i\alpha} \Circ r_i$, where $e_i, r_i$ realise $U_i$ as a retract of $\Omeka$.

\medskip\noindent
Proof:\\
First recall that we denote the morphisms realising $U_i\Times\ol{U_i}$ as a
retract of $T_\Cc$ by $\tilde e_i$ and $\tilde r_i$, and note that the morphism 
$e_Z \Circ r_C \Circ( \id_{A\times\one} \oti \tilde e_i)$ can be rewritten as
  \eqpic{eq:phiS-tensor} {170} {45} {
  \put(140,-2)   {\IncludeourniceMediumpicture 35b}
  \put(0,47)     {$ e_Z \circ r_C \circ(\id_{A\times\one}\oti\tilde e_i) ~= $}
  \put(135,-10)	  {\sse $(A\!\oti\! U_i)\!\!\times\!\!\ol\one$}
  \put(178,-10)	  {\sse $\one\!\!\times\!\!\ol U_i$}
  \put(135,108)  {\sse $(A\!\oti\! K)\!\!\times\!\!\ol\one$}
  \put(178,108)  {\sse $\one\!\!\times\!\!\ol K$}
  \put(158,85)	  {\sse $e_i$}
  \put(182,84)	  {\sse $\ol r_i$}
  \put(170,35)	  {\sse $A$}
  }
where first the explicit form \erf{eq:eZ-rZ-def} of $e_Z$ is inserted, and then
$e_C \Circ r_C$ is replaced by the projector $P^l_A$. One then uses that
$\tilde r_i \Circ \tilde e_i \eq \id_{U_i \times \ol{U_i}}$ and that objects
of the form $V\Times \one$ are transparent to objects of the form $\one \Times 
\ol W$. With the help of \erf{eq:phiS-tensor} we obtain
  \eqpic{pic_fjfrs2_36} {410} {55} {
      \put(0,5){
  \put(86,-5)   {\IncludeourniceMediumpicture 36a}
  \put(231,-5)  {\IncludeourniceMediumpicture 36b}
  \put(0,52)    {$ \phi_\text{cl}^A ~=~ \gamma^2 \dsty\sum_{i,\alpha,\beta} $}
  \put(163,52)  {$ =~ \gamma^2\dsty\sum_{i,\alpha,\beta} $}
  \put(297,52)  {$ =~ \gamma^2\dsty\sum_{i,\alpha}\phi_{i\alpha}'
                   \oti\phi_{i\alpha}''~.$}
  \put(95,-13)   {\sse $B_l{\times}\overline{\one}$}
  \put(122,-13)  {\sse $\one{\times}\overline{B_r}$}
  \put(77,115)   {\sse $(A\!\oti\! K){\times}\overline{K}$}
  \put(122,115)  {\sse $\one{\times}\overline{K}$}
  \put(107,20)   {\sse $B_l$}
  \put(107,38)   {\sse $B_r$}
  \put(109,60)   {\sse $U_i$}
  \put(132,28)   {\sse $\overline{B_r}$}
  \put(132,70)   {\sse $\overline{U_i}$}
  \put(104,9)    {\tiny $p_\alpha'$}
  \put(126,9)    {\tiny $\overline{p}_\alpha''$}
  \put(95,28.3)  {\sse $\Phi_\sfc$}
  \put(103,50.5) {\tiny $r_{i\beta}$}
  \put(126,50.5) {\tiny $\overline{e}_{i\beta}$}
  \put(105,94)   {\tiny $e_i$}
  \put(128,93.5) {\tiny $\overline{r}_i$}
  \put(233,-13)  {\sse $B_l{\times}\overline{\one}$}
  \put(260,-13)  {\sse $\one{\times}\overline{B_r}$}
  \put(215,115)  {\sse $(A\!\oti\! K){\times}\overline{K}$}
  \put(260,115)  {\sse $\one{\times}\overline{K}$}
  \put(245,20)   {\sse $B_l$}
  \put(245,38)   {\sse $B_r$}
  \put(247,60)   {\sse $B_r$}
  \put(247,80)   {\sse $U_i$}
  \put(270,28)   {\sse $\overline{B_r}$}
  \put(270,80)   {\sse $\overline{U_i}$}
  \put(242,9)    {\tiny $p_\alpha'$}
  \put(233,28.3) {\sse $\Phi_\sfc$}
  \put(242,50.5) {\tiny $p_\alpha''$}
  \put(241,72)   {\tiny $r_{i\beta}$}
  \put(265,72)   {\tiny $\overline{e}_{i\beta}$}
  \put(243,94)   {\tiny $e_i$}
  \put(267,93.5) {\tiny $\overline{r}_i$}
  } }
Here in the first equality \erf{eq:phiS-tensor} is substituted, in the second 
equality lemma \ref{lem:shift-hom} is applied for the case $f \eq p_\alpha'' \iN
\Hom(B_r,B_r)$, and lemma \ref{lem:leave-proj} (which applies because
by lemma \ref{lem:PhiS-HomAA}\,(i) $\Phi_\sfc$ is an intertwiner of bimodules)
is used to omit the $A$-loop.
The final equality amounts to lemma \ref{lem:PhiS-HomAA}\,(ii). 
\qed


\subsection{$\varphi_\text{cl}^A$ is a monomorphism}\label{sec:phicl-mono}

Denote by $D$ the world sheet such that $\widetilde D/\langle \iota \rangle$ is
the unit \disc. Cutting
$D$ as $q{:}\ \Xr_\eta \,{\sqcup}\, \Xr_\eps \To D$ shows that
  \eqpic{eq:CorD-B3} {430} {26} {
  \put(212,2)    {\Includeourbeautifulpicture 28{}}
  \put(-10,34)   {$ \cor_D\,=\,\bl(q)\circ\big(\cor_{\Xr_\eta}\oti\cor_{\Xr_\eps}
                    \big)~=~\tftC${\Huge(} }
  \put(288,34)   { {\Huge)}$~=~\gamma^2\dim(A)\;\tftC(B^3)\,. $}
  \put(255,39)   {\sse $A$}
  \put(244.8,22) {\sse $\gamma{\cdot}\eta$}
  \put(245.5,57) {\sse $\gamma{\cdot}\varepsilon$}
  } 

Since $A$ is special it follows in particular that $\cor_D \,{\ne}\, 0$. Next 
consider the cylindrical word sheet $\Xr_{Bp}$ from figure \ref{fig:fund-world}
and define $\tilde p_{\sfc}$ to be the unique element of $\Hom_{\Cc\boxtimes
\ol\Cc} (\BlxBr,\BlxBr)$ such that upon expanding $\tilde p_{\sfc}\eq\sum_\alpha
\tilde p_{\sfc,\alpha}' \oti \tilde p_{\sfc,\alpha}''$ we have
  \be
  \cor_{\Xr_{Bp}} = \sum_\alpha \tftC(F(\Xr_{Bp};
  \tilde p_{\sfc,\alpha}',\tilde p_{\sfc,\alpha}'') \,,
  \labl{eq:cor-X_Bp-p_BS} 
where $F(\Xr_{Bp};\,\cdot\,,\,\cdot\,)$ is the corresponding cobordism
from figure \ref{fig:fund-cobord}.
\\[-2em]

\dtl{Lemma}{lem:pBS=p} 
The endomorphisms $\tilde p_{\sfc}$ and $p_{\sfc} \eq e_\sfc \Circ r_\sfc$ are 
equal, $\,\tilde p_{\sfc} \eq p_{\sfc}$.

\medskip\noindent
Proof:\\
By the non-degeneracy of the sphere two-point function (i.e.\ property (iii) 
in theorem \ref{thm:unique}), $\tilde p_{\sfc}$ is invertible on the image of 
the idempotent $p_{\sfc}$. By definition, $\text{Im}(p_{\sfc})\eq\Hcl$, so that 
$r_\sfc \Circ \tilde p_{\sfc} \Circ e_\sfc$ is an invertible element of
$\Hom_{\Cc\boxtimes\ol\Cc}(\Hcl,\Hcl)$. Analogously as in formula \erf{CorXp},
by cutting the world sheet $\Xr_{Bp}$ along its circumference via a 
morphism $q {:}\ \Xr_{Bp} \,{\sqcup}\, \Xr_{Bp} \To \Xr_{Bp}$ one obtains the 
identity $\cor_{\Xr_{Bp}} \eq \bl(q) \Circ \big( \cor_{\Xr_{Bp}} \oti
\cor_{\Xr_{Bp}} \big)$. Together with \erf{eq:cor-X_Bp-p_BS} it follows that
  \be
  \tilde p_{\sfc}
  \equiv \sum_\alpha \tilde p_{\sfc,\alpha}' \oti \tilde p_{\sfc,\alpha}''
  = \sum_{\alpha,\beta} (\tilde p_{\sfc,\alpha}' \Circ \tilde p_{\sfc,\beta}')
  \oti (\tilde p_{\sfc,\beta}'' \Circ \tilde p_{\sfc,\alpha}'') 
  = \tilde p_{\sfc} \Circ \tilde p_{\sfc} \,,
  \ee
i.e.\ $\tilde p_{\sfc}$ is an idempotent. Furthermore, since the right hand side
of \erf{eq:cor-X_Bp-p_BS} is in $\bl(\Xr_{Bp})$, by definition of $\bl$ it 
follows that $p_{\sfc}\Circ\tilde p_{\sfc}\Circ p_{\sfc} \eq \tilde p_{\sfc}$.
Hence we have, using also $e_\sfc \eq p_{\sfc} \Circ e_\sfc$,
  \be
  (r_\sfc\Circ \tilde p_{\sfc}\Circ e_\sfc) \circ (r_\sfc\Circ \tilde p_{\sfc}
  \Circ e_\sfc)
  = r_\sfc \Circ \tilde p_{\sfc} \Circ p_{\sfc} \Circ \tilde p_{\sfc} \Circ e_\sfc
  = r_\sfc \Circ \tilde p_{\sfc} \Circ e_\sfc \,.
  \ee 
Since $r_\sfc \Circ \tilde p_{\sfc} \Circ e_\sfc$ is invertible, it follows that
in fact $r_\sfc \Circ \tilde p_{\sfc} \Circ e_\sfc \eq \id_{\Hcl}$. Finally,
composing with $e_\sfc$ and $r_\sfc$ yields $\tilde p_{\sfc} \eq p_{\sfc}$. 
\qed

Next we analyse the properties of the correlator on the world sheet 
  \eqpic{eq:phi-mono-aux1} {180} {44} {
  \put(45,0)    {\Includeourbeautifulpicture 29{}}
  \put(0,50)    {$ \Yr~= $}
  \put(170,89)  {$\alpha$}
  \put(160,12)  {$\beta$}
  \put( 74,44)  {\footnotesize $in$}
  \put(112,44)  {\footnotesize $out$}
  }
i.e.\ on a \disc\ with an additional in-going and an additional out-going 
closed state boundary. The dashed lines in the picture indicate two cuttings 
$\alpha$ and $\beta$ of $\Yr$. Consider the cobordisms
  \eqpic{pic_fjfrs2_30} {420} {58} {
  \put(33,0)    {\Includeourbeautifulpicture 30a}
  \put(288,0)   {\Includeourbeautifulpicture 30b}
  \put(-20,60)  {$ \tilde{\Mr}^\text{vac}~:= $}
  \put(186,60)  {and~~~$\mathrm N(\psi,\psi')~:= $}
  \put(77,98)   {\sse $U_i$}
  \put(107,98)  {\sse $B_l$}
  \put(80,30)   {\sse $U_j$}
  \put(107,30)  {\sse $B_r$}
  \put(95,92)   {\sse $K$}
  \put(95,15)   {\sse $K$}
  \put(56,92)   {\sse \begin{turn}{-10}$w_{\!K}^{}$\end{turn}}
  \put(55,25)   {\sse \begin{turn}{-10}$w_{\!K}^{}$\end{turn}}
  \put(375,93)  {\sse $B_l$}
  \put(377,30)  {\sse $B_r$}
  \put(326,81)  {\sse $B_l$}
  \put(325,43)  {\sse $B_r$}
  \put(327,106) {\sse $U_i$}
  \put(327,16)  {\sse $U_j$}
  \put(326,101) {\sse \begin{turn}{-135}$\psi$\end{turn}}
  \put(326,32)  {\sse \begin{turn}{135}$\psi'$\end{turn}}
  }
and set $\tilde P^\text{vac} \,{:=}\, \tftC(\tilde \Mr^\text{vac})$.
For $\varphi \eq \sum_\alpha \varphi_\alpha' \oti \varphi_\alpha''$
with $\varphi \iN \Hom_{\Cc\boxtimes\ol\Cc}( U_i\Times\ol{U_j},\BlxBr)$
define further
  \be
  R_\varphi := \sum_\alpha \tftC(\mathrm N(\varphi_\alpha',\varphi_\alpha''))\,.
  \ee
Note that the cobordism defining $\tilde P^\text{vac}$ differs from the one
defining $P^\text{vac}_{\sew_\alpha,\Yr|_\alpha}$ (see \erf{pic-fjfrs2_16})
only in the labeling of ribbons. Consider now the cobordism for (each term of 
the sum in) the composition $\tilde P^\text{vac} \Circ R_\varphi$. By moving 
the coupons labeled $\varphi_\alpha'$ and $\varphi_\alpha''$ through the 
annular $K$-rib\-bons one verifies the equality $\tilde P^\text{vac} \Circ 
R_\varphi \eq  R_ \varphi \Circ P^\text{vac}_{\sew_\alpha,\Yr|_\alpha}$.
Suppose now further that $p_{\sfc} \Circ \varphi \eq \varphi$. Then
together with \erf{eq:proj-vac} and \erf{eq:CorD-B3} it follows that
  \eqpicc{pic_fjfrs2_31} {420} {66} {
  \put(88,-6)   {\Includeourbeautifulpicture 31a}
  \put(327,-6)  {\Includeourbeautifulpicture 31b}
  \put(-20,140) {$ \Lambda_{\sfc}\, \tilde P^\text{vac} \circ R_\varphi \circ
                   \cor_\Yr ~=~ R_\varphi  \circ E^\text{vac}_{\sew_\alpha,
                   \Yr|_\alpha} \circ \cor_{\!D \sqcup \Xr_{Bp}} $}
  \put(-10,56)  {$ =~\gamma^{2}\dim(A) \dsty\sum_{\alpha,\beta} $}
  \put(231,56)  {$ =~\gamma^{2}\dim(A) \dsty\sum_{\alpha} $}
  \put(175.6,89){\sse$ B_l $}
  \put(177,23)  {\sse$ B_r $}
  \put(147,79.5){\sse \begin{turn}{-90}$ p_\beta' $\end{turn}}
  \put(147,45)  {\sse \begin{turn}{-90}$ p_\beta'' $\end{turn}}
  \put(123,96)  {\sse \begin{turn}{-135}$ \varphi_\alpha' $\end{turn}}
  \put(123,24)  {\sse \begin{turn}{135}$ \varphi_\alpha'' $\end{turn}}
  \put(363,96)  {\sse \begin{turn}{-135}$ \varphi_\alpha' $\end{turn}}
  \put(363,24)  {\sse \begin{turn}{135}$ \varphi_\alpha'' $\end{turn}}
  \put(126,104.5){\sse$ U_i $}
  \put(126,9)   {\sse$ U_j $}
  \put(365,104.5){\sse$ U_i $}
  \put(364.6,9) {\sse$ U_j $}
  \put(415,89)  {\sse$ B_l $}
  \put(417,21)  {\sse$ B_r $}
  }
where it is understood that $\tftC$ is applied to each cobordism.
The last expression in this chain of equalities is zero iff $\varphi\eq0$. We 
conclude that $R_\varphi \Circ \cor_\Yr \eq 0$ implies $\varphi \eq 0$.

Next consider the cutting $\beta$ in \erf{eq:phi-mono-aux1}. We find
  \eqpic{eq:phi-mono-aux2} {395} {48} {
  \put(253,0)           {\Includeourbeautifulpicture 32{}}
  \put(-32,50)          {$ R_\varphi\circ\cor_\Yr~=~
                           R_\varphi\circ\bl\big((\sew_\beta,\id)\big)\circ
                           \cor_{\Yr|_\beta}~=~R_\varphi\circ\tftC${\Huge(} }
  \put(398,50)          {\Huge)}
  \put(294,67)          {\sse $B_l$}
  \put(311,15)          {\sse $B_r$}
  \put(346,34)          {\sse $B_r$}
  \put(348,86)          {\sse $B_l$}
  \put(315,58)          {\sse $A$}
  \put(343,65)          {\sse $A$}
  \put(273,47)          {\sse \begin{turn}{-50}$\gamma\,\eta$\end{turn}}
  \put(301.7,50)        {\tiny \begin{turn}{-50}$\Phi_\sfc$\end{turn}}
  \put(336,57)          {\tiny \begin{turn}{-50}$\tilde{\Phi}_\sfc $\end{turn}}
  \put(366.1,68)        {\sse \begin{turn}{-50}$\gamma\,\varepsilon$\end{turn}}
  } 
Here the morphism $\tilde\Phi_\sfc$ on the right hand side is analogous to
$\Phi_\sfc$, but with an out-going closed state boundary instead of an in-going
one. Combining \erf{eq:phi-mono-aux2} with the information that $R_\varphi\Circ
\cor_\Yr \eq 0$ implies $\varphi \eq 0$ we obtain
  \eqpic{eq:phi-mono-aux3} {270} {44} {
  \put(-20,0)		{\Includeournicesmallpicture 33{}}
  \put(-48,46)   	{$\dsty\sum_{\alpha}$}
  \put(-7,-7)		{\sse $U_i$}
  \put(-8.8,19.5)	{\sse $\varphi_\alpha'$}
  \put(-4,31)		{\sse $B_l$}
  \put(-15,43.5)	{\sse $\Phi_\sfc$}
  \put(-4,56)		{\sse $B_r$}
  \put(-8.8,69.5)	{\sse $\varphi_\alpha''$}
  \put(-20,95)		{\sse $A$}
  \put(-6,95)		{\sse $U_j$}
  \put(10,46)		{$~=~0 $~~for all~$i,j\iN\Ic ~~~\Longrightarrow~~~
                          R_\varphi\circ \cor_\Yr \,=\, 0~~~\Longrightarrow~~~
                          \varphi \,=\, 0 ~.$}
  } 
We have now gathered all ingredients needed to prove the following
\\[-2.3em]

\dtl{Lemma}{lem:phicl-mono} 
The morphism $\varphi_\text{cl}^A$ defined in \erf{eq:phi_cl-def} is a 
monomorphism.

\medskip\noindent
Proof:\\
We will show that $e_Z \Circ \varphi_\text{cl}^A \Circ \psi \eq 0$ implies 
$\psi\eq0$ for any $\psi \iN \Hom_{\Cc\boxtimes\ol\Cc}(U_i\Times\ol{U_j},\Hcl)$
and $i,j\iN\Ic$; this implies that $\varphi_\text{cl}^A$ is a monomorphism.
\\
Decompose $\varphi \,{:=}\, e_\sfc \Circ \psi \iN \Hom_{\Cc\boxtimes\ol\Cc}
( U_i\Times\ol{U_j},\BlxBr)$ as $\varphi \eq \sum_\alpha \varphi_\alpha' \oti 
\varphi_\alpha''$. Using lemma \ref{lem:phiS-tensor} to rewrite the combination
$\phi_\text{cl}^A \eq e_Z \Circ \varphi_\text{cl}^A \Circ r_\sfc$ appearing in
$e_Z \Circ \varphi_\text{cl}^A \Circ \psi\eq e_Z \Circ \varphi_\text{cl}^A \Circ 
r_\sfc \Circ e_\sfc \Circ \psi$ gives
  \eqpic{eq:pic-fjfrs2_34} {240} {64} {
  \put(70,0)		{\Includeournicesmallpicture 34a{}}
  \put(220,0)		{\Includeournicesmallpicture 34b{}}
  \put(-80,68)		{$e_Z \circ \varphi_\text{cl}^A \circ \psi ~=~
                          \dsty\gamma^2\, \sum_\alpha  \sum_{k,\beta}$}
  \put(158,68)		{$\dsty=~\gamma^2\, \sum_\alpha$}
  \put(79,-6)		{\sse $U_i{\times}\overline{\one}$}
  \put(113,-6)		{\sse $\one{\times}\overline{U_j}$}
  \put(85,19)		{\sse $\varphi_\alpha'$}
  \put(116.5,19)	{\sse $\overline{\varphi}_\alpha''$}
  \put(90,33)		{\sse $B_l{\times}\overline{\one}$}
  \put(74,46)		{\sse $\Phi_\sfc$}
  \put(122,45)		{\sse $\one{\times}\overline{B_r}$}
  \put(90,61)		{\sse $B_r{\times}\overline{\one}$}
  \put(84.5,75.5)	{\sse $r_{k\beta}$}
  \put(116,75.5)	{\sse $\overline{e}_{k\beta}$}
  \put(92,88)		{\sse $U_k{\times}\overline{\one}$}
  \put(122,88)		{\sse $\one{\times}\overline{U_k}$}
  \put(86,104)		{\sse $e_k$}
  \put(117.5,103.5)	{\sse $\overline{r}_k$}
  \put(63,130)		{\sse $A{\times}\overline{\one}$}
  \put(81,130)		{\sse $\Omeka{\times}\overline{\one}$}
  \put(113,130)		{\sse $\one{\times}\overline{\Omeka}$}
    \put(20,0){
  \put(209,-6)		{\sse $U_i{\times}\overline{\one}$}
  \put(243,-6)		{\sse $\one{\times}\overline{U_j}$}
  \put(215,19)		{\sse $\varphi_\alpha'$}
  \put(220,33)		{\sse $B_l{\times}\overline{\one}$}
  \put(204,46)		{\sse $\Phi_\sfc$}
  \put(220,61)		{\sse $B_r{\times}\overline{\one}$}
  \put(214.5,75.5)	{\sse $\varphi_\alpha''$}
  \put(222,88)		{\sse $U_j{\times}\overline{\one}$}
  \put(216.5,105)	{\sse $e_j$}
  \put(248,104)		{\sse $\overline{r}_j$}
  \put(193,130)		{\sse $A{\times}\overline{\one}$}
  \put(211,130)		{\sse $\Omeka{\times}\overline{\one}$}
  \put(243,130)		{\sse $\one{\times}\overline{\Omeka}$}
    }
  }
where the second equality holds owing to lemma \ref{lem:shift-hom}, applied to 
$f \eq \varphi_\alpha'' \iN \Hom(B_r,U_j)$. Since $\id_{A \times \one} \oti 
( e_j \Times \ol{r_j})$ is a monomorphism, $e_Z \Circ \varphi_\text{cl}^A
\Circ \psi \eq 0$ implies that the morphism displayed
on the left hand side of \erf{eq:phi-mono-aux3} is zero. This, in turn, again by
\erf{eq:phi-mono-aux3}, implies that $\varphi \eq e_\sfc \circ \psi \eq 0$. 
Finally, since $e_\sfc$ is a monomorphism as well, we arrive at $\psi\eq0$. 
\qed


\subsection{$\varphi_\text{cl}^A$ is an epimorphism}\label{sec:phicl-epi}

The following assertion is the algebraic analogue of the statement that the 
torus partition function of a rational CFT is a modular invariant combination 
of characters. We denote by $\hM$ the $|\Ic|{\times}|\Ic|$-matrix with entries 
$\hM_{ij} \iN \Zb_{\ge 0}$ defined by the decomposition 
$\Hcl \Cong \bigoplus_{i,j} (U_i\Times\ol{U_j})^{\oplus \hM_{ij}}$. Then we have
\\[-2.3em]

\dtl{Lemma}{lem:mod-inv} 
The matrix $\hM$ obeys $[s,\hM]\eq0$ and $[t,\hM]\eq0$, where $s$ is 
the $|\Ic|{\times}|\Ic|$-matrix given in \erf{eq:s-mat-def}, and $t$ is the 
$|\Ic|{\times}|\Ic|$-matrix with entries $t_{ij} \eq \theta_i\,\delta_{i,j}$.

\medskip\noindent
Proof:\\
We show $[s,\hM]\eq0$; that $[t,\hM]$ is zero as well is seen in a similar
manner, and we skip the details.
\\
Let $\Yr$ be the world sheet such that $\widetilde\Yr\eq T^2\,{\sqcup}\,({-}T^2)$ 
is the union of two tori with opposite orientation and $\iota_\Yr$ the involution
that exchanges the two tori. On $\Yr$ we consider the two cuttings
  \eqpic{pic_fjfrs2_53} {120} {32} {
  \put(42,0)     {\Includeourbeautifulpicture 53{}}
  \put(0,33)     {$ \Yr ~= $}
  \put(64,15)    {$\beta$}
  \put(85,53)    {$\gamma$}
  }
For $c_\gamma{:}\ \Xr_{Bp}\To \Yr$ a realisation of the cutting $\gamma$ we have
  \eqpic{eq:mod-inv-aux1} {350} {68} {
  \put(180,0)    {\Includeourbeautifulpicture 54a }
  \put(275,12)   {\Includeourbeautifulpicture 54b }
  \put(-28,68)   {$ \cor_\Yr ~=~ \bl(c_\gamma) \circ \cor_{\Xr_{Bp}}
                             ~=~ \dsty\sum_\alpha~\tftC\Big( $}
  \put(353,65)   {$ \Big) $}
  \put(237,54)   {\sse $B_l$}
  \put(237,111)  {\sse $B_l$}
  \put(232,73)   {\sse $p_\alpha'$}
  \put(282,43)   {\sse $B_r$}
  \put(282,100)  {\sse $B_r$}
  \put(289,82)   {\sse $p_\alpha''$}
  }
The manifolds shown on the right hand side are two solid tori in the `wedge 
presentation' for three-manifolds with boundary, i.e.\ the top and bottom faces
are identified, as are the two side faces drawn in dashed lines (for more 
details on the wedge presentation see section 5.1 of \cite{tft5}); also, 
unlike in figure \ref{fig:fund-world}, here and below we suppress the symbol 
`$\sqcup$' indicating the disjoint sum of the two components. In the second step
of \erf{eq:mod-inv-aux1} $\bl(c_\gamma)$ and $\cor_{\Xr_{Bp}}$ are replaced by 
their representations in terms of cobordisms, and also \erf{eq:cor-X_Bp-p_BS} 
and lemma \ref{lem:pBS=p} are used. The morphisms $p_\alpha'$ and $p_\alpha''$ 
are again those appearing in the expansion of $p_{\sfc} \eq 
e_\sfc \circ r_\sfc$ as $p_{\sfc}\eq\sum_\alpha p_\alpha' \oti p_\alpha''$.
\\[2pt]
Let $(U_i\Times\ol{U_j},e^\nu_{ij},r^\nu_{\!ij})$ for $\nu\eq 1,2,...\,,\hM_{ij}$ 
be realisations of the simple subobjects of $\Hcl$ as retracts, so that
$r^\mu_{\!ij} \Circ e^\nu_{ij} \eq \delta_{\mu,\nu}\, \id_{U_i \times \ol{U_j}}$
and $\id_{\Hcl} \eq \sum_{i,j,\nu} e^\nu_{ij} \Circ r^\nu_{\!ij}$. Defining 
$\tilde e^\nu_{ij} \,{:=}\, e_\sfc \Circ e^\nu_{ij}$ and $\tilde r^\nu_{\!ij} 
\,{:=}\, r^\nu_{\!ij} \Circ r_\sfc$ and expanding
$\tilde e^\nu_{ij} = \sum_\alpha \tilde e^\nu_{ij,\alpha}\!\!\!\!\!\phantom{|}'\,
\oti \tilde e^\nu_{ij,\alpha}\!\!\!\!\!\phantom{|}''\,$ and
$\tilde r^\nu_{\!ij} = \sum_\beta \tilde r^\nu_{\!ij,\beta}\!\!\!\!\!\phantom{|}'
\, \oti \tilde r^\nu_{\!ij,\beta}\!\!\!\!\!\phantom{|}''\,$ allows us to write
  \be
  p_{\sfc} = e_\sfc \circ r_\sfc = \sum_{i,j,\nu,\alpha,\beta} \big(
  \tilde e^\nu_{ij,\alpha}\!\!\!\!\phantom{|}'\,
  \Circ \tilde r^\nu_{\!ij,\beta}\!\!\!\!\phantom{|}'\,\,\big)
  \otic \big(\tilde e^\nu_{ij,\alpha}\!\!\!\!\!\phantom{|}''\, \Circ
  \tilde r^\nu_{\!ij,\beta}\!\!\!\!\!\phantom{|}''\,\,\big) \,.
  \ee
Substituting into \erf{eq:mod-inv-aux1} we get
\void{
  \eqpiczz{eq:mod-inv-aux2} {410} {160} {
  \put(-7,-8)    {\Includeourbeautifulpicture 54a }
  \put(80,18)    {\Includeourbeautifulpicture 54b }
  \put(-35,72)   {$ \dsty\sum_\alpha $}
  \put(173,72)   {$ = \dsty\sum_{i,j,\nu,\alpha,\beta} $}
  \put(223,-10)  {\Includeourbeautifulpicture 51a }
  \put(323,20)   {\Includeourbeautifulpicture 51b }
     \put(-23,0){
  \put(73.2,46)  {\sse $B_l$}
  \put(73.2,104) {\sse $B_l$}
  \put(110.5,50) {\sse $B_r$}
  \put(110.5,107){\sse $B_r$}
  \put(68,65)    {\sse $p_\alpha'$}
  \put(118,87)   {\sse $p_\alpha''$}
     }
  \put(285,38)   {\sse $B_l$}
  \put(285,110)  {\sse $B_l$}
  \put(288,67)   {\sse $U_i$}
  \put(338,45)   {\sse $B_r$}
  \put(338,117)  {\sse $B_r$}
  \put(338,86)   {\sse $U_j$}
  \put(276.5,53) {\tiny $\tilde{r}_{ij,\beta}^\nu\!\!\!\!\!{}'$}
  \put(276.5,84) {\tiny $\tilde{e}_{ij,\alpha}^\nu\!\!\!\!\!{}'$}
  \put(343,73)   {\tiny $\tilde{e}_{ij,\alpha}^\nu\!\!\!\!\!{}''$}
  \put(343,105)  {\tiny $\tilde{r}_{ij,\beta}^\nu\!\!\!\!\!{}''$}
  } {
  \put(28,-10)   {\Includeourbeautifulpicture 51a }
  \put(128,20)   {\Includeourbeautifulpicture 51b }
  \put(-23,72)   {$ =\dsty\sum_{i,j,\nu,\alpha,\beta} $}
  \put(227,72)   {$ =~\dsty\sum_{i,j}\hM_{ij} $}
  \put(283,-10)  {\Includeourbeautifulpicture 51c }
  \put(383,20)   {\Includeourbeautifulpicture 51d }
  \put(91,37)    {\sse $U_i$}
  \put(91,110)   {\sse $U_i$}
  \put(92,66)    {\sse $B_l$}
  \put(145,47)   {\sse $U_j$}
  \put(145,119)  {\sse $U_j$}
  \put(144,84)   {\sse $B_r$}
  \put(81.5,84)  {\tiny $\tilde{r}_{ij,\beta}^\nu\!\!\!\!\!{}'$}
  \put(81.5,53)  {\tiny $\tilde{e}_{ij,\alpha}^\nu\!\!\!\!\!{}'$}
  \put(148,105)  {\tiny $\tilde{e}_{ij,\alpha}^\nu\!\!\!\!\!{}''$}
  \put(148,73)   {\tiny $\tilde{r}_{ij,\beta}^\nu\!\!\!\!\!{}''$}
  \put(346,75)   {\sse $U_i$}
  \put(400,82)   {\sse $U_j$}
  }
}
\begin{equation*}
  \raisebox{-80pt}{
  \begin{picture}(410,170)
  \put(-7,-8)    {\Includeourbeautifulpicture 54a }
  \put(80,18)    {\Includeourbeautifulpicture 54b }
  \put(-35,72)   {$ \dsty\sum_\alpha $}
  \put(173,72)   {$ = \dsty\sum_{i,j,\nu,\alpha,\beta} $}
  \put(223,-10)  {\Includeourbeautifulpicture 51a }
  \put(323,20)   {\Includeourbeautifulpicture 51b }
     \put(-23,0){
  \put(73.2,46)  {\sse $B_l$}
  \put(73.2,104) {\sse $B_l$}
  \put(110.5,50) {\sse $B_r$}
  \put(110.5,107){\sse $B_r$}
  \put(68,65)    {\sse $p_\alpha'$}
  \put(118,87)   {\sse $p_\alpha''$}
     }
  \put(285,38)   {\sse $B_l$}
  \put(285,110)  {\sse $B_l$}
  \put(288,67)   {\sse $U_i$}
  \put(338,45)   {\sse $B_r$}
  \put(338,117)  {\sse $B_r$}
  \put(338,86)   {\sse $U_j$}
  \put(276.5,53) {\tiny $\tilde{r}_{ij,\beta}^\nu\!\!\!\!\!{}'$}
  \put(276.5,84) {\tiny $\tilde{e}_{ij,\alpha}^\nu\!\!\!\!\!{}'$}
  \put(343,73)   {\tiny $\tilde{e}_{ij,\alpha}^\nu\!\!\!\!\!{}''$}
  \put(343,105)  {\tiny $\tilde{r}_{ij,\beta}^\nu\!\!\!\!\!{}''$}
  \end{picture}}
\end{equation*}
\begin{equation}
  \begin{picture}(410,170)(0,-10)
  \put(28,-10)   {\Includeourbeautifulpicture 51a }
  \put(128,20)   {\Includeourbeautifulpicture 51b }
  \put(-23,72)   {$ =\dsty\sum_{i,j,\nu,\alpha,\beta} $}
  \put(227,72)   {$ =~\dsty\sum_{i,j}\hM_{ij} $}
  \put(283,-10)  {\Includeourbeautifulpicture 51c }
  \put(383,20)   {\Includeourbeautifulpicture 51d }
  \put(91,37)    {\sse $U_i$}
  \put(91,110)   {\sse $U_i$}
  \put(92,66)    {\sse $B_l$}
  \put(145,47)   {\sse $U_j$}
  \put(145,119)  {\sse $U_j$}
  \put(144,84)   {\sse $B_r$}
  \put(81.5,84)  {\tiny $\tilde{r}_{ij,\beta}^\nu\!\!\!\!\!{}'$}
  \put(81.5,53)  {\tiny $\tilde{e}_{ij,\alpha}^\nu\!\!\!\!\!{}'$}
  \put(148,105)  {\tiny $\tilde{e}_{ij,\alpha}^\nu\!\!\!\!\!{}''$}
  \put(148,73)   {\tiny $\tilde{r}_{ij,\beta}^\nu\!\!\!\!\!{}''$}
  \put(346,75)   {\sse $U_i$}
  \put(400,82)   {\sse $U_j$}
  \end{picture}
  \label{eq:mod-inv-aux2}
\end{equation}
where it is again understood that $\tftC$ is applied to each cobordism, 
and in the second step the morphisms $e'$ and $e''$ are moved along the 
vertical direction through the identification region so as to appear below $r'$ 
and $r''$, respectively. The third step amounts to the identity 
$\tilde r^\nu_{\!ij} \Circ \tilde e^\nu_{ij} \eq r^\nu_{\!ij} \Circ r_\sfc \Circ 
e_\sfc \Circ e^\nu_{ij} \eq \id_{U_i \times \ol{U_j}}$, summed over 
$\nu\eq1,...\,,\hM_{ij}$.
\\[2pt]
Carrying out the same calculation for the cutting $\beta$ in 
\erf{eq:mod-inv-aux1} leads to the cobordism 
  \eqpic{eq:mod-inv-aux3} {350} {50} {
  \put(100,0)    {\Includeourbeautifulpicture 55a }
  \put(225,5)    {\Includeourbeautifulpicture 55b }
  \put(0,50)     {$ \cor_\Yr~=~\dsty\sum_{i,j}~\hM_{ij} $}
  \put(168,50)   {\sse $U_i$}
  \put(272,57)   {\sse $U_j$}
  } 
for $\cor_\Yr$. Composing the expression for $\cor_\Yr$ obtained in 
\erf{eq:mod-inv-aux2} with the linear form
  \eqpic{pic_fjfrs2_56} {260} {50} {
  \put(0,10)     {\Includeourbeautifulpicture 56a }
  \put(97,0)     {\Includeourbeautifulpicture 56b }
  \put(190,50)   {$ \in~\Htft{(T^2 \,{\sqcup}\, {-}T^2)}^*_{} $}
  \put(11,48)    {\sse $U_k$}
  \put(153,65)   {\sse $U_l$}
  } 
results in two copies of $S^2 \Times S^1$, with invariant 
$\sum_{i,j} \delta_{k,i} \hM_{ij} \delta_{j,l} \eq \hM_{k,l}$. On the
other hand, performing the same manipulation on the expression in
\erf{eq:mod-inv-aux3} yields two copies of $S^3$ with embedded Hopf links
(one with labels $k,i$ and one with labels $j,\bar l$),
resulting in the invariant $\text{Dim}(\Cc)^{-1}\! \sum_{i,j} s_{ki} \hM_{ij} 
s_{j\bar l} \eq \text{Dim}(\Cc)^{-1} (s\,\hM\,s)_{k\bar l} $.
For more details on the invariants resulting from glueing tori see appendix A.3
of \cite{tft2}; the factor $\text{Dim}(\Cc)^{-1}$ appears as a consequence of 
$\tftC(S^3) \eq \text{Dim}(\Cc)^{-1/2}$. Comparing the two results we get 
$\hM_{jl} \eq \text{Dim}(\Cc)^{-1} (s\hM s)_{j\bar l}$.
Using further that $\sum_{b \in \Ic} s_{a,b} s_{b,c} \eq \text{Dim}(\Cc)\,
\delta_{a,\bar c}$ then yields $\hM\,s \eq s\,\hM$ and hence proves the claim.
\qed

\medskip

Now recall that by $A$ we denote the normalised algebra of open states for 
$\sfc$ and that $A$ is special and absolutely
simple. Further, as objects in $\CxCb$ we have
$Z(A) \Cong \bigoplus_{i,j} (U_i \Times \ol{U_j})^{\oplus z_{ij}}$ for some 
$z_{ij}\,{\equiv}\,z(A)_{ij}\iN \Zb_{\ge 0}$. By combining equation (3.5) and 
appendix A of \cite{rffs} with theorem 5.1 of \cite{tft1} and remark 2.8\,(i) 
of \cite{ffrs} it follows that $[s,z(A)]\eq 0$ and that 
$z(A)_{00}\eq 1$. Comparison with the previous analyses then leads to
\\[-2.3em]

\dtl{Lemma}{lem:phicl-epi}
The morphism $\varphi_\text{cl}^A$ defined in \erf{eq:phi_cl-def} is an 
epimorphism.

\medskip\noindent
Proof:\\
Since by lemma \ref{lem:phicl-mono} there is a monomorphism from $\Hcl$ to 
$Z(A)$, the integers $\hM_{ij}$ defined in lemma \ref{lem:mod-inv} satisfy 
$\hM_{ij} \,{\le}\, z_{ij}$ for all $i,j \iN \Ic$. By proposition 
\ref{prop:open-alg-simple} and uniqueness of the closed state vacuum
(property (i) in theorem \ref{thm:unique}), $A$ is absolutely simple and hence 
$z_{00}\eq1$. The matrix $D \,{:=}\, z\,{-}\,\hM$ thus obeys $D_{00}\eq0$,
$D_{kl} \,{\ge}\, 0$ for all $k,l\iN\Ic$, and $[s,D]\eq 0$. It follows that
  \be
  0 = D_{00} = (\text{Dim}\,\Cc)^{-1}\!\sum_{k,l\in\Ic} s_{0k}\,D_{kl}\,s_{l0}
  = \sum_{k,l\in\Ic} \dim(U_k)\, \dim(U_l)\, D_{kl} \,.
  \labl{eq:Z-is-H-aux}
Since $D_{kl} \,{\le}\, z_{kl}$, the sum is only over those pairs
$k,l$ for which $U_k \Times \ol U_l$ is a subobject of $Z(A)$.
Combining \erf{eq:Z-is-H-aux} with the positivity assumption in 
condition (iv) of theorem \ref{thm:unique}, it follows that each coefficient
$D_{kl}$ in the sum on the right hand side vanishes, i.e.\ that
$D \,{\equiv}\,0$. Thus $\hM_{ij} \eq z_{ij}$ for all $i,j \iN \Ic$, 
and hence the monomorphism $\varphi_\text{cl}^A$ is also an epimorphism. 
\qed


\subsection{Equivalence of solutions}\label{sec:equiv-of-sol}

As data in theorem \ref{thm:unique} we are given a solution $\sfc \eq 
(\Cc,\Hop,\Hcl,B_l,B_r,e_\sfc,r_\sfc,\cor)$ to the sewing constraints. Using 
the normalised algebra $A$ of open states of $\sfc$ we obtain another solution
$\sfc(\Cc,A) \equiv (\Cc,A,Z(A),A\oti\Omeka,\Omeka,e_Z,r_Z,\corA)$
via theorem \ref{thm:Frob-SCA}.

We will show that these two solutions are actually equivalent in the sense of
definition \ref{def:sol-equiv}. More specifically, an equivalence between $\sfc$ 
and $\sfc(\Cc,A)$ is provided by the isomorphism 
$\varphi_\text{cl}^A \iN \Hom_{\Cc\boxtimes\ol\Cc}(\Hcl,Z(A))$ studied in 
sections \ref{sec:phicl-def}\,--\,\ref{sec:phicl-epi} together with
  \be
  \varphi_\text{op}^A := \id_A ~\in \Hom(\Hop,A) \,,
  \ee
with the number $\gamma$ 
appearing in definition \ref{def:sol-equiv} given as in \erf{eq:norm-open-state}. 
Let us abbreviate $\Alpha \,{\equiv}\, \Alpha(\id_A,\varphi_\text{cl}^A)$, as 
well as $\bl \,{\equiv}\, \bl(\Cc,\Hop,\Hcl,B_l,B_r,e,r)$ and 
$\bl^A \,{\equiv}\, \bl(\Cc,A,Z(A),A\oti\Omeka,\Omeka,e_Z,r_Z)$, and similarly 
for $P^\text{vac}_{\sew,\Xr}$ and $E^\text{vac}_{\sew,\Xr}$ versus 
$P^{\text{vac},A}_{\sew,\Xr}$ and $E^{\text{vac},A}_{\sew,\Xr}$. For having an
equivalence, by lemma \ref{lem:compare} it suffices to establish that
  \be
  \cor_\Xr = \gamma^{2\chi(\Xr)} \, \Alpha_\Xr \circ \corA_\Xr
  \labl{eq:equivalence-check} 
for the selection of fundamental world sheets given in figure 
\ref{fig:fund-world}. Below we describe how to obtain \erf{eq:equivalence-check}
for each of these world sheets.
\\[5pt]
\nxt $\Xr_\eta$ and $\Xr_\eps$:\\[1pt]
According to \erf{eq:norm-open-state} the unit morphisms $\eta$ of $A$
and $\eta_\sfc$ of $\Hop$ are related by $\eta_\sfc\eq\gamma\eta$. Using also 
that $\chi(\Xr_\eta)\eq\tfrac12$ and $\Alpha_{\Xr_\eta}\eq\id_{\bl(\Xr_\eta)}$,
we thus immediately have
  \be
  \cor_{\Xr_\eta} = \tftC\big( \Fr(\Xr_\eta;\gamma \eta) \big)
  = \gamma \,\tftC\big( \Fr(\Xr_\eta;\eta) \big)
  = \gamma^{2\chi(\Xr_\eta)}\, \corA_{\Xr_\eta} \,.
  \ee 
The calculation for $\Xr_\eps$ 
follows analogously from $\eps_\sfc\eq\gamma\eps$.
\\[5pt]
\nxt $\Xr_m$ and $\Xr_\Delta$:\\[1pt]
We have $m_\sfc\eq\gamma^{-1}m$,
$\chi(\Xr_m) \eq {-}\tfrac12$ and $\Alpha_{\Xr_m} \eq \id_{\bl(\Xr_m)}$, so that
  \be
  \cor_{\Xr_m} = \tftC\big( \Fr(\Xr_m;\gamma^{-1} m) \big)
  = \gamma^{-1} \, \tftC\big( \Fr(\Xr_m;m) \big)
  = \gamma^{2\chi(\Xr_m)}\, \corA_{\Xr_m} \,.
  \ee 
For $\Xr_\Delta$ the calculation is analogous.
\\[5pt]
\nxt $\Xr_{Bb}$:\\[1pt]
We have $\chi(\Xr_{Bb}) \eq {-}1$; the natural transformation 
\erf{eq:alphaX-def} is given by $\Alpha_{\Xr_{Bb}} \eq \gamma^2 \sum_{i,\alpha}
\tftC(N_{Bb,i\alpha})$ with the cobordisms
  \eqpic{pic-fjfrs2_37} {180} {57} {
  \put(63,0)     {\Includeourbeautifulpicture 37{}}
  \put(0,61)     {$ N_{Bb,i\alpha}~= $}
  \put(120.5,25) {\sse $\phi_{i\alpha}''$}
  \put(120.5,100.5){\sse $\phi_{i\alpha}'$}
  \put(84,68)    {\sse $A$}
  \put(162,68)   {\sse $A$}
  \put(113,11)   {\sse $B_r$}
  \put(115,35)   {\sse $\Omeka$}
  \put(99,88.5)  {\sse $A\oti \Omeka$}
  \put(114,113)  {\sse $B_l$}
  }  
where $\gamma^2 \sum_{i,\alpha}\phi_{i\alpha}' \oti \phi_{i\alpha}'' \eq
\phi_\text{cl}^A \eq e_Z \Circ \varphi_\text{cl}^A \Circ r_\sfc$ as in lemma 
\ref{lem:phiS-tensor}. With this information one verifies that
  \be
  \Alpha_{\Xr_{Bb}} \circ \corA_{\Xr_{Bb}}
  = \gamma^2 \sum_{i,\alpha} \tftC(N_{Bb,i\alpha}) \circ \tftC(
    \Fr(\Xr_{Bb};\Phi_{\!A}))
  = \gamma^2 \, \tftC( \Fr(\Xr_{Bb};u)) \,,
  \ee 
where the morphism $u \iN \Hom(A \oti B_l,A \oti B_r)$ is given by
  \eqpic{pic_fjfrs2_39} {390} {91} {
\void{
  \put(67,13)     {\Includeourbeautifulpicture 39a }
  \put(165,13)    {\Includeourbeautifulpicture 39b }
  \put(270,13)    {\Includeourbeautifulpicture 39c }
  \put(0,95)      {$ u~=~\dsty\sum_{i,\alpha} $}
  \put(113,95)    {$ =~\dsty\sum_{i,\alpha} $}
  \put(239,95)    {$ = $}
  \put(350,95)    {$ =~\Phi_\sfc~. $}
  \put(72,101)    {\sse $\Phi_A$}
  \put(81.5,145.5){\sse $\phi_{i\alpha}''$}
  \put(81.5,56.5) {\sse $\phi_{i\alpha}'$}
  \put(197,46)    {\sse $\Phi_\sfc$}
  \put(203,74)    {\sse $r_{i\alpha}$}
  \put(205,96)    {\sse $e_i$}
  \put(206,150)   {\sse $r_i$}
  \put(203,171)   {\sse $e_{i\alpha}$}
  \put(302,46)    {\sse $\Phi_\sfc$}
  \put(61,5)      {\sse $A$}
  \put(82,5)      {\sse $B_l$}
  \put(173,5)     {\sse $A$}
  \put(203,5)     {\sse $B_l$}
  \put(278,5)     {\sse $A$}
  \put(308,5)     {\sse $B_l$}
  \put(63,194)    {\sse $A$}
  \put(82,194)    {\sse $B_r$}
  \put(175,194)   {\sse $A$}
  \put(203.5,194) {\sse $B_r$}
  \put(279.5,194) {\sse $A$}
  \put(309,194)   {\sse $B_r$}
  \put(86,82)     {\sse $A\oti K$}
  \put(85,119)    {\sse $K$}
  \put(208,57)    {\sse $B_r$}
  \put(210,83)    {\sse $U_i$}
  \put(200,117)   {\sse $K$}
  \put(210,159.5) {\sse $U_i$}
  \put(218,117)   {\sse $A$}
  \put(295,60)    {\sse $A$}
  \put(325,83)    {\sse $A$}
  }
  \put(67,13)     {\Includeourbeautifulpicture 39a }
  \put(165,13)    {\Includeourbeautifulpicture 39b }
  \put(270,13)    {\Includeourbeautifulpicture 39c }
  \put(0,95)      {$ u~=~\dsty\sum_{i,\alpha} $}
  \put(113,95)    {$ =~\dsty\sum_{i,\alpha} $}
  \put(239,95)    {$ = $}
  \put(350,95)    {$ =~\Phi_\sfc~. $}
  \put(72,100.5)  {\sse $\Phi_A$}
  \put(81,145.5)  {\sse $\phi_{i\alpha}''$}
  \put(81,56) {\sse $\phi_{i\alpha}'$}
  \put(197,46)    {\sse $\Phi_\sfc$}
  \put(203,74)    {\sse $r_{i\alpha}$}
  \put(205.5,95.6){\sse $e_i$}
  \put(205.5,149.6){\sse $r_i$}
  \put(203,171.5) {\sse $e_{i\alpha}$}
  \put(302,46)    {\sse $\Phi_\sfc$}
  \put(63,5)      {\sse $A$}
  \put(83,5)      {\sse $B_l$}
  \put(175,5)     {\sse $A$}
  \put(205,5)     {\sse $B_l$}
  \put(280,5)     {\sse $A$}
  \put(310,5)     {\sse $B_l$}
  \put(64,194)    {\sse $A$}
  \put(83,194)    {\sse $B_r$}
  \put(176,194)   {\sse $A$}
  \put(204,194)   {\sse $B_r$}
  \put(280,194)   {\sse $A$}
  \put(310,194)   {\sse $B_r$}
  \put(86,82)     {\sse $A\oti K$}
  \put(85,119)    {\sse $K$}
  \put(208,57)    {\sse $B_r$}
  \put(210,83)    {\sse $U_i$}
  \put(200,117)   {\sse $K$}
  \put(210,159.2) {\sse $U_i$}
  \put(218,117)   {\sse $A$}
  \put(295,66)    {\sse $A$}
  \put(325,83)    {\sse $A$}
  }

Here in the second step the expressions for $\phi'_{i\alpha}$ and 
$\phi''_{i\alpha}$ as given in lemma \ref{lem:phiS-tensor} is substituted, as 
well as the expression \erf{eq:PhiA} for $\Phi_{\!A}$. In the third step the 
various embedding and restriction morphisms are canceled. The Frobenius and 
unit properties of $A$ is used to replace the encircled coproduct by 
$(m \oti \id_A) \Circ (\id_A \oti (\Delta \Circ \eta))$; the resulting 
multiplication is moved upwards past the top multiplication morphism.
In the final step the $A$-loop is omitted using \erf{eq:leave-proj}, and then 
the encircled multiplication morphism is moved past $\Phi_\sfc$ using that, 
according to lemma \ref{lem:PhiS-HomAA}\,(i), $\Phi_\sfc$ is an intertwiner 
of bimodules. Thus altogether we obtain
  \be
  \gamma^{-2} \, \Alpha_{\Xr_{Bb}}\circ \corA_{\Xr_{Bb}}
  = \tftC( \Fr(\Xr_{Bb};\Phi_\sfc)) = \cor_{\Xr_{Bb}} \,,
  \ee 
in accordance with \erf{eq:equivalence-check}.
\\[5pt]
\nxt $\Xr_{Bp}$:\\[1pt]
(This is not among the fundamental world sheets listed in figure 
\ref{fig:fund-world}, but below we will need the equivalence of correlators on 
$\Xr_{Bp}$.) Combining lemma \ref{lem:pBS=p} and equation \erf{eq:cor-X_Bp-p_BS}
we see that $\cor_{\Xr_{Bp}} \eq \sum_\alpha\tftC(F(\Xr_{Bp};p_\alpha',p_\alpha
''))$, 
where $e_\sfc \Circ r_\sfc \eq p_\sfc \eq \sum_\alpha p'_\alpha \oti p''
_\alpha$. By writing out the explicit form of the cobordisms, one checks that
  \be
  \Alpha_{\Xr_{Bp}} \circ \corA_{\Xr_{Bp}}
  = \sum_\beta \tftC(F(\Xr_{Bp};q_{\beta}',q_{\beta}'')) \,,
  \ee 
where $q \eq \sum_\beta q'_\beta \oti q''_\beta$ is given by
  \be
  q = e_\sfc \circ (\varphi_\text{cl}^A)^{-1} \circ r_Z \circ e_Z \circ
  \varphi_\text{cl}^A \circ r_\sfc = e_\sfc \circ r_\sfc = p_\sfc \,.
  \ee 
Thus $\cor_{\Xr_{Bp}} \eq \Alpha_{\Xr_{Bp}} \Circ \corA_{\Xr_{Bp}}$.
\\[5pt]
\nxt $\Xr_{B(\ell)}$:\\[1pt]
Denote by $\Xr_{B(\ell)}$ the world sheet given by a sphere with $\ell$ in-going 
closed state boundaries. Let $\Yr$ be a \disc\ with $\ell$ in-going closed 
state boundaries and consider the following two cuttings:
  \eqpic{pic_fjfrs2_43} {310} {67} {
  \put(47,0)    {\Includeourbeautifulpicture 43{}}
  \put(0,80)    {$ \Yr~= $}
  \put(147,2)   {$ \alpha $}
  \put(98,146)  {$ \beta $}
  \put(96.2,72) {\small$ in $}
  \put(145.3,72){\small$ in $}
  \put(237,72)  {\small$ in $}
  }
$\Yr|_\alpha$ is isomorphic to $\Xr_\eta \,{\sqcup}\, \Xr_{Bb} \,{\sqcup}\, 
\cdots \,{\sqcup}\, \Xr_{Bb}\,{\sqcup}\, \Xr_\eps$. By the previous results and
lemma \ref{lem:compare} we can thus conclude that
  \be
  \cor_\Yr = \gamma^{2-2\ell} \, \Alpha_\Yr \circ \corA_\Yr \,.
  \labl{eq:sol-equiv-aux1} 
Now apply $P^\text{vac}_{\!\sew_\beta,\Yr|_\beta}$ to both sides of this 
equality. Using proposition \ref{prop:proj-vac} one finds
  \be
  P^\text{vac}_{\!\sew_\beta,\Yr|_\beta} \circ \cor_\Yr
  = \Lambda_\sfc^{-1} E^\text{vac}_{\sew_\beta,\Yr|_\beta} \circ
  \cor_{\fill_{\sew_\beta}(\Yr|_\beta)}
  = \Lambda_\sfc^{-1} E^\text{vac}_{\sew_\beta,\Yr|_\beta} \circ \bl(q) \circ
  \cor_{D \sqcup \Xr_{B(\ell)}}
  \ee 
for the left hand side, where $q \eq (\emptyset,f){:}\ D{\sqcup}\Xr_{B(\ell)}\To 
\fill_{\sew_\beta}(\Yr|_\beta)$ is an isomorphism
of world sheets, while for the right hand side we get
  \bea
  \gamma^{2-2\ell} P^\text{vac}_{\!\sew_\beta,\Yr|_\beta} \circ
  \Alpha_\Yr \circ \corA_\Yr
  = \gamma^{2-2\ell}\, \Alpha_\Yr \circ P^{\text{vac},A}_{\sew_\beta,\Yr|_\beta}
  \circ \corA_\Yr
  \\{}\\[-.7em] \hspace*{7em}
  = \gamma^{2-2\ell} \Lambda_A^{-1}\, \Alpha_\Yr \circ
  E^{\text{vac},A}_{\sew_\beta,\Yr|_\beta} \circ
  \corA_{\fill_{\sew_\beta}(\Yr|_\beta)}
  \\{}\\[-.7em] \hspace*{7em}
  = \gamma^{2-2\ell} \dim(A)^{-1} E^{\text{vac}}_{\sew_\beta,\Yr|_\beta} \circ
  \Alpha_{\fill_{\sew_\beta}(\Yr|_\beta)} \circ \bl^A(q) \circ
  \corA_{D \sqcup \Xr_{B(\ell)}}
  \\{}\\[-.7em] \hspace*{7em}
  = \gamma^{2-2\ell}\dim(A)^{-1} E^{\text{vac}}_{\sew_\beta,\Yr|_\beta}\circ \bl(q)
  \circ \Alpha_{D \sqcup \Xr_{B(\ell)}} \circ \corA_{D \sqcup \Xr_{B(\ell)}} \,.
  \end{array}
  \ee
These equalities are obtained by using \erf{eq:Pvac-alpha-comm},
\erf{eq:alpha-past-Evac} and the fact that, as follows from evaluating 
\erf{eq:LamS-def} for the solution $\sfc(\Cc,A)$), $\Lambda_A \eq \dim(A)$.
Using further that, according to lemma \ref{lem:Evac-inject}, 
$E^{\text{vac}}_{\sew_\beta,\Yr|_\beta}$ is injective, and that
since $q\eq(0,f)$ is an isomorphism, so is $\bl(q)$, we can conclude that
  \be
  \Lambda_\sfc^{-1}\, \cor_{D \sqcup \Xr_{B(\ell)}}
  = \gamma^{2-2\ell} \dim(A)^{-1}\,
  \Alpha_{D \sqcup \Xr_{B(\ell)}} \circ \corA_{D \sqcup \Xr_{B(\ell)}} \,.
  \ee 
Now $\cor_{D} \eq \gamma^{2} \corA_D \eq \gamma^{2} \dim(A)\, \tftC(B^3)$ so 
that, using also that $\cor$ and $\corA$ are monoidal,
  \be
  \Lambda_\sfc^{-1}\gamma^{2} \dim(A)\, \tftC(B^3) \oti \cor_{\Xr_{B(\ell)}}
  = \gamma^{2-2\ell}\, \tftC(B^3) \oti \big(
  \Alpha_{\Xr_{B(\ell)}} {\circ}\, \corA_{\Xr_{B(\ell)}} \big) \,.
  \ee 
Since $\tftC(B^3) \,{\ne}\, 0$ and $\chi(\Xr_{B(\ell)}) \eq 2{-}\ell$, we can 
finally write
  \be
  \cor_{\Xr_{B(\ell)}} = \frac{\Lambda_\sfc}{\gamma^4\dim(A)}
  \,\gamma^{2\chi(\Xr_{B(\ell)})}\,
  \Alpha_{\Xr_{B(\ell)}} \circ \corA_{\Xr_{B(\ell)}} \,.
  \ee 
In order to establish \erf{eq:equivalence-check} it remains to be shown that 
  \be
  \Lambda_\sfc = \gamma^4\,\dim(A) \,.
  \ee
This will be done below; for the moment we keep 
$\mu\,{:=}\, \Lambda_\sfc / (\gamma^4\dim(A))$ as a parameter.
\\[5pt]
\nxt $\Xr_{oo}$:\\[1pt]
To make the calculation below more transparent, we introduce the notations
$\Xr_{io} \eq \Xr_{oi} \eq \Xr_{Bp}$ and $\Xr_{ii} \eq \Xr_{B(2)}$, by which 
the symbols `$i$' and `$o$' indicate the in-going and out-going closed state 
boundaries on the sphere. Consider the world sheet $\Xr_{oo} \,{\sqcup}\, 
\Xr_{ii} \,{\sqcup}\, \Xr_{oo}$. There are morphisms
  \be
  \begin{array}{ll}
  l_1 :\quad \Xr_{oo} \sqcup \Xr_{ii} \to \Xr_{oi} \,,\quad&
  l_2 :\quad \Xr_{oi} \sqcup \Xr_{oo} \to \Xr_{oo} \,,
  \\{}\\[-.6em]
  r_1 :\quad \Xr_{ii} \sqcup \Xr_{oo} \to \Xr_{io} \,,&
  r_2 :\quad \Xr_{oo} \sqcup \Xr_{io} \to \Xr_{oo} 
  \end{array}
  \ee
such that on $\Xr_{oo}\,{\sqcup}\,\Xr_{ii}\,{\sqcup}\,\Xr_{oo}$ one has
  \be
  l_2 \circ (l_1 \,{\sqcup}\, \id_{\Xr_{oo}})
  = r_2 \circ (\id_{\Xr_{oo}} \,{\sqcup}\, r_1) \,.
  \labl{l2l1r2r1}
We can then write
  \be
  \cor_{\Xr_{oo}}
  = \bl(l_2) \circ \big( \cor_{\Xr_{oi}} \oti \cor_{\Xr_{oo}} \big) \,.
  \labl{eq:XBo2-aux1}
For $\Xr_{oi}$ we have already established \erf{eq:equivalence-check}, so that
  \be
  \begin{array}{ll}
  \cor_{\Xr_{oi}} = \Alpha_{\Xr_{oi}} \circ \corA_{\Xr_{oi}} \!\!
  &= \Alpha_{\Xr_{oi}} \circ \bl^A(l_1) \circ \big(
     \corA_{\Xr_{oo}} \oti \corA_{\Xr_{ii}} \big)
  \\{}\\[-.6em]
  &= \bl(l_1) \circ \Alpha_{\Xr_{oo}\sqcup\Xr_{ii}} \circ \big(
     \corA_{\Xr_{oo}} \oti \corA_{\Xr_{ii}} \big)
  \\{}\\[-.6em]
  &= \bl(l_1) \circ \big((\Alpha_{\Xr_{oo}} \Circ \corA_{\Xr_{oo}}) \oti
     (\Alpha_{\Xr_{ii}} \Circ \corA_{\Xr_{ii}}) \big)
  \\{}\\[-.6em]
  &= \bl(l_1) \circ \big((\Alpha_{\Xr_{oo}} \Circ \corA_{\Xr_{oo}}) \oti
     (\mu^{-1} \cor_{\Xr_{ii}}) \big) \,.
  \end{array}
  \ee 
Substituting this result into the right hand side of \erf{eq:XBo2-aux1} gives
  \be
  \cor_{\Xr_{oo}}
  = \mu^{-1} \bl(l_2) \circ \big( \bl(l_1) \oti \id_{\bl(\Xr_{oo})} \big)
  \circ \big( (\Alpha_{\Xr_{oo}} \Circ \corA_{\Xr_{oo}}) \oti
    \cor_{\Xr_{ii}} \oti \cor_{\Xr_{oo}} \big) \,.
  \labl{eq:XBo2-aux2} 
At this point we can use the defining condition \erf{l2l1r2r1} for the morphisms
$l_{1,2}$ and $r_{1,2}$ to obtain
  \be
  \begin{array}{ll}
  \bl(l_2) \circ \big( \bl(l_1) \oti \id_{\bl(\Xr_{oo})} \big) \!\! 
  &= \bl\big(l_2 \circ ( l_1 \sqcup \id_{\Xr_{oo}} )\big)
  \\{}\\[-.8em]
  &= \bl\big(r_2 \circ (\id_{\Xr_{oo}}\sqcup r_1 )\big)
  = \bl(r_2) \circ \big( \id_{\bl(\Xr_{oo})} \oti \bl(r_1) \big) \,.
  \end{array}
  \ee 
Substituting into \erf{eq:XBo2-aux2} yields
  \be
  \begin{array}{ll}
  \cor_{\Xr_{oo}} \!\!\!
  &= \mu^{-1} \bl(r_2) \Circ \big( \id_{\bl(\Xr_{oo})} \oti \bl(r_1) \big)
  \Circ \big( (\Alpha_{\Xr_{oo}}{\circ}\,\corA_{\Xr_{oo}}) \oti
    \cor_{\Xr_{ii}} \oti \cor_{\Xr_{oo}} \big)
  \\{}\\[-.6em]
  &= \mu^{-1}\, \bl(r_2) \circ
    \big( (\Alpha_{\Xr_{oo}} {\circ}\, \corA_{\Xr_{oo}}) \oti
    ( \bl(r_1) \circ \cor_{\Xr_{ii} \sqcup \Xr_{oo}} ) \big)
  \\{}\\[-.6em]
  &= \mu^{-1}\, \bl(r_2) \circ
    \big( (\Alpha_{\Xr_{oo}} {\circ}\, \corA_{\Xr_{oo}}) \oti
    \cor_{\Xr_{io}}  \big)
  \\{}\\[-.6em]
  &= \mu^{-1}\, \bl(r_2) \circ
    \big( (\Alpha_{\Xr_{oo}} {\circ}\, \corA_{\Xr_{oo}}) \oti
    (\Alpha_{\Xr_{io}} {\circ}\, \corA_{\Xr_{io}} ) \big)
  \\{}\\[-.6em]
  &= \mu^{-1}\, \bl(r_2) \circ \Alpha_{\Xr_{oo} \sqcup \Xr_{io}}
    \circ \corA_{\Xr_{oo} \sqcup \Xr_{io}}
  = \mu^{-1}\, \Alpha_{\Xr_{oo}}
    \circ \bl^A(r_2) \circ \corA_{\Xr_{oo} \sqcup \Xr_{io}} \,,
  \end{array}
  \ee 
so that altogether
  \be
  \cor_{\Xr_{oo}} = \mu^{-1}\, \Alpha_{\Xr_{oo}} \circ \corA_{\Xr_{oo}} \,.
  \labl{eq:X_oo-equiv}
%
Consider now a world sheet $\Yr$ which is a \disc\ with one in-going closed 
state boundary,
  \eqpic{pic_fjfrs2_44} {140} {44} {
  \put(45,0)    {\Includeourbeautifulpicture 44{}}
  \put(0,50)    {$ \Yr~= $}
  \put(170.5,51){$ \alpha $}
  \put(94,44)   {\small$ in $}
  }
The cutting $\alpha$ shows that there is a morphism $\xm{:}\ \Xr_\eta 
\,{\sqcup}\, \Xr_{Bb} \sqcup \Xr_\eps \To \Yr$, and hence by lemma 
\ref{lem:compare} we see that $\cor_\Yr \eq \Alpha_\Yr \Circ \corA_\Yr$. Next
consider a world sheet $\Xr$ in the form of an annulus with cuttings $\alpha$
and $\beta$ as follows:
  \eqpic{pic_fjfrs2_42} {140} {43} {
  \put(45,0)    {\Includeourbeautifulpicture 42{}}
  \put(0,50)    {$ \Xr~= $}
  \put(0,50)    {$ \Xr~= $}
  \put(170.3,42.6)  {$\alpha$}
  \put(158,80)  {$\beta$}
  }
The cutting $\alpha$ shows that there is a morphism $\xm{:} \Xr_p \To \Xr$, so 
that again by lemma \ref{lem:compare} we know that
  \be
  \cor_\Xr = \Alpha_\Xr \circ \corA_\Xr \,.
  \labl{eq:cor-annulus} 
Resulting from the cutting $\beta$ there exists a morphism 
$q{:}\ \Yr \,{\sqcup}\, \Xr_{oo} \,{\sqcup}\, \Yr \To \Xr$. Applying this 
to rewrite $\cor_\Xr$ and invoking \erf{eq:X_oo-equiv} results in
  \be
  \begin{array}{ll}
  \cor_\Xr \!\!
  &= \bl(q) \circ
  \big( \cor_{\Yr} \oti \cor_{\Xr_{oo}} \oti \cor_{\Yr} \big)
  \\{}\\[-.8em]
  &= \mu^{-1}\, \bl(q) \circ
  \big( \Alpha_{\Yr} \otic \Alpha_{\Xr_{oo}}
    \otic \Alpha_{\Yr} \big) \circ
  \big( \corA_{\Yr} \oti \corA_{\Xr_{oo}} \oti \corA_{\Yr} \big)
  \\{}\\[-.8em]
  &= \mu^{-1}\,\Alpha_\Xr \circ \bl(q) \circ \corA_{\Yr\sqcup\Xr_{oo}\sqcup\Yr}
  = \mu^{-1}\, \Alpha_\Xr \circ \corA_\Xr \,.
  \end{array}
  \ee 
Comparing to \erf{eq:cor-annulus} and using that $\corA_\Xr\,{\ne}\,0$ (as is 
seen by explicit calculation according to the construction in section 
\ref{sec:frob-to-sew}), we conclude that indeed $\mu \eq 1$, as required.

\medskip

The list of world sheets for which we have by now established 
\erf{eq:equivalence-check} includes
  \be
  \Xr_{\eta} \,,\quad
  \Xr_{\eps} \,,\quad
  \Xr_{m} \,,\quad
  \Xr_{\Delta} \,,\quad
  \Xr_{Bb} \,,\quad
  \Xr_{B(1)} \,,\quad
  \Xr_{B(3)} \,,\quad
  \Xr_{oo} \,.
  \ee 
Every world sheet can be obtained as a sewing of world sheets in this list, 
and hence by lemma \ref{lem:compare}, equation \erf{eq:equivalence-check} holds 
in fact for all world sheets. This completes the proof of
theorem \ref{thm:unique}.


\sect{CFT on world sheets with metric}\label{sec:wswm}

Having completed the proof of the uniqueness theorem \ref{thm:unique},
we now return to the issue of passing from the results for correlators on
topological world sheets to conformal field theory on conformal world sheets,
where one deals with actual correlation {\em functions\/} of the locations of 
field insertions and of the moduli of the world sheet.

\subsection{Conformal world sheets and conformal blocks}

The basic picture is that
the construction of a rational CFT, or more specifically, of a consistent set of
correlation functions, proceeds in two steps. The first is complex-analytic and 
consists of evaluating the restrictions imposed by the chiral symmetries of the
theory, which include in particular the Virasoro algebra. The second step then 
consists of imposing the non-chiral consistency requirements. This step, to 
which we refer as {\em solving the sewing constraints\/}, has been the subject 
of sections \ref{sec:oc-sewing}\,--\,\ref{sec:unique-proof}.
As we have seen, it can be discussed in a purely algebraic and combinatoric
framework, without reference to the complex-analytic considerations, and in 
particular we need to consider the CFT only on topological world sheets.

The correlators are elements of suitable vector spaces of conformal blocks.
In the combinatoric setting, a space of conformal blocks is just an abstract 
finite-dimensional complex vector space. In contrast, for CFT on conformal 
world sheets each space of conformal blocks is given more concretely as the 
fiber of a vector bundle, equipped with a projectively flat connection, over a 
moduli space of decorated complex curves (the complex doubles of the world 
sheets). This bundle, in turn, is determined through the chiral symmetry algebra
$\Vc$ and the $\Vc$-representations that are carried by the field insertions.
The chiral symmetries can be formalised in the structure of a conformal 
vertex algebra $\Vc$. Then the space of conformal blocks for a correlator with
field insertions carrying $\Vc$-representations
$\lambda_1,\lambda_2,...\,,\lambda_m$ can be described as a certain
$\Vc$-invariant subspace in the space of multilinear maps from
$\lambda_1 \Times \lambda_2 \Times \cdots \Times \lambda_m$ to $\Cb$.
(To describe how these vector spaces fit together to form the total space of
a vector bundle of the relevant moduli space one must study sheaves of 
conformal vertex algebras \cite{FRbe}.)

For a rational CFT, the representation category $\RepVc$
is ribbon \cite{hule3.5} and even modular \cite{huan21}.
Motivated by the path-integral formulation in the case of Chern-Simons
theories \cite{witt27,frki2} one identifies the spaces of states that
the 3-d TFT associated to the modular tensor category $\RepVc$ assigns to surfaces
with fibers of the bundles of conformal blocks. The 3-d TFT should then
encode the behaviour of conformal blocks under sewing as well as the
action of the mapping class group. (This is known to be true for genus zero
and genus one if $\Vc$ obeys the conditions of theorem 2.1 in \cite{huan21},
but for higher genus it still remains open.)
Note that in the second step, i.e.\ for solving the sewing constraints,
the only input needed is the category $\RepVc$ as a ribbon category.

\medskip

Our aim is now to analyse CFT on conformal world sheets with the help of
categories, functors between them and natural transformations that are
analogous to those that appeared in the combinatorial setting above. To begin 
with, it does not suffice to endow the topological world sheet with a conformal 
structure, but we must also specify a metric in the conformal equivalence 
class. Thus an (oriented open/closed) {\em world sheet with metric\/}, or 
{\em world sheet\/}, for short, is a surface of the type shown in figure 
\ref{fig:oc-world} (p.\,\pageref{fig:oc-world}), except that the specification 
of an orientation must be supplemented by the specification of a metric. We 
denote world sheets with metric by $\Xrc$,\label{def:wsc} where $\Xr$ is the 
underlying topological world sheet. When discussing CFT on world sheets with 
metric we can draw from descriptions used in string theory (see e.g.\ 
\cite{frsh2,vafa0,bcdcd}), from the study of sewing constraints 
\cite{card3,cale,lewe3} and from aspects of the axiomatics of
\cite{sega8,sega8b}. (For recent treatments of open/closed CFT from similar 
points of view see e.g.\ \cite{huan14,huko,hukr3}.) What we are interested in 
here could more precisely be referred to as compact oriented open/closed CFT; 
the qualifier `compact' refers to a discreteness condition on the relevant 
spaces of states.


\subsection{Consistency conditions for correlation functions}%
\label{sec:intro-consist}

We regard world sheets $\Xrc$ as the objects of a category $\Worc$ and,
analogously as in the case of the category $\Wor$, as morphisms of $\Worc$
we allow for both isomorphisms of world sheets and for sewings, and for 
combinations of the two. An isomorphism between world sheets $\Xrc$ and $\Yrc$ 
is an orientation preserving isometry $f$ from $\Xrc$ to $\Yrc$ that is 
compatible with the parametrisation of the state boundaries. A sewing $\sew$ 
is analogous to a sewing in $\Wor$, but only such sewings are allowed which 
lead to a smooth metric on the sewed world sheet.
One can also define a tensor product on
$\Worc$ by taking disjoint unions; the tensor unit is the empty set.

The defining data of a (compact, oriented) open/closed CFT are the 
{\em boundary\/} and {\em bulk state spaces\/}, i.e.\ spaces of `open' and 
`closed' states, and a collection $\Cor$ of {\em correlation functions\/} 
-- analogues of the corresponding combinatorial data that enter the definition 
of the block functor $\bl$. The state spaces are complex vector spaces which
come with a hermitian inner product and are discretely $\Rb$-graded by the 
eigenvalues of the dilation operator; we denote them
by $\Hop$ and $\Hcl$, respectively, and their graded duals by $\Hop^\vee$ and 
$\Hcl^\vee$. Given a world sheet $\Xrc\iN \Obj(\Worc)$, 
we denote by $F(\Xrc)$ the vector space of multilinear functions
  \be
  f :\quad \Hop^{|\text{o-in}|}
  \Times (\Hop^\vee)^{|\text{o-out}|} \Times \Hcl^{|\text{c-in}|}
  \Times (\Hcl^\vee)^{|\text{c-out}|} \longrightarrow \Cb
  \labl{eq:Fdef}
subject to a suitable boundedness condition; note that $F(\Xrc)$ does not 
depend on the metric. A collection of correlation functions assigns to every
$\Xrc\iN \Obj(\Worc)$ a multilinear function $\Cor_{\Xrc} \iN F(\Xrc)$, called 
the correlation function on the world sheet $\Xrc$. (The boundedness condition 
on $F(\Xrc)$ is imposed to ensure that $\Cor_{\Xrc}$ is a bounded multilinear 
function on the relevant product of state spaces.) Given a sewing $\sew$ of 
$\Xrc$, we introduce the operation of {\em partial evaluation\/} 
$F(\sew)$ on $F(\Xrc)$; it consists of evaluating, for each pair in $\sew$, 
the corresponding pair $\Hop$ and $\Hop^\vee$, respectively $\Hcl$ and 
$\Hcl^\vee$, for the in- and out-going state boundaries of that pair in $\sew$.

\medskip

To define an open/closed CFT, we now demand that the collection $\Cor$
of correlation functions possesses the following properties:
  \begin{itemize}
\item[C1 --] {\em Sewing\/}: $\Cor_{\sew(\Xrc)} \eq F(\sew)\big(\Cor_{\Xrc}\big)\,$ 
             for every sewing $\sew$ of a world sheet $\Xrc$.
\item[C2 --] {\em Isomorphism\/}: If two world sheets $\Xrc$ and $\Yrc$ are
             isomorphic, then $\,\Cor_{\Xrc} \eq \Cor_{\Yrc}$.
\item[C3 --] {\em Disjoint union\/}:
             $\Cor_{\Xrc \sqcup \Yrc} \eq \Cor_{\Xrc} \Cdot \Cor_{\Yrc}$.
\item[C4 --] {\em Weyl transformations\/}:
             If $\Xrc$ and ${\Xrc}'$ have the same underlying topological world
             sheet $\Xr$ and their metrics $g$ and $g'$ are conformally related, i.e.\
             obey $g'_p \eq {\rm e}^{\sigma(p)} g_p$ with some smooth function 
             $\sigma\colon\,\Xr \To \Rb$, then
             $\Cor_{\Xrc}\eq {\rm e}^{cS[\sigma]}\Cor_{{\Xrc}'}$,
             where $c$ is the conformal central charge and
             $S[\sigma]$ is the Liouville action (see \cite{gawe2,gawe13} for details).
  \end{itemize}

We will now argue that the conditions C1\,--\,C3 amount to requiring $\Cor$ to 
be a monoidal natural transformation between suitable functors, analogously
as imposing $\cor$ to be a monoidal natural transformation yields a solution
to the sewing constraints for correlators on topological world sheets.
We denote by $\Mult(\Hop,\Hcl)$ the category whose objects are spaces of 
multilinear maps $\Hop^m\Times(\Hop^\vee)^n\Times\Hcl^r\Times(\Hcl^\vee)^s 
\,{\longrightarrow}\, \Cb$ for $m,n,r,s \iN \Zb_{\ge 0}$ and whose morphisms
are suitable linear maps between these spaces, which include in particular
the partial evaluations.\,%
  \footnote{~For characterising the allowed maps one must also address some
  convergence issues. We refrain from going into any details in the present
  discussion.}
Via the product of multilinear functions (i.e., setting $(f{\cdot} g)(x,y) 
\,{:=}\, f(x)g(y)$), one turns $\Mult(\Hop,\Hcl)$ into a symmetric strict 
monoidal category. The tensor unit is given by the multilinear maps from zero 
copies of the spaces $\Hop,...\,,\Hcl^\vee$ to $\Cb$, which we identify with
the space $M_0$ of linear maps from $\Cb$ to $\Cb$.

The assignment $F$ defined through formula \erf{eq:Fdef} for world sheets 
and through partial evaluation for sewings becomes a strict 
monoidal functor from $\Worc$ to $\Mult(\Hop,\Hcl)$ by complementing 
it to act as 
  \be
  \Xrc \Mapsto F(\Xrc), \quad \sew \Mapsto F(\sew) \quad {\rm and}\quad 
  f \Mapsto F(f) := \id_{F(\Xrc)} 
  \labl{def:F}
for world sheets, sewings, and isomorphisms $f{:}\ \Xrc \To \Yrc$ of world 
sheets, respectively. Note that if there exists an isomorphism 
$f{:}\ \Xrc \To \Yrc$, then $F(\Xrc) \eq F(\Yrc)$.\,%
  \footnote{~In order for $F$ to be monoidal,
  for non-connected world sheets we must define $F(\Xrc)$ to be the tensor
  product (in the category $\Mult(\Hop,\Hcl)$)
  $F(\XrC_1) \oti \cdots \oti F(\XrC_m)$, with $\Xr_i$ the connected components
  of $\Xr$. In contrast, were we to take e.g.\ $F(\XrC_1{\sqcup}\XrC_2)$ to
  consist of all multilinear maps as well, then $F(\XrC_1)\oti F(\XrC_2)$ would
  typically only be a subspace of $F(\XrC_1{\sqcup}\XrC_2)$, e.g.\ not every
  bilinear function from $\Hop\Times\Hop$ to $\Cb$ can be written as a finite
  sum of products $f \Cdot g$ of linear functions $f$ and $g$
  from $\Hop$ to $\Cb$. }
We also need a `trivial' monoidal functor 
  \be
  \Triv :\quad \Worc \to \Mult(\Hop,\Hcl)
  \labl{def:Triv}
analogous to \erf{eq:triv-def}; we set $\Triv(\Xr)\,{:=}\,\id_\Cb \iN M_{0}$
and $\Triv(\sew) \,{:=}\,\id_{M_0}$ as well as $\Triv(f) \,{:=}\, \id_{M_0}$.

\medskip

The conditions C1\,--\,C3 are equivalent to the statement that $\Cor$ is a 
monoidal natural transformation from $\Triv$ to $F$.\label{def:Cor} Indeed, by 
definition, a natural transformation furnishes a family $\{\Cor_{\Xrc}\}$ of 
linear maps
  \be
  \Cor_{\Xrc}:\quad \Triv(\Xrc) \to F(\Xrc) 
  \labl{CorXr}
for $\Xrc \iN \Obj(\Worc)$, in a way consistent with composition. Thus for any
sewing $\sew{:}\ \Xrc \To \sew(\Xrc)$ and any isomorphism $f{:}\ \Xrc \To \Yrc$
the diagrams 
  \be
  \xymatrix@C=3.8em{
  \Triv(\Xrc) \ar[d]^{\Cor_{\Xrc}}
  \ar[r]^{\phantom{~~~~~~~}\Triv(\sew)\phantom{~~~~~~~~~}} &
    \Triv(\sew(\Xrc)) \ar[d]^{\Cor_{\sew(\Xrc)}}   \\
  F(\Xrc) \ar[r]^{F(\sew)} & F(\sew(\Xrc))  }
  \qquad\mbox{\raisebox{-1.8em}{and}}\qquad
  \xymatrix@C=3.8em{
  \Triv(\Xrc) \ar[d]^{\Cor_{\Xrc}}
    \ar[r]^{\phantom{~~~~~~~~~}\Triv(f)\phantom{~~~~~~~~~}} &
    \Triv(\Yrc) \ar[d]^{\Cor_{\Yrc}}   \\
  F(\Xrc) \ar[r]^{F(f)} & F(\Yrc)  }
  \ee
commute. The first diagram precisely amounts to condition C1, and the second
to C2. Finally, that $\Cor$ is monoidal is, by definition of the tensor product 
in the category $\Mult(\Hop,\Hcl)$, nothing but the statement of C3.
Thus it is indeed justified to interpret the linear maps \erf{CorXr} as
the correlation functions. 

\bigskip

As outlined in section 6.1, to find solutions to C1\,--\,C4 it is 
useful to first specify a minimal symmetry one desires the final 
theory to possess, and then analyse how that symmetry constrains the possible 
solutions to C1\,--\,C4. This approach effectively amounts to selecting a 
subspace $\Bl(\Xrc) \,{\subset}\, F(\Xrc)$ for each
$\Xrc \iN \Obj(\Worc)$. Afterwards one tries to find
a consistent set $\Cor_{\Xrc}$ of correlation functions in the restricted
spaces $\Bl(\Xrc)$. For suitable choices of
symmetry the spaces $\Bl(\Xrc)$ are finite-dimensional, so that
passing from $F$ to $\Bl$ is a significant simplification.
    
To proceed, one would now like to employ the natural transformation $\cor$ from 
$\triv$ to $\bl$ discussed in sections 
\ref{sec:oc-sewing}\,--\,\ref{sec:unique-proof} to obtain also a natural 
transformation $\Cor$ from $\Triv$ 
to $\Bl$ which, as just seen, is the same as giving a solution to C1\,--\,C3.
Condition C4 is then taken care of automatically by the fact that an element in
$\bl(\Xrc)$ corresponds to an appropriate section in the bundle over the moduli 
space of world sheets in $\Worc$ of the same topological type as 
$\Xrc$, whose fibers are given by $\Bl(\,\cdot\,)$.
How this will work out exactly is, however, still unclear, and thus it seems
fair to say that a precise and detailed understanding of the relation between 
the complex-analytic and the combinatorial part of the construction of RCFT
correlation functions is still lacking.


\vskip4em

\noindent {\bf Acknowledgements} \\[1pt]
We thank Terry Gannon, Liang Kong, Hendryk Pfeiffer, Karl-Henning Rehren and 
Gerard Watts for useful discussions, 
and Natalia Potylitsina-Kube for help with the illustrations. JFj received 
partial support from the EU RTN ``Quantum Spaces-Noncommutative Geometry",
JFu from VR under project no.\ 621--2003--2385, IR from the EPSRC First Grant 
EP/E005047/1 and the PPARC rolling grant PP/C507145/1, and
CS from the Collaborative Research Centre 676 ``Particles, Strings and the 
Early Universe - the Structure of Matter and Space-Time''. 

\newpage

 \def\Bi              {\bibitem}
 \newcommand\J[5]     {{\sl #5\/}, {#1} {#2} ({#3}) {#4} }
 \newcommand\K[6]     {{\sl #6\/}, {#1} {#2} ({#3}) {#4} \,{[#5]}}
 \newcommand\Prep[2]  {{\sl #2\/}, pre\-print {#1}}
 \newcommand\BOOK[4]  {{\sl #1\/} ({#2}, {#3} {#4})}
 \newcommand\inBO[7]  {{\sl #7\/}, in:\ {\sl #1}, {#2}\ ({#3}, {#4} {#5}), p.\ {#6}}
 \newcommand\inBOq[7] {{\sl #7\/}  in:\ {\sl #1} {#2}\ ({#3}, {#4} {#5}), p.\ {#6}}
 \newcommand\wb{\,\linebreak[0]} \def\wB {$\,$\wb}

\def\jf            {J.\ Fuchs}
\def\cfts          {conformal field theories}
\def\q             {quantum }
\def\Q             {Quantum }
\def\twodim        {two-di\-men\-sio\-nal}

 \def\adma  {Adv.\wb Math.}
 \def\anma  {Ann.\wb Math.}
 \def\anop  {Ann.\wb Phys.}
 \def\aspm  {Adv.\wb Stu\-dies\wB in\wB Pure\wB Math.}
 \def\aste  {Ast\'e\-ris\-que}
 \def\cocm  {Com\-mun.\wb Con\-temp.\wb Math.}
 \def\coma  {Con\-temp.\wb Math.}
 \def\comp  {Com\-mun.\wb Math.\wb Phys.}
 \def\cpma  {Com\-pos.\wb Math.}
 \newcommand\dgmd[2] {\inBO{{\rm Proceedings of the} XXth International Conference on
           Differential Geometric Methods in Theoretical Physics} {S.\ Catto and A.\ Rocha,
            eds.} \WS\Si{1991} {{#1}}{{#2}}}
 \def\fiic  {Fields\wB Institute\wB Commun.}
 \def\foph  {Fortschritte\wB d.\wb Phys.}
 \def\ijmp  {Int.\wb J.\wb Mod.\wb Phys.\ A}
 \def\ihes  {Publ.\wb Math.\wB I.H.E.S.}
 \def\inma  {Invent.\wb math.}
 \def\jgap  {J.\wb Geom.\wB and\wB Phys.}
 \def\jhep  {J.\wb High\wB Energy\wB Phys.}
 \def\jktr  {J.\wB Knot\wB Theory\wB and\wB its\wB Ramif.}
 \def\jlms  {J.\wB London\wB Math.\wb Soc.}
 \def\joal  {J.\wB Al\-ge\-bra}
 \def\jomp  {J.\wb Math.\wb Phys.}
 \def\jopa  {J.\wb Phys.\ A}
 \def\josp  {J.\wb Stat.\wb Phys.}
 \def\jpaa  {J.\wB Pure\wB Appl.\wb Alg.}
 \def\lemp  {Lett.\wb Math.\wb Phys.}
 \def\maan  {Math.\wb Annal.}
 \def\mams  {Memoirs\wB Amer.\wb Math.\wb Soc.}
 \def\mpla  {Mod.\wb Phys.\wb Lett.\ A}
 \newcommand\ncmp[2] {\inBO{IXth International Congress on
            Mathematical Physics} {B.\ Simon, A.\ Truman, and I.M.\ Davis,
            eds.} {Adam Hilger}{Bristol}{1989} {{#1}}{{#2}} } 
 \def\nuci  {Nuovo\wB Cim.}
 \def\nupb  {Nucl.\wb Phys.\ B} 
 \def\phlb  {Phys.\wb Lett.\ B}
 \newcommand\phgt[2] {\inBO{Physics, Geometry, and Topology}
            {H.C.\ Lee, ed.} \PL\NY{1990} {{#1}}{{#2}} }
 \def\phrd  {Phys.\wb Rev.\ D}
 \def\phrl  {Phys.\wb Rev.\wb Lett.}
 \def\phrp  {Phys.\wb Rep.}
 \def\pnas  {Proc.\wb Natl.\wb Acad.\wb Sci.\wb USA}
 \def\ptrs  {Phil.\wb Trans.\wb Roy.\wb Soc.\wB Lon\-don}
 \newcommand\qfsm[2] {\inBO{\Q Fields and Strings: A Course for Mathematicians} 
            {P.\ Deligne et al., eds.} \AMS\PR{1999} {{#1}}{{#2}}}
 \def\rvmp  {Rev.\wb Math.\wb Phys.}
 \def\taac  {Theory\wB and\wB Appl.\wB Cat.}
 \def\tams  {Trans.\wb Amer.\wb Math.\wb Soc.}
 \newcommand\tgqf [2] {\inBO{Topology, Geometry and Quantum Field Theory
            {\rm [London Math.\ Soc.\ Lecture Note Series \# 308]}}
            {U.\ Tillmann, ed.} \CUP\Ca{2004} {{#1}}{{#2}}}
 \newcommand\tgqfq [2] {\inBOq{Topology, Geometry and Quantum Field Theory
            {\rm [London Math.\ Soc.\ Lecture Note Series \# 308]}}
            {U.\ Tillmann, ed.} \CUP\Ca{2004} {{#1}}{{#2}}}
 \def\toap  {Topology\wB Applic.}
 \def\trgr  {Trans\-form.\wB Groups}
 \def\AMS    {{American Mathematical Society}}
 \def\AP     {{Academic Press}}
 \def\BIR    {{Birk\-h\"au\-ser}}
 \def\CUP    {{Cambridge University Press}}
 \def\EMS    {{European Mathematical Society}}
 \def\IPC    {{International Press Company}}
 \def\NH     {{North Holland Publishing Company}}
 \def\KLU    {{Kluwer Academic Publishers}}
 \def\PUP    {{Princeton University Press}}
 \def\PL     {{Plenum Press}}
 \def\SV     {{Sprin\-ger Ver\-lag}}
 \def\WI     {{Wiley Interscience}}
 \def\WS     {{World Scientific}}
 \def\Ad     {{Amsterdam}}
 \def\Be     {{Berlin}}
 \def\Bo     {{Boston}}
 \def\Ca     {{Cambridge}}
 \def\Do     {{Dordrecht}}
 \def\pR     {{Princeton}}
 \def\PR     {{Providence}}
 \def\Si     {{Singapore}}
 \def\NY     {{New York}}

\end{document}